\documentclass[aps,floats,amssymb,amsmath,prb,twocolumn,superscriptaddress]{revtex4-1}
\usepackage{calc}
\usepackage{psfrag}
\usepackage{graphicx}
\usepackage{color}
\usepackage[dvipsnames]{xcolor}
\usepackage[unicode=true,pdfusetitle,
 bookmarks=true,bookmarksnumbered=false,bookmarksopen=false,
 breaklinks=false,pdfborder={0 0 0},backref=false,colorlinks=true]
 {hyperref}

\pdfpageattr {/Group << /S /Transparency /I true /CS /DeviceRGB>>} %necessary for transparent plots

\newcommand{\up}{\uparrow}
\newcommand{\dn}{\downarrow}

\newcommand{\bPsi}{\boldsymbol{\Psi}}

\begin{document}

\title{TRILEX and $GW$+EDMFT approach to $d$-wave superconductivity in the Hubbard model}
\author{J. Vu\v ci\v cevi\'c}
\affiliation{Institut de Physique Th\'eorique (IPhT), CEA, CNRS, UMR 3681, 91191 Gif-sur-Yvette, France}
\affiliation{Scientific Computing Laboratory, Center for the Study of Complex Systems, Institute of Physics Belgrade,
University of Belgrade, Pregrevica 118, 11080 Belgrade, Serbia}
\author{T. Ayral}
\affiliation{Department of Physics and Astronomy, Rutgers University, Piscataway, NJ 08854, USA}
\affiliation{Institut de Physique Th\'eorique (IPhT), CEA, CNRS, UMR 3681, 91191 Gif-sur-Yvette, France}
\author{O. Parcollet}
\affiliation{Institut de Physique Th\'eorique (IPhT), CEA, CNRS, UMR 3681, 91191 Gif-sur-Yvette, France}

\begin{abstract}
We generalize the recently introduced TRILEX approach (TRiply Irreducible Local EXpansion) to superconducting phases. 
The method treats simultaneously Mott and spin-fluctuation physics
using an Eliashberg theory supplemented by local vertex corrections determined by a self-consistent quantum impurity model. 
We show that, in the two-dimensional Hubbard model, at strong coupling, 
TRILEX yields a $d$-wave superconducting dome as a function of doping.
Contrary to the standard cluster dynamical mean field theory (DMFT) approaches, TRILEX can
capture $d$-wave pairing using only a single-site effective impurity model.
We also systematically explore the dependence of the superconducting temperature on the bare dispersion at weak coupling,
which shows a clear link between strong antiferromagnetic (AF) correlations and the onset of superconductivity.
We identify a combination of hopping amplitudes particularly favorable to
superconductivity at intermediate doping.
Finally, we study within $GW$+EDMFT the low-temperature $d$-wave superconducting phase at strong coupling in a region of parameter space 
with reduced AF fluctuations.
\end{abstract}

\maketitle

Strongly-correlated electrons systems such as high-temperature superconductors pose a difficult challenge to condensed-matter theory.

One class of theoretical approaches for this problem 
focuses on the effect of long-range spin-fluctuations\cite{Chubukov2002,Efetov2013,Wang2014,Metlitski2010,Onufrieva2009,Onufrieva2012}.
They neglect vertex corrections
in an Eliashberg-like approximation for the electronic self-energy and predict a $d$-wave superconducting order.

Another class of approaches focuses, following the seminal work of
Anderson\cite{Anderson1987}, on the fact that high-temperature superconductors are doped Mott insulators.
In the recent years, progress has been made in this direction
with cluster extensions\cite{Hettler1998,Hettler1999,Lichtenstein2000,Kotliar2001,Maier2005a} of dynamical mean field theory (DMFT) \cite{Georges1996}.
These methods have been shown to capture the essential aspects
of cuprate physics, such as Mott insulating, pseudogap and
$d$-wave superconducting phases\cite{Kyung2009,Sordi2012,Civelli2008,Ferrero2010,Gull2013,Macridin2004,Maier2004,Maier2005,Maier2006,Gull2010,Yang2011,Macridin2008,Macridin2006,Jarrell2001,Bergeron2011,Kyung2004,Kyung2006a,Okamoto2010,Sordi2010,Sordi2012a,Civelli2005,Parcollet2004, Ferrero2008,Ferrero2009,Gull2009,Chen2015,Chen2016}.
Cluster DMFT methods can be converged with respect to the cluster size
at relatively high temperature\cite{Leblanc2015, Gunnarsson2008}, 
including in the pseudogap region \cite{Wu2016},
but not at lower temperatures and in particular in the superconducting phase.

Several approaches beyond cluster DMFT have been proposed recently
\cite{Rubtsov2008,Rubtsov2011,VanLoon2014,Stepanov2015, Toschi2007,Katanin2009,Schafer2014,Valli2014,Li2015a,Rohringer2016, Ayral2016, Biermann2003,Sun2002,Sun2004,Ayral2012,Ayral2013,Biermann2014,Ayral2017,Rohringer2017}.
In Refs~\onlinecite{Ayral2015, Ayral2015c}, the Triply Irreducible Local Expansion (TRILEX) approach
was introduced. It consists in a local approximation of the electron-boson
vertex extracted from a quantum impurity model
with a self-consistently determined bath and interaction,
in the spirit of DMFT.
TRILEX interpolates between
DMFT at strong interaction and
the weak-coupling Eliashberg-like spin-fluctuation approximation
at weak interaction. 
It is able to simultaneously describe Mott physics
and the effect of long-range bosonic fluctuations.
Hence, it unifies the two theoretical approaches mentioned above in the same formalism.

The main purpose of this paper is to study $d$-wave 
superconductivity in
the Hubbard model within the single-site TRILEX approach.
Contrary to DMFT, where $d$-wave superconducting correlations can by
construction be captured only within multi-site (cluster) impurity models,
here we only need to solve a \emph{single-site} impurity model.
We also compare TRILEX to two simpler approaches, $GW$+EDMFT and $GW$, which can be viewed as further approximations
of the electron-boson vertex in TRILEX.
We show that TRILEX yields a $d$-wave superconducting dome
at strong coupling.

We also study the dependence of the superconducting critical temperature $T_c$ on the choice
of the tight-binding parameters at weak coupling using the $GW$ method.
While $T_c$ is enhanced by strong antiferromagnetic fluctuations, 
we find a region of parameter space where the superconducting transition occurs
at a higher temperature than the antiferromagnetic instability of the method.
At this point, we stabilize and study a superconducting solution below $T_c$ within $GW$+EDMFT.
We also identify a choice of dispersion where,
at $16\%$ doping, we have a pronounced maximum of $T_c$ in the space of hopping parameters, 
which seems to persist even at strong coupling.

The paper is organized as follows:
In Section \ref{sec:model}, we describe the Hubbard model studied in this paper.
In Section \ref{sec:formalism}, we generalize the TRILEX equations to superconducting phases via the Nambu formalism, and 
discuss their simplifications $GW$ and $GW$+EDMFT.
In Section \ref{sec:methods}, we describe the numerical methods and details used to solve the equations.
In Section \ref{sec:results}, we turn to the results. We first describe the phase diagram (subsection \ref{sec:phase_diagram}) within TRILEX and $GW$+EDMFT, and then focus  on the weak-coupling regime
(subsection \ref{sec:weak_coupling}) 
where, using the $GW$ method, we scan the space of the nearest and next-nearest-neighbor hopping parameters in search of dispersions with a weak antiferromagnetic instability where it is possible to reach a paramagnetic superconducting phase.
The two dispersions which we thus identify are investigated in more detail at strong coupling with $GW$+EDMFT in subsections \ref{sec:sc_phase} and \ref{sec:ptA}.

\section{Model} \label{sec:model}

\begin{figure}[!ht]
\centering{}\includegraphics[width=2.8in, trim=2.0cm 1.0cm 2.0cm 0.0cm]{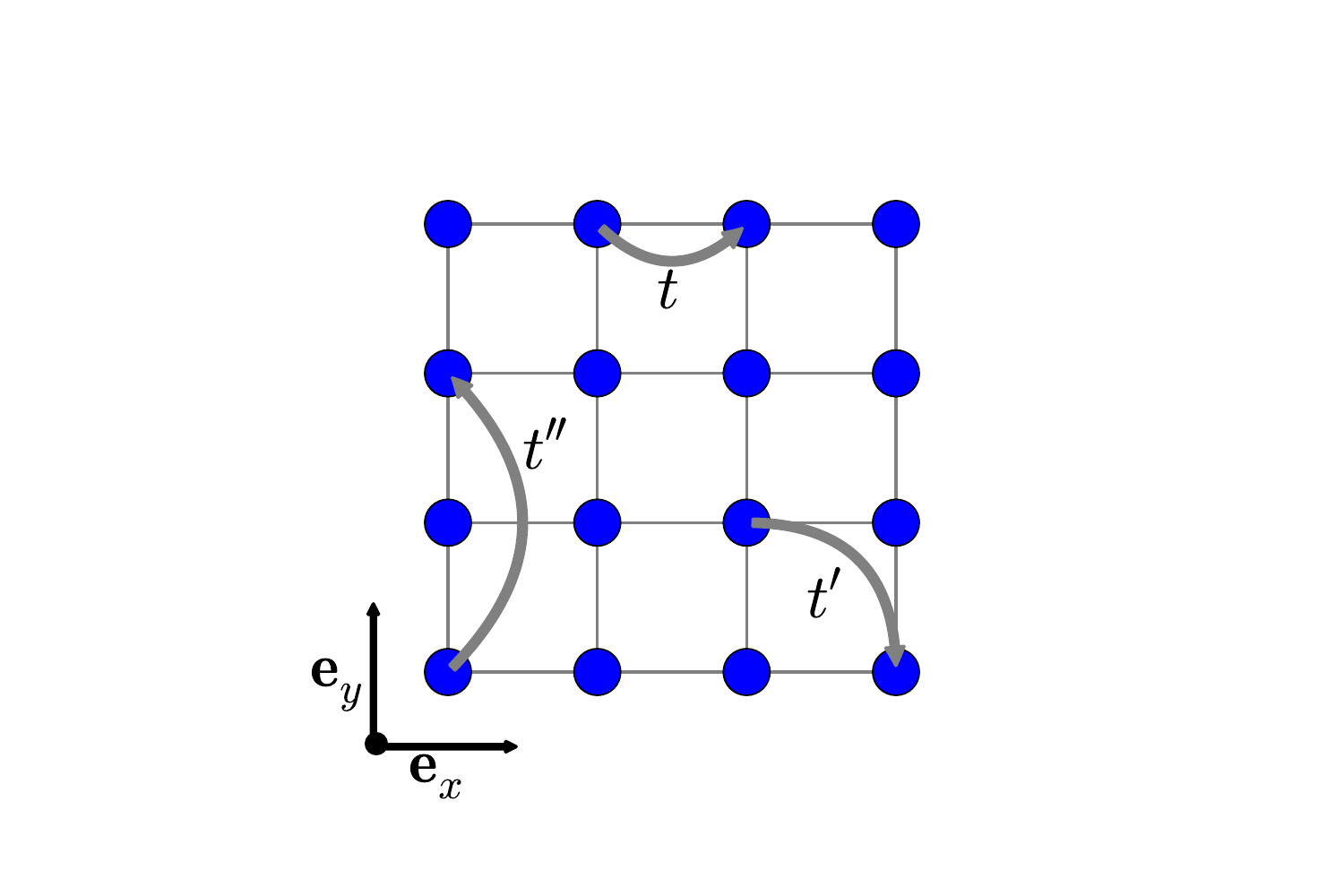}
\caption{Definition of the tight-binding parameters on the square lattice.\label{fig:hoppings} }
\end{figure}

We solve the Hubbard model on the square lattice with longer range
hoppings, defined by the Hamiltonian:
\begin{equation}
H=\sum_{ij\sigma}t_{ij}c_{i\sigma}^{\dagger}c_{j\sigma}-\mu\sum_{i\sigma}n_{i\sigma}+U\sum_{i}n_{i\uparrow}n_{i\downarrow}\label{eq:Hubbard}
\end{equation}
with $i,j$ indexing lattice sites. $c_{\sigma i}^{\dagger}$($c_{\sigma i}$) denote creation (annihilation)
operators, $n_{\sigma i}=c_{\sigma i}^{\dagger}c_{\sigma i}$ the
density operator, $\mu$ the chemical potential, and $U$ the on-site
Hubbard interaction. The hopping amplitudes, depicted on Fig.~\ref{fig:hoppings}, are given by 
\begin{equation}
t_{ij}=\left\{ \begin{array}{cc}
t, & \mathbf{r}_{i}=\mathbf{r}_{j}\pm\mathbf{e}_{x,y}\\
t', & \mathbf{r}_{i}=\mathbf{r}_{j}\pm\mathbf{e}_{x}\pm\mathbf{e}_{y}\\
t'', & \mathbf{r}_{i}=\mathbf{r}_{j}\pm2\mathbf{e}_{x,y}\\
0, & \mathrm{otherwise.}
\end{array}\right.
\end{equation}
where $\mathbf{e}_{x,y}$ are the lattice vectors in the $x$ and
$y$ directions. The bare dispersion is therefore 
\begin{align}
\varepsilon_{\mathbf{k}} & =2t(\cos k_{x}+\cos k_{y})+4t'\cos k_{x}\cos k_{y}\nonumber \\
 & \;\;+2t''(\cos2k_{x}+\cos2k_{y})\label{eq:bare_dispersion}
\end{align}
When $t'=t''=0$, the half-bandwidth is $D=4|t|$, but non zero $t',t''$
in general make the bandwidth larger. Hereinafter, we express all
quantities in units of $D$, unless stated differently.

%%%%%%%%%%%%%%%%%%%%%%%%%%%%%%%%%%%%%%%%%%%%%%%%%%%%%%%%%%%%%%%%%%%%%%%%%%%%%%%%%%%%%%%%
%%%%%%%%%%%%%%%%%%%%%%%%%%%%%%%%%%%%%%%%%%%%%%%%%%%%%%%%%%%%%%%%%%%%%%%%%%%%%%%%%%%%%%%%
%%%%%%%%%%%%%%%%%%%%%%%%%%%%%%%%%%%%%%%%%%%%%%%%%%%%%%%%%%%%%%%%%%%%%%%%%%%%%%%%%%%%%%%%

\section{Formalism} \label{sec:formalism}

The main goal of this paper is to study the superconducting (SC) phase of the two-dimensional Hubbard model
within the TRILEX approach introduced in Refs.~\onlinecite{Ayral2015} and \onlinecite{Ayral2015c}.
TRILEX is based on a bosonic decoupling of the interaction and a self-consistent
approximation of the electron-boson vertex $\Lambda$ with a quantum
impurity model. The decoupling of the on-site interaction is done
by an exact Hubbard-Stratonovich transformation, leading to a model of non-interacting
electrons coupled to some auxiliary bosonic modes representing charge
and spin fluctuations.

We also study two methods which can be regarded as simplifications of the TRILEX method, namely $GW$+EDMFT\cite{Sun2002,Biermann2003,Sun2004,Ayral2012,Ayral2013,Biermann2014} and $GW$\cite{Hedin1965,Hedin1999}.
In $GW$+EDMFT, vertex corrections are neglected in the non-local part of the self-energy and polarization.
As both decay to zero, this additional approximation is negligible at very long distances.
Due to the full treatment of
the local vertex corrections, $GW$+EDMFT can capture the Mott transition, and we use it to obtain superconducting results
in the doped Mott insulator regime.
In the $GW$ method, vertex corrections are neglected altogether, and the self-energy and polarization are entirely calculated
from bold bubble diagrams. The $GW$ equations do not require the solution of an auxiliary quantum impurity model,
and are therefore less costly to solve. This additional approximation is justified only at weak coupling (see e.g. Ref.~\onlinecite{Ayral2012} for an illustration of its failure at large $U$),
and there we use it to explore a large region of $(t',t'',T,n_\sigma)$ parameter space ($T$ denotes temperature, $n_\sigma$ occupancy per spin).

Finally, let us stress that, in this paper, we use only
\textit{single-site} impurity models. 
Cluster extensions of TRILEX are discussed in our different work, Ref.\onlinecite{Ayral2017c}. They naturally incorporate
the effect of short-range antiferromagnetic exchange $J$ and give a quantitative control on the accuracy of the solution.

\subsection{Superconducting Hedin equations}
In this section, we derive the Hedin equations\cite{Hedin1965,Hedin1999,Aryasetiawan2008} which give the self-energy and polarization as functions of the three-leg vertex function. The derivation holds in superconducting phases and is relevant for fluctuations not only in the charge channel\cite{Linscheid2015}, but also in the longitudinal and transversal spin channels.

\subsubsection{The electron boson action}

The starting point of the TRILEX method, as described in Ref.~\onlinecite{Ayral2015c}, is
the following electron-boson action: 
\begin{align}
  S_{\mathrm{eb}}[c,c^*,\phi] & =c^{*}_\mu\left[-G_{0}^{-1}\right]_{\mu\nu}c_\nu+\frac{1}{2}\phi_{\alpha}\left[-W_{0}^{-1}\right]_{\alpha\beta}\phi_{\beta}\label{eq:eb_action}\\
 & \;+\lambda_{\mu\nu\alpha}c^{*}_\mu c_\nu \phi_{\alpha}\nonumber 
\end{align}
where $c^{*}_{\mu}$ and $c_{\nu}$ are Grassmann fields describing fermionic
degrees of freedom, while $\phi_{\alpha}$ is a real bosonic field
describing bosonic degrees of freedom. Indices $\mu,\nu$ stand for space,
time, spin, and possibly other (e.g. band) indices $\mu\equiv(\mathbf{r}_{\mu},\tau_{\mu},\sigma_{\mu},\dots)$,
where $\mathbf{r}_{\mu}$ denotes a site of the Bravais lattice, $\tau_{\mu}$
denotes imaginary time and $\sigma_{\mu}$ is a spin index ($\sigma_{\mu}\in\{\uparrow,\downarrow\}$).
Indices $\alpha,\beta$ denote $\alpha\equiv(\mathbf{r}_{\alpha},\tau_{\alpha},I_{\alpha},\dots)$,
where $I_{\alpha}$ indexes the bosonic channels. 
Repeated indices are summed over. Summation $\sum_{\mu}$ is shorthand for $\sum_{\mathbf{r}\in\mathrm{BL}}\sum_{\sigma}\int_{0}^{\beta}\mathrm{d}\tau$.
$G_{0,\mu\nu}$ (resp. $W_{0,\alpha\beta}$) is the non-interacting
fermionic (resp. bosonic) propagator. 

Action (\ref{eq:eb_action}) can result from the exact Hubbard-Stratonovich
decoupling of the Hubbard interaction of Eq. (\ref{eq:Hubbard}) with
bosonic fields $\phi$, but it can also
simply describe an electron-phonon coupling problem.

In the present work, we are interested in a generalization of TRILEX able to accommodate superconducting order. To this purpose, we rederive the TRILEX equations starting from a more general action, written in terms of Nambu four-component spinors. The departure from the usual two-component Nambu-spinor formalism is necessary to allow for spin-flip electron-boson coupling in the action. Such terms do appear in the Heisenberg decoupling of the Hubbard interaction (see Section \ref{subsec:hubbard_model}).

We define a four-component Nambu-Grassmann spinor field as a column-vector
\begin{equation}\label{eq:Nambu_spinor}
 \boldsymbol{\Psi}_{i}(\tau)\equiv
 \left[
 \begin{array}{c}
  c_{\up i}^*(\tau) \\ c_{\dn i}(\tau) \\ c_{\dn i}^*(\tau) \\ c_{\up i}(\tau)
 \end{array}
 \right]
\end{equation}
where $i$ stands for the lattice-site $\mathbf{r}_i$.
In combined indices, analogously to \eqref{eq:eb_action}, a general electron-boson action can be written as
\begin{align}
 S^\mathrm{Nambu}_{\mathrm{eb}}[\bPsi,\phi] & =\frac{1}{2}\bPsi_{u}\left[-\boldsymbol{G}_{0}^{-1}\right]_{uv}\bPsi_{v}-\frac{1}{2}\phi_{\alpha}\left[W_{0}^{-1}\right]_{\alpha\beta}\phi_{\beta}\nonumber \\
  & \;\;+\frac{1}{2}\phi_{\alpha}\bPsi_{u}\boldsymbol{\lambda}_{uv\alpha}\bPsi_{v}\label{eq:S_eb}
\end{align}
where $u,v$ is a combined index $u\equiv(\mathbf{r}_{u},\tau_{u}, a_{u},\dots)$, with $a,b,c,... \in \{0,1,2,3\}$ a Nambu index comprising the spin degree of freedom. The sum is redefined to go over all Nambu indices $\sum_{u} \equiv \sum_{\mathbf{r}\in\mathrm{BL}}\sum_{a}\int_{0}^{\beta}\mathrm{d}\tau$. Bold symbols are used for Nambu-index-dependent quantities.

This action does {\it not} depend on the conjugate field of $\bPsi$, because $\bPsi_i$ already contains all the degrees of freedom of the action \eqref{eq:eb_action} at the site $i$. 
The partition function corresponding to the bare fermionic part of the action has the following form\cite{Zinn-Justin2002}
\begin{equation}
  \int {\cal D}[\boldsymbol \Psi] e^{\frac{1}{2} \boldsymbol{\Psi}_u \boldsymbol{A}_{uv} \boldsymbol{\Psi}_v} = \left(\det \boldsymbol{A}\right)^\frac{1}{2}
\end{equation}
which is valid for any anti-symmetric matrix $\boldsymbol{A}$. Due to the unusual form of the action (no conjugated fields), the right-hand side is not the determinant of $\boldsymbol{A}$, but its square root, i.e. the Pfaffian.
%In this functional formalism there is no distinction between a particle and an anti-particle, hence the term ``Majorana''.
We can redefine the propagators/correlation functions of interest as
\begin{align}
\boldsymbol{G}_{uv} & \equiv - \Big\langle\boldsymbol{\Psi}_u\boldsymbol{\Psi}_v\Big\rangle \\
W_{\alpha\beta} & \equiv -\langle(\phi_{\alpha}-\langle\phi_{\alpha}\rangle)(\phi_{\beta}-\langle\phi_{\beta}\rangle)\rangle,\label{eq:W_def_generic}\\
\boldsymbol{\chi}^{3,\mathrm{conn}}_{uv\alpha} & \equiv \Big\langle\bPsi_u\bPsi_v\phi_\alpha\Big\rangle  -\Big\langle\bPsi_u\bPsi_v\Big\rangle\Big\langle\phi_\alpha\Big\rangle \label{eq:chi3conn_def}
\end{align}
The ``conn'' superscript denotes the connected part of the correlation function.
The renormalized vertex is defined by
\begin{equation}
\boldsymbol{\Lambda}_{uv\alpha} \equiv
\left[\boldsymbol{G}^{-1}\right]_{uw}\left[\boldsymbol{G}^{-1}\right]_{xv}\left[W^{-1}\right]_{\alpha\beta} \boldsymbol{\chi}^{3,\mathrm{conn}}_{wx\beta} \label{eq:Lambda_def}
\end{equation}

Actions \eqref{eq:S_eb} and \eqref{eq:eb_action} are physically equivalent, namely their partition functions coincide:
\begin{equation}
  Z = \int {\cal D}[\boldsymbol{\Psi}, \phi] e^{-S_{\mathrm{eb}}^{\mathrm{Nambu}}[\boldsymbol{\Psi},\phi]}=  \int {\cal D}[c, c^*, \phi] e^{-S_{\mathrm{eb}}[c, c^*,\phi]}\label{eq:Z_with_eb_Nambu}
\end{equation}
for an appropriate choice of $\boldsymbol{G}_0$ and $\boldsymbol{\lambda}$. Yet, they are not formally identical to each other, i.e. one cannot reconstruct \eqref{eq:S_eb} from \eqref{eq:eb_action} by mere relabeling $c\rightarrow\bPsi$, $\mu\nu\rightarrow uv$ (note the absence of Grassmann conjugation and the additional prefactors in the Nambu action). 
Therefore, one must rederive the Hedin equations which connect the self-energy and polarization with the full propagators $\boldsymbol{G}$ and $W$ and the renormalized vertex $\boldsymbol{\Lambda}$. We present the full derivation using equations of motion in Appendix~\ref{sec:eom}; here we just present the final result:
\begin{subequations}\label{eq:Hedin_with_indices}
\begin{eqnarray}\label{eq:Hedin_Sigma}
\boldsymbol{\Sigma}_{uv}&=& -\boldsymbol{\lambda}_{uw\alpha}\boldsymbol{G}_{wx}W_{\alpha\beta}\boldsymbol{\Lambda}_{xv\beta} \\ \nonumber
                        &&  +\frac{1}{2}\boldsymbol{\lambda}_{uv\alpha}W_{0,\alpha\beta}\langle\bPsi_{y}\boldsymbol{\lambda}_{yz\beta}\bPsi_{z}\rangle \\
\label{eq:Hedin_P} 
P_{\alpha\beta}&=&\frac{1}{2}\boldsymbol{\lambda}_{uw,\alpha}\boldsymbol{G}_{xu}\boldsymbol{G}_{wv}\boldsymbol{\Lambda}_{vx,\beta}
\end{eqnarray}
\end{subequations}
Compared to the expressions in the normal case, there are extra factors $\frac{1}{2}$ in the Hartree term (second line in \eqref{eq:Hedin_Sigma}) and polarization \eqref{eq:Hedin_P}. These factors come from the fact that with four-spinors, the summation over spin is performed twice. 
Note that the Hartree term can in principle have a frequency dependence if the bare electron-boson vertex has a dynamic part. On the other hand, the term beyond Hartree may as well contribute to the static part of the self-energy, if the bosonic propagator and the bare electron-boson vertex contain a static part. In all the calculations in this paper, the Hartree term is static and is the sole contributor the static part of self-energy. We will thus henceforth omit the Hartree term, as it can be absorbed in the chemical potential.

\subsubsection{Connection to the Hubbard model\label{subsec:hubbard_model}}

In this section, we specify the bare propagators and vertices such that action \eqref{eq:S_eb} corresponds to the Hubbard model Eq.(\ref{eq:Hubbard}). We then rewrite the Hedin equations under the assumption of spatial and temporal translational symmetry.

The Hubbard-Stratonovich transformation leading from Eq.(\ref{eq:Hubbard})
to an action of the form Eq.(\ref{eq:eb_action}) relies on decomposing the Hubbard interaction
as follows
\begin{equation}
Un_{i\uparrow}n_{i\downarrow}=\frac{1}{2}\sum_{I}U^{I}n_{i}^{I}n_{i}^{I}\label{eq:Fierz_rewriting}
\end{equation}
with $n_{I}\equiv\sum_{\sigma\sigma'}c_{\sigma}^{\dagger}\sigma_{\sigma\sigma'}^{I}c_{\sigma'}$,
and $I$ running within $\{0,z\}$ (``Ising decoupling'') or $\{0,x,y,z\}$
(``Heisenberg decoupling'') ($\sigma^{0}$ is the $2\times2$ identity
matrix, $\sigma^{x/y/z}$ are the usual Pauli matrices). This identity
is verified, up to a density term, whenever\begin{subequations}
\begin{align}
U^{\mathrm{ch}}-U^{\mathrm{sp}} & =U\label{eq:Fierz_Ising}
\end{align}
in the Ising decoupling, or
\begin{align}
U^{\mathrm{ch}}-3U^{\mathrm{sp}} & =U\label{eq:Fierz_Heisenberg}
\end{align}
\end{subequations} in the Heisenberg decoupling. We have defined
$U^{\mathrm{ch}}\equiv U^{0}$ and $U^{\mathrm{sp}}\equiv U^{x}=U^{y}=U^{z}$.
Eqs (\ref{eq:Fierz_Ising}-\ref{eq:Fierz_Heisenberg}) leave a degree
of freedom in the choice of $U^{\mathrm{ch}}$ and $U^{\mathrm{sp}}$.
Here, the choice $U^{x}=U^{y}=U^{z}$ stems from the isotropy of the Heisenberg decoupling (contrary to the Ising decoupling); it can describe SU(2) symmetry-broken phases.
In the rest of the paper, we denote all quantities diagonal in the channel index with the channel as a superscript.

To make contact with the results of Ref.~\onlinecite{AokiPRB2015}, for $GW$ we will  use the Ising decoupling with\begin{subequations}
\begin{align}
 U^\mathrm{ch}  = U/2, \;\;\; U^\mathrm{sp} =  -U/2
\end{align}
while in TRILEX and $GW$+EDMFT (unless stated differently) we will use the Heisenberg decoupling with
\begin{align}
 U^\mathrm{ch} = U/2, \;\;\;
 U^\mathrm{sp} = -U/6.
\end{align}
\end{subequations}
because the AF instabilities discussed in Sec.~\ref{sec:AF_instability}, which violate the Mermin-Wagner theorem, are weaker in this scheme.

The equivalence of the action \eqref{eq:S_eb} with the Hubbard model is accomplished by setting 
\begin{subequations}
\begin{equation}\label{eq:G0_matrix}
 \boldsymbol{G}_{0,ij}(\tau) = \left[ \begin{array}{cccc}
          0 & 0 & 0 & -G_{0,ji}(-\tau)\\
          0 & 0 & G_{0,ij}(\tau) & 0 \\
          0 & -G_{0,ji}(-\tau) & 0 & 0\\          
          G_{0,ij}(\tau) & 0 & 0 & 0\\
        \end{array}\right]
\end{equation}
where $i,j$ denote lattice sites, and
\begin{align} \nonumber
G_{0,ij}(\tau) & = \sum_{i\omega,\mathbf{k}} e^{-i(\omega\tau-(\mathbf{r}_i-\mathbf{r}_j)\cdot\mathbf{k})}G_{0\mathbf{k}}(i\omega) \\ 
G_{0\mathbf{k}}(i\omega) & =\frac{1}{i\omega+\mu-\varepsilon_{\mathbf{k}}}\label{eq:G0_def}
\end{align}
\end{subequations}

The $4\times 4$ matrices are written in Nambu indices. 
The bare vertex reads:\begin{subequations}
\begin{eqnarray}
 \boldsymbol{\lambda}_{uv\alpha} &=& \delta_{\mathbf{r}_u\mathbf{r}_\alpha}\delta_{\mathbf{r}_u\mathbf{r}_v}\delta_{\tau_u\tau_\alpha} 
 [ \boldsymbol{\delta}_{\tau_u,\tau_v} \cdot \boldsymbol{\lambda}^{I_\alpha} ]_{a_ua_v}
\end{eqnarray}
with
\begin{eqnarray}
 \boldsymbol{\delta}_{\tau_u,\tau_v} =
       \left[ \begin{array}{cccc}
           \delta_{\tau_u,\tau_v^+}&&& \\
           &\delta_{\tau_u^+,\tau_v}&& \\
           &&\delta_{\tau_u,\tau_v^+}& \\
           &&&\delta_{\tau_u^+,\tau_v} 
           \end{array}\right] 
\end{eqnarray}
and:
\begin{equation}
\boldsymbol{\lambda}^{I}=\left[\begin{array}{cccc}
     & \sigma_{\uparrow\downarrow}^{I} &  & \sigma_{\uparrow\uparrow}^{I}\\
     -\sigma_{\uparrow\downarrow}^{I} &  & -\sigma_{\downarrow\downarrow}^{I}\\
      & \sigma_{\downarrow\downarrow}^{I} &  & \sigma_{\downarrow\uparrow}^{I}\\
      -\sigma_{\uparrow\uparrow}^{I} &  & -\sigma_{\downarrow\uparrow}^{I}
  \end{array}\right]
\end{equation}
\end{subequations}
Thus, this vertex is local and static. The bare bosonic propagators are also local and static, as well as diagonal in the channel index:
\begin{equation}
W^I_{0,ij}(\tau) = \delta_{ij}\delta_\tau U^I
\end{equation}
Our Hubbard lattice Nambu action reads (in explicit indices)
\begin{eqnarray} \label{eq:latt_eb_action_Psi}
 &  & S_{\mathrm{eb}}^{\mathrm{Nambu}}[\boldsymbol{\Psi},\phi]=\nonumber \\
 &  & \frac{1}{2} \sum_{i,j,a,b}\iint\mathrm{d}\tau\mathrm{d}\tau'
                  \boldsymbol{\Psi}_{ia}(\tau)
                  [-\boldsymbol{G}_{0}^{-1}]_{ia,jb}(\tau-\tau')
                  \boldsymbol{\Psi}_{jb}(\tau')\nonumber \\
 &  & +\frac{1}{2}\sum_{i}\sum_{I}\int\mathrm{d}\tau\phi_{i}^{I}(\tau)[-(U^I)^{-1}]\phi_{i}^{I}(\tau)\\ \nonumber 
 &  & +\frac{1}{2}\sum_{i}\sum_{I}\int\mathrm{d}\tau\phi_{i}^{I}(\tau)\boldsymbol{\Psi}_{ia}(\tau)\boldsymbol{\lambda}_{ab}^{I}\boldsymbol{\Psi}_{ib}(\tau)
\end{eqnarray}

\subsubsection{Translational invariance, singlet pairing and SU(2) symmetry \label{subsec:trilex_dwave}}

In this paper, we restrict ourselves to phases with no breaking of translational invariance.
With translational invariance in time and space, the propagators depend on frequency and momentum, and are matrices only in the Nambu index. We rewrite the Hedin equations derived above in the special case of the Hubbard action \label{eq:latt_eb_action_Psi}
\begin{subequations} \label{eq:Nambu_Hedin}
\begin{align}
\boldsymbol{\Sigma}_{ab,\mathbf{k}}(i\omega) & =-\sum_{\mathbf{q},i\Omega}\sum_{c,d}\sum_{I}\boldsymbol{\lambda}_{ac}^{I}\boldsymbol{G}_{cd,\mathbf{k}+\mathbf{q}}(i\omega+i\Omega)\nonumber \\
 & \;\;\;\;\;\times W_{\mathbf{q}}^{I}(i\Omega)\boldsymbol{\Lambda}_{db, \mathbf{kq}}^{I}(i\omega,i\Omega),\\
P_{\mathbf{q}}^{I}(i\Omega) & =\frac{1}{2}\sum_{\mathbf{k},i\omega}\sum_{a,b,c,d}\boldsymbol{\lambda}_{ac}^{I}\boldsymbol{G}_{ba,\mathbf{k}+\mathbf{q}}(i\omega+i\Omega)\nonumber \\
 & \;\;\;\;\;\times\boldsymbol{G}_{cd,\mathbf{k}}(i\omega)\boldsymbol{\Lambda}_{db, \mathbf{kq}}^{I}(i\omega,i\Omega).
\end{align}
\end{subequations}
Similarly (see Appendix \ref{app:translational} for details),
\begin{eqnarray} \nonumber
\boldsymbol{\Lambda}^{I}_{\mathbf{kq},ab}(i\omega,i\Omega) &=& \sum_{cd}
  [\boldsymbol{G}_{\mathbf{k}+\mathbf{q}}^{-1}(i\omega+i\Omega)]_{ac}[\boldsymbol{G}_\mathbf{k}^{-1}(i\omega)]_{db} \\ \label{eq:Lambda_translational}
 && \times \big(W^{I}_{\mathbf{q}}(i\Omega)\big)^{-1} \boldsymbol{\chi}^{3,\mathrm{conn},I}_{\mathbf{kq},cd}(i\omega,i\Omega)
\end{eqnarray}

Furthermore, we restrict ourselves to SU(2) symmetric phases, and allow only for singlet pairing,
therefore
\begin{equation}
\langle c^*_\up(\tau) c^*_\up(0)\rangle = \langle c^*_\dn(\tau) c^*_\dn(0)\rangle = 0
\end{equation}
We allow no emergent mixing of spin
\begin{equation}
\langle c^*_\up(\tau) c_\dn(0)\rangle = \langle c^*_\dn(\tau) c_\up(0)\rangle = 0
\end{equation}
These assumptions simplify the structure of the Green's function in Nambu space
\begin{align}
\boldsymbol{G}_{\mathbf{k}}(i\omega) =
\left[ \begin{array}{cccc}
  & & -F_{\mathbf{k}}(i\omega)& -G_{\mathbf{k}}^{*}(i\omega) \\
 &&G_{\mathbf{k}}(i\omega)& -F^{*}_{\mathbf{k}}(i\omega) \\
F_{\mathbf{k}}(i\omega)& -G_{\mathbf{k}}^{*}(i\omega)& &\\
G_{\mathbf{k}}(i\omega)&F^{*}_{\mathbf{k}}(i\omega)& & \\
\end{array}\right] \\ \nonumber
\end{align}
where  the normal and anomalous Green's functions read 
\begin{align}
G_{ij}(\tau-\tau')&\equiv-\langle c_{\up i}(\tau) c^*_{\up j}(\tau')\rangle\\
F_{ij}(\tau-\tau')&\equiv-\langle c^*_{\dn i}(\tau) c^*_{\up j}(\tau')\rangle
\end{align}
Under the present assumptions $G_\mathbf{k}(\tau)$ is real, therefore $G_\mathbf{k}(-i\omega)=G_\mathbf{k}^*(i\omega)$.
Here note that SU(2) symmetry and lattice inversion symmetry imply $F_{ij}(\tau)=F_{ij}(-\tau)=F_{ji}(\tau)$ (this can be proven by rotating
$c_\sigma \rightarrow (-)^{\delta_{\up,\sigma}} c_{\bar\sigma}$).
Therefore, if $F_{ij}(\tau)$ is real, $F_\mathbf{k}(i\omega)$ is also purely real. In this paper we consider only purely real $F_{ij}(\tau)$.

Similarly, the block structure of the self-energy is given by:
\begin{align}
\boldsymbol{\Sigma}_{\mathbf{k}}(i\omega) =
\left[ \begin{array}{cccc}
  & & S^{*}_{\mathbf{k}}(i\omega)&\Sigma_{\mathbf{k}}(i\omega) \\
 &&-\Sigma_{\mathbf{k}}^{*}(i\omega)& S_{\mathbf{k}}(i\omega) \\
-S^{*}_{\mathbf{k}}(i\omega)&\Sigma_{\mathbf{k}}(i\omega)& & \\
-\Sigma_{\mathbf{k}}^{*}(i\omega)&-S_{\mathbf{k}}(i\omega)& & \\
\end{array}\right] \\ \nonumber
\end{align}
$\Sigma$ and $S$ are the normal and anomalous self-energies defined by the Nambu-Dyson equation
\begin{equation}
  {\boldsymbol G}^{-1}_\mathbf{k}(i\omega) =  {\boldsymbol G}^{-1}_{0,\mathbf{k}}(i\omega) - {\boldsymbol\Sigma}_\mathbf{k}(i\omega)
\end{equation}
where the inverse is assumed to be the matrix inverse in Nambu indices.
Component-wise, under the present assumptions, the Nambu-Dyson equation reads
\begin{subequations}
\begin{align}
G_{\mathbf{k}}(i\omega) & =\frac{\left(G_{0\mathbf{k}}^{-1}(i\omega)-\Sigma_{\mathbf{k}}(i\omega)\right)^{*}}{|G_{0\mathbf{k}}^{-1}(i\omega)-\Sigma_{\mathbf{k}}(i\omega)|^{2}+|S_{\mathbf{k}}(i\omega)|^{2}}\label{eq:Dyson_Nambu_G}\\
F_{\mathbf{k}}(i\omega) & =\frac{-S_{\mathbf{k}}(i\omega)}{|G_{0\mathbf{k}}^{-1}(i\omega)-\Sigma_{\mathbf{k}}(i\omega)|^{2}+|S_{\mathbf{k}}(i\omega)|^{2}}\label{eq:Dyson_Nambu_F}
\end{align}
\end{subequations}

Furthermore, due to SU(2) symmetry, the full bosonic propagator will be identical in the $x$, $y$ and $z$ channels, so we define 
\begin{equation}
\eta(I) = \left\{ \begin{array}{c} 
                   \mathrm{ch}, \;\;\; I=0 \\
                   \mathrm{sp}, \;\;\; I=x,y,z
                  \end{array}
                  \right.
\end{equation}
and have $W^x=W^y=W^z=W^\mathrm{sp}$, and similarly for the renormalized vertex. This will simplify the calculation of the self-energy in the Heisenberg decoupling scheme, as the contribution coming from $x$ and $y$ bosons is the same as the one coming from the $z$ boson.
The bosonic Dyson equation is then always solved in only two channels
\begin{align}
W_{\mathbf{q}}^{\eta}(i\Omega) & =\frac{U^{\eta}}{1-U^{\eta}P_{\mathbf{q}}^{\eta}(i\Omega)}\label{eq:W_dyson}
\end{align}

\subsection{TRILEX, $GW$+EDMFT and $GW$ equations}
\subsubsection{Single-site TRILEX approximation for $d$-wave superconductivity\label{subsec:trilex_dwave}}

The single-site TRILEX method consists in approximating the renormalized vertex by a local quantity, obtained from an effective single-site impurity model
\begin{eqnarray} \label{eq:imp_eb_action_Psi}
 &  & S_{\mathrm{imp},\mathrm{eb}}^{\mathrm{Nambu}}[\bPsi,\phi]=\nonumber \\
 &  & \frac{1}{2}\iint\mathrm{d}\tau\mathrm{d}\tau'\boldsymbol{\Psi}_{a}(\tau)[-\boldsymbol{\cal G}^{-1}]_{a,b}(\tau-\tau')\boldsymbol{\Psi}_{b}(\tau')\nonumber \\
 &  & +\frac{1}{2}\sum_{I}\iint\mathrm{d}\tau\mathrm{d}\tau'\phi^{I}(\tau)[-({\cal U}^{I})^{-1}](\tau-\tau')\phi^{I}(\tau')\\
 &  & +\frac{1}{2}\sum_{I}\int\mathrm{d}\tau\phi^{I}(\tau)\boldsymbol{\Psi}_a(\tau)\boldsymbol{\lambda}_{ab}^{I}\boldsymbol{\Psi}_{b}(\tau)\nonumber 
\end{eqnarray}
Solving the TRILEX equations amounts to finding $\boldsymbol{\mathcal{G}}(i\omega)$ and
$\mathcal{U}(i\Omega)$ such that the full propagators in the effective impurity problem \eqref{eq:imp_eb_action_Psi} coincide with the local components of the ones obtained on the lattice, namely, we want to satisfy
\begin{subequations}
\begin{align}
\sum_{\mathbf{k}}\boldsymbol{G}_{\mathbf{k}}(i\omega)[\boldsymbol{\mathcal{G}},\mathcal{U}] & =\boldsymbol{G}_{\mathrm{imp}}(i\omega)[\boldsymbol{\mathcal{G}},\mathcal{U}]\label{eq:G_sc}\\
\sum_{\mathbf{q}}W_{\mathbf{q}}^{\eta}(i\Omega)[\boldsymbol{\mathcal{G}},\mathcal{U}] & =W_{\mathrm{imp}}^{\eta}(i\Omega)[\boldsymbol{\mathcal{G}},\mathcal{U}]\label{eq:W_sc}
\end{align}
\end{subequations}
where the vertex of Eq.~\eqref{eq:Nambu_Hedin} is approximated by the impurity vertex:
\begin{equation}
  \boldsymbol{\Lambda}_\mathbf{kq} = \boldsymbol{\Lambda}_\mathrm{imp}[\boldsymbol{\mathcal{G}},\mathcal{U}]
\end{equation}

In this paper, we allow only strictly $d$-wave superconducting pairing. Thus
\begin{equation}
\sum_{\mathbf{k}}F_{\mathbf{k}}(i\omega)=0\label{eq:vanishing_local_F}
\end{equation}
which means that the anomalous components of the local Green's function $\boldsymbol{G}_\mathrm{loc}$ will be zero. 
Therefore, at self-consistency (Eq.~\eqref{eq:G_sc}), the impurity's Green's function is normal and thus the anomalous components of the bare propagator on the impurity must vanish
\begin{equation}
\boldsymbol{\mathcal{G}}_{02/20/13/31}=0\label{eq:vanishing_anom_Weiss}
\end{equation}
This means that the impurity problem will be identical to the one in the normal-phase calculations, which can be expressed in terms of the original Grassmann fields
\begin{eqnarray} \label{eq:imp_eb_action_c}
 &  & S_{\mathrm{imp},\mathrm{eb}}[c^*,c,\phi]=\nonumber \\
 &  & \sum_\sigma\iint\mathrm{d}\tau\mathrm{d}\tau'c^*_{\sigma}(\tau)[-{\cal G}^{-1}](\tau-\tau')c_{\sigma}(\tau')\nonumber \\
 &  & +\frac{1}{2}\sum_{I}\iint\mathrm{d}\tau\mathrm{d}\tau'\phi^{I}(\tau)[-({\cal U}^{I})^{-1}](\tau-\tau')\phi^{I}(\tau')\\
 &  & +\sum_{I,\sigma,\sigma'}\int\mathrm{d}\tau\phi^{I}(\tau)c^*_\sigma(\tau)\lambda_{\sigma\sigma'}^{I}c_{\sigma'}(\tau)\nonumber 
\end{eqnarray}
where the bare vertices (slim symbols denote the impurity quantities) are given by Pauli matrices $\lambda^I_{\sigma\sigma'} = \sigma^I_{\sigma\sigma'}$.
After integrating out the bosonic degrees of freedom, one obtains an electron-electron action with retarded interactions
\begin{eqnarray} \nonumber
S_{\mathrm{imp},\mathrm{ee}}[c^*,c] & =&\iint_{\tau,\tau'}\sum_{\sigma}c^{*}_{\sigma}(\tau)\left[-\mathcal{G}^{-1}(\tau-\tau')\right]c_{\sigma}(\tau')\\ \label{eq:S_imp}
 &&+\frac{1}{2}\iint_{\tau\tau'}\sum_{I}n^{I}(\tau)\mathcal{U}^{I}(\tau-\tau')n^{I}(\tau')
\end{eqnarray}
This single-site impurity problem is solved using the numerically exact hybridization-expansion continuous-time quantum Monte Carlo (CTHYB or HYB-CTQMC\cite{Werner2006,Werner2007}), employing the segment algorithm.
The transverse spin-spin interaction term is dealt with in an interaction-expansion manner\cite{Otsuki2013}. See Ref.~\onlinecite{Ayral2015c} for details.

Under the present assumptions, the approximation for the renormalized vertex entering the Hedin equations Eq.~\eqref{eq:Nambu_Hedin} is
\newcommand{\ls}{\Big(\Lambda_\mathrm{imp}^{\eta(I)}\Big)^* }
\newcommand{\lns}{\Lambda_\mathrm{imp}^{\eta(I)} }
\begin{align} \label{eq:boldLambda_imp}
 \boldsymbol{\Lambda}_{\mathbf{kq}}^{I}(i\omega,i\Omega) & \approx  \boldsymbol{\Lambda}_{\mathrm{imp}}^{I}(i\omega,i\Omega) \\ \nonumber
 =  \boldsymbol{\lambda}^I \circ &
 \left[
 \begin{array}{cccc}
   &\lns&&\lns \\
   \ls&&\ls& \\
   &\lns&&\lns \\
   \ls&&\ls& \\   
 \end{array}
 \right](i\omega,i\Omega)
\end{align}
where $\circ$ denotes the elementwise product $[A\circ B]_{ij} = A_{ij} B_{ij}$ (see Appendix \ref{app:Lambda_imp} for details).

We obtain $\Lambda_\mathrm{imp}^{\eta}$ from the three-point correlation function on the impurity using
\begin{align}
 & \Lambda_{\mathrm{imp}}^{\eta}(i\omega,i\Omega)\label{eq:vertex_def}\\
 & \equiv\frac{\tilde{\chi}^{3,\eta,\mathrm{conn}}_\mathrm{imp}(i\omega,i\Omega) }{G_{\mathrm{imp}}(i\omega)G_{\mathrm{imp}}(i\omega+i\Omega)(1-\mathcal{U}^{\eta}(i\Omega)\chi_{\mathrm{imp}}^{\eta}(i\Omega))}\nonumber 
\end{align}
where
\begin{align}
 & \tilde{\chi}^{3,\eta,\mathrm{conn}}_\mathrm{imp}(i\omega,i\Omega) \equiv \iint_{\tau\tau'}e^{i\tau\omega+i\tau'\Omega}\times\label{eq:chi3tilde_def}\\
 & \;\;\;\times\Big(\langle c_{\up}(\tau)c_{\up}^{*}(0)n^{\eta}(\tau')\rangle_{\mathrm{imp}}+G_{\mathrm{imp}}(\tau)\langle n^{\eta}\rangle_{\mathrm{imp}}\Big). 
\end{align}
and
\begin{align}
G_{\mathrm{imp}}(i\omega) & \equiv-\int_{0}^{\beta}\mathrm{d}\tau e^{i\tau\omega}\langle c_\up(\tau)c_\up^*(0)\rangle_{\mathrm{imp}}\label{eq:G_def}\\
W^\eta_{\mathrm{imp}}(i\Omega) & \equiv-\int_{0}^{\beta}\mathrm{d}\tau e^{i\tau\Omega}\langle(\phi(\tau)-\langle\phi\rangle)(\phi(0)-\langle\phi\rangle)\rangle_{\mathrm{imp}}\label{eq:W_def}\\
 & =\mathcal{U}(i\Omega)-\mathcal{U}(i\Omega)\chi_{\mathrm{imp}}^{\eta}(i\Omega)\mathcal{U}(i\Omega)
\end{align}
\begin{equation}
\chi_{\mathrm{imp}}^{\eta}(i\Omega)\equiv\int_{0}^{\beta}\mathrm{d}\tau e^{i\tau\Omega}\left(\langle n^{\eta}(\tau)n^{\eta}(0)\rangle_{\mathrm{imp}}-\langle n^{\eta}\rangle_{\mathrm{imp}}^{2}\right)\label{eq:chi_def}
\end{equation}

We can now write the final expressions for the self-energy and polarization: 
\begin{subequations}
\begin{eqnarray}
& &\Sigma_{\mathbf{k}}(i\omega)  =\label{eq:Sigma_Hedin}\\ \nonumber
 &   & \;\;\;\;-\sum_{\eta}m_{\eta}\sum_{\mathbf{q},i\Omega}G_{\mathbf{k}+\mathbf{q}}(i\omega+i\Omega)W_{\mathbf{q}}^{\eta}(i\Omega)\Lambda_{\mathrm{imp}}^{\eta}(i\omega,i\Omega)\\
 &  & S_{\mathbf{k}}(i\omega)=\label{eq:S_Nambu}\\
 &  & \;\;\;\;-\sum_{\eta}(-)^{p_{\eta}}m_{\eta}\sum_{\mathbf{q},i\Omega}F_{\mathbf{k}+\mathbf{q}}(i\omega+i\Omega)W_{\mathbf{q}}^{\eta}(i\Omega)\Lambda_{\mathrm{imp}}^{\eta}(i\omega,i\Omega)\nonumber \\
 &  & P_{\mathbf{q}}^{\eta}(i\Omega)=\label{eq:P_Nambu}\\
 &  & \;\;\;\;2\sum_{\mathbf{k},i\omega}G_{\mathbf{k}+\mathbf{q}}(i\omega+i\Omega)G_{\mathbf{k}}(i\omega)\Lambda_{\mathrm{imp}}^{\eta}(i\omega,i\Omega)\nonumber \\
 &  & \;\;\;\;+(-)^{p_{\eta}}2\sum_{\mathbf{k},i\omega}F_{\mathbf{k}+\mathbf{q}}(i\omega+i\Omega)F_{\mathbf{k}}(i\omega)\Lambda_{\mathrm{imp}}^{\eta}(i\omega,i\Omega)\nonumber 
\end{eqnarray}
\end{subequations}
with $p_{\mathrm{ch}}=1$, $p_{\mathrm{sp}}=0$, $m_{\mathrm{ch}}=1$. These equations hold in both the Heisenberg ($m_{\mathrm{sp}}=3$) and Ising ($m_{\mathrm{sp}}=1$) decoupling schemes. 
In the expression for the polarization (Eq.~\eqref{eq:P_Nambu}) we have used lattice inversion symmetry and the symmetries of $\Lambda$ and $\boldsymbol{G}$. Under the present assumptions, $P$ is purely real (see Appendix \ref{app:realness_of_P} for details).

\subsubsection{$GW$+EDMFT}

The $GW$+EDMFT approximation can be regarded as a simplified version
of TRILEX where, in the calculation of the non-local ($\mathbf{r}\neq0$) part of self-energy and polarization (second line of Eqs. (\ref{eq:Sigma_Hedin_split}),(\ref{eq:S_Nambu_split}) and (\ref{eq:P_Nambu_split}) below), an additional approximation is made:
\begin{equation}
\Lambda_{\mathrm{imp}}^{\eta}(i\omega,i\Omega)\approx1\label{eq:GW_EDMFT_approx}.
\end{equation}

The efficiency is gained because one need not measure the three-point correlator $\tilde{\chi}^{3,\eta,\mathrm{conn}}$ in the impurity model. The local self-energy and polarization still have vertex-corrections, but are identical to $\Sigma$ and $P$ on the impurity, which can be computed from only two-point correlators. Furthermore, the calculation of the non-local parts of the self-energy and polarization can now be performed in imaginary time, as opposed to the explicit summation over frequency needed in Eqs. (\ref{eq:Sigma_Hedin_split}),
(\ref{eq:S_Nambu_split}) and (\ref{eq:P_Nambu_split}).

\subsubsection{$GW$}

If we approximate the renormalized vertex by unity even in the calculation of the local part of self-energies,
we obtain an approximation similar to the $GW$ approximation, with the important difference that we are using a decoupling in both charge and spin channels, unlike the conventional $GW$ approaches which are limited to the charge channel.
This additional approximation eliminates the need for solving an impurity
problem, as now even the local self-energy and polarization are calculated by
the bubble diagrams Eq.~\eqref{eq:Sigma_Hedin}, \eqref{eq:S_Nambu} and \eqref{eq:P_Nambu}, simplified by Eq.~\eqref{eq:GW_EDMFT_approx}.

\begin{figure*}[!ht]
\centering{}\includegraphics[width=5.4in,trim=0cm 0cm 0cm 0cm]{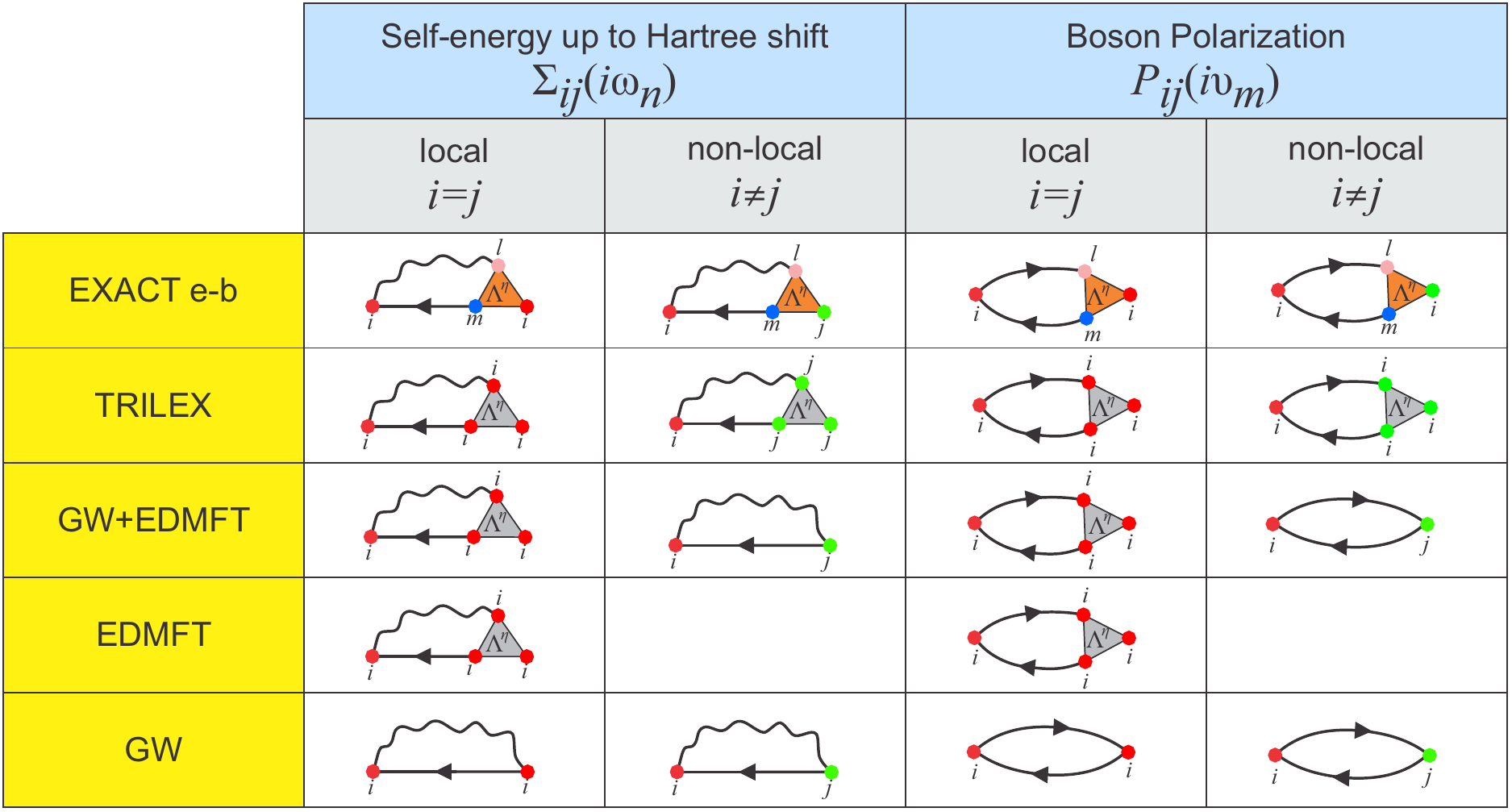}
\caption{ Self-energy/polarization approximations in various methods based
on a Hubbard-Stratonovich decoupling, compared to the exact expression. The renormalized
electron-boson vertex is either approximated by a local dynamical
quantity, or by the bare vertex. Orange triangle denotes the exact renormalized vertex, with full spatial dependence; gray triangle denotes the local approximation of the vertex. Colored circles denote terminals of the propagators and the vertex, and the (local) bare vertex at a given site; different colors denote different lattice sites $ijlm$. Internal site-indices are summed over, but when the vertex is local, only a single term in the summation survives.
\label{fig:selfenergy_approximations} }
\end{figure*}

To summarize, the exact expressions for the self-energy and boson
polarization are compared to the approximate ones in $GW$, EDMFT,
$GW$+EDMFT, and TRILEX in Fig.~\ref{fig:selfenergy_approximations}. 

\subsubsection{Normal phase calculation}

In the normal phase, the further simplification is that $F_\mathbf{k}(i\omega)=0$. Therefore, $S_\mathbf{k}(i\omega)=0$ and the Dyson equation \eqref{eq:Dyson_Nambu_G} reduces to the familiar form
 \begin{align}
 G_{\mathbf{k}}(i\omega) & =\frac{1}{i\omega+\mu-\varepsilon_{\mathbf{k}}-\Sigma_{\mathbf{k}}(i\omega)}\label{eq:nomral_G_dyson}
\end{align}

%%%%%%%%%%%%%%%%%%%%%%%%%%%%%%%%%%%%%%%%%%%%%%%%%%%%%%%%%%%%%%%%%%%%%%%%%%%%%%%%%%%%%%%%
%%%%%%%%%%%%%%%%%%%%%%%%%%%%%%%%%%%%%%%%%%%%%%%%%%%%%%%%%%%%%%%%%%%%%%%%%%%%%%%%%%%%%%%%
%%%%%%%%%%%%%%%%%%%%%%%%%%%%%%%%%%%%%%%%%%%%%%%%%%%%%%%%%%%%%%%%%%%%%%%%%%%%%%%%%%%%%%%%
%%%%%%%%%%%%%%%%%%%%%%%%%%%%%%%%%%%%%%%%%%%%%%%%%%%%%%%%%%%%%%%%%%%%%%%%%%%%%%%%%%%%%%%%
%%%%%%%%%%%%%%%%%%%%%%%%%%%%%%%%%%%%%%%%%%%%%%%%%%%%%%%%%%%%%%%%%%%%%%%%%%%%%%%%%%%%%%%%
%%%%%%%%%%%%%%%%%%%%%%%%%%%%%%%%%%%%%%%%%%%%%%%%%%%%%%%%%%%%%%%%%%%%%%%%%%%%%%%%%%%%%%%%
%%%%%%%%%%%%%%%%%%%%%%%%%%%%%%%%%%%%%%%%%%%%%%%%%%%%%%%%%%%%%%%%%%%%%%%%%%%%%%%%%%%%%%%%
%%%%%%%%%%%%%%%%%%%%%%%%%%%%%%%%%%%%%%%%%%%%%%%%%%%%%%%%%%%%%%%%%%%%%%%%%%%%%%%%%%%%%%%%

\section{Methods}
\label{sec:methods}
\subsection{Numerical implementation of the Hedin equations}

As shown in Ref.~\onlinecite{Ayral2015c}, it is numerically advantageous
to perform the computation in real space and to split the self-energy
and polarization in the following way:
\begin{subequations}
\begin{align}
 & \Sigma_{\mathbf{r}}(i\omega)=\delta_{\mathbf{r}}\Sigma_{\mathrm{imp}}(i\omega)\label{eq:Sigma_Hedin_split}\\
 & \;\;-\sum_{\eta}m_{\eta}\sum_{i\Omega}\tilde{G}_{\mathbf{r}}(i\omega+i\Omega)\tilde{W}_{\mathbf{r}}^{\eta}(i\Omega)\Lambda_{\mathrm{imp}}^{\eta}(i\omega,i\Omega)\nonumber \\
 &   S_{\mathbf{r}}(i\omega)=\label{eq:S_Nambu_split}\\
  &   \;\;\;\;-\sum_{\eta}(-)^{p_{\eta}}m_{\eta}\sum_{i\Omega}\tilde{F}_{\mathbf{r}}(i\omega+i\Omega)\tilde{W}_{\mathbf{r}}(i\Omega)\Lambda_{\mathrm{imp}}^{\eta}(i\omega,i\Omega)\nonumber \\
  &   P_{\mathbf{r}}^{\eta}(i\Omega)=\delta_\mathbf{r}P_{\mathrm{imp}}^{\eta}(i\Omega)\label{eq:P_Nambu_split}\\
  &   \;\;\;\;+2\sum_{i\omega}\tilde{G}_{\mathbf{r}}(i\omega+i\Omega)\tilde{G}_{-\mathbf{r}}(i\omega)\Lambda_{\mathrm{imp}}^{\eta}(i\omega,i\Omega)\nonumber \\
  &   \;\;\;\;+(-)^{p_{\eta}}2\sum_{i\omega}\tilde{F}_{\mathbf{r}}(i\omega+i\Omega)\tilde{F}_{-\mathbf{r}}(i\omega)\Lambda_{\mathrm{imp}}^{\eta}(i\omega,i\Omega)\nonumber 
\end{align}
\end{subequations} where $\tilde{X}_{\mathbf{r}}(i\omega)\equiv (1-\delta_\mathbf{r})X_{\mathbf{r}}(i\omega)$. In the presence of lattice inversion symmetry, $X_\mathbf{r}=X_{-\mathbf{r}}$. The impurity's self-energy and polarization are defined as\begin{subequations}
\begin{align}
\Sigma_{\mathrm{imp}}(i\omega) & \equiv\mathcal{G}^{-1}(i\omega)-G_{\mathrm{imp}}^{-1}(i\omega)\label{eq:Dyson_G_imp}\\
P_{\mathrm{imp}}^{\eta}(i\Omega) & \equiv\left[\mathcal{U}^{\eta}(i\Omega)\right]^{-1}-\left[W_{\mathrm{imp}}^{\eta}(i\Omega)\right]^{-1}\nonumber \\
 & =\frac{-\chi_{\mathrm{imp}}^{\eta}(i\Omega)}{1-\mathcal{U}^{\eta}\chi_{\mathrm{imp}}^{\eta}(i\Omega)}\label{eq:Dyson_W_imp}
\end{align}
\end{subequations}

\subsection{Solution by forward recursion}\label{sec:forward_recursion}

In practice, the TRILEX, $GW$+EDMFT and $GW$ equations can be solved
by forward recursion:
\begin{enumerate} 
\item Start with a given $\Sigma_{\mathbf{k}}(i\omega)$ and $P^\eta_{\mathbf{q}}(i\Omega)$, and (for SC phase only) $S_{\mathbf{k}}(i\omega)$
and (for TRILEX and $GW$+EDMFT only) $\Sigma_{\mathrm{imp}}(i\omega)$
and $P^\eta_{\mathrm{imp}}(i\Omega)$ (for instance set them to zero, or
use EDMFT results)
\item Compute the new $G_{\mathbf{k}}(i\omega)$ and $W_{\mathbf{q}}^{\eta}(i\Omega)$ and  (for SC phase only) $F_{\mathbf{k}}(i\omega)$ 
  from Eqs. (\ref{eq:Dyson_Nambu_G}, \ref{eq:W_dyson}, \ref{eq:Dyson_Nambu_F}). 
\item (TRILEX/$GW$+EDMFT only) Impose the self-consistency condition Eq.~(\ref{eq:G_sc}, \ref{eq:W_sc}) by reversing the impurity Dyson equations
(\ref{eq:Dyson_G_imp}, \ref{eq:Dyson_W_imp}), such that\begin{subequations}
\begin{align}
\mathcal{G}(i\omega) & =\left[\left\{ \sum_{\mathbf{k}}G_{\mathbf{k}}(i\omega)\right\} ^{-1}+\Sigma_{\mathrm{imp}}(i\omega)\right]^{-1}\label{eq:Weiss_G}\\
\mathcal{U}^{\eta}(i\Omega) & =\left[\left\{ \sum_{\mathbf{q}}W_{\mathbf{q}}^{\eta}(i\Omega)\right\} ^{-1}+P_{\mathrm{imp}}^{\eta}(i\Omega)\right]^{-1}\label{eq:Weiss_U}
\end{align}
\end{subequations}
\item (TRILEX/$GW$+EDMFT only) Solve the impurity model with the above bare fermionic and bosonic propagators: compute $G_{\mathrm{imp}}$,
$\chi_{\mathrm{imp}}^{\eta}$, $\langle n^{\eta}\rangle_{\mathrm{imp}}$
and (for TRILEX only) $\tilde{\chi}^{3,\eta,\mathrm{conn}}$
and from them 
$\Sigma_{\mathrm{imp}}$ (Eq.~\ref{eq:Dyson_G_imp}), $P_{\mathrm{imp}}^{\eta}$ (Eq.~\ref{eq:Dyson_W_imp}) and (TRILEX only) $\Lambda_{\mathrm{imp}}^{\eta}$ (Eq.~\ref{eq:vertex_def});
\item Compute $\Sigma_{\mathbf{k}}(i\omega)$ and $P_{\mathbf{q}}^{\eta}(i\Omega)$ and (for SC phase only) $S_{\mathbf{k}}(i\omega)$
with Eqs. (\ref{eq:Sigma_Hedin_split}, \ref{eq:P_Nambu_split}, \ref{eq:S_Nambu_split}); 
\item Go back to step 2 until convergence is reached.
\end{enumerate}

\subsection{Superconducting temperature $T_c$} \label{sec:lge}

In order to determine the superconducting transition temperature $T_{c}$, we solve
a linearized gap equation (LGE). 
At $T=T_c$, the anomalous part of the self-energy $S$ vanishes. Linearizing  Eq.~\eqref{eq:Dyson_Nambu_F}
with respect to $S$ and plugging it into Eq.~\eqref{eq:S_Nambu_split} leads to an implicit equation for $T_c$,
featuring only the normal component of the Green's function
\begin{align} \nonumber
\label{eq:trilex_lge}
S_\mathbf{r}(i\omega) & =  -\sum_{\eta,i\Omega}(-)^{\delta_{\eta,\mathrm{ch}}}F_\mathbf{r}(i\omega+i\Omega)W_\mathbf{r}^{\eta}(i\Omega)\Lambda^{\mathrm{imp},\eta}(i\omega,i\Omega)\\
F_{\mathbf{k}}(i\omega_{n}) & = -S_{\mathbf{k}}(i\omega_{n})|G_{\mathbf{k}}(i\omega_{n}))|^{2} 
\end{align}
Using four-vector notation $k\equiv(\mathbf{k},i\omega)$, we obtain
\begin{align}
  &A_{kk'} \equiv \sum_{\eta=\mathrm{ch},\mathrm{sp}} (-)^{p_\eta}  m_\eta |G(k')|^2 W^\eta_{k-k'} \Lambda^{\mathrm{imp},\eta}_{k,k-k'}\\
 &A_{kk'} S_{k'} = S_k      
\end{align}
This is an eigenvalue problem for $S$.
In practice, it is more convenient to consider the spectrum of the operator $A$, 
\begin{align}
  A_{kk'} S_{k'}^\lambda &= \lambda S_k^\lambda      
\end{align}
The eigenvalues $\lambda$ and the eigenvectors $S_k^\lambda$ 
depend on the temperature $T$. 
The critical temperature $T_c$ is therefore given by $$\lambda_m(T_c) =1$$ 
where $\lambda_m$ is the largest eigenvalue of $A$.
In other words, $T=T_c$ when the first eigenvalue crosses 1.
In addition, the symmetry of the superconducting instability is given by the 
$k$ dependence of $S$ for the corresponding eigenvector.

In practice, we first solve the normal-phase equations, and
then solve the LGE Eq.~\eqref{eq:trilex_lge} by forward
substitution.
Starting from an initial simple $d_{x^2-y^2}$-wave form 
\begin{equation}
S_{\mathbf{k}}(i\omega_{n})=(\delta_{n,0}+\delta_{n,-1})(\cos k_{x}-\cos k_{y})
\end{equation}
we use the power method \cite{MisesZAMM1929} to compute the leading eigenvalue of the operator $A$.
We do this in a select range of temperature
for the given parameters $(U,n,t,t',t'')$ and monitor the leading
eigenvalue $\lambda_m(T)$.
If we observe a $T_c$ ($\lambda_m(T)>1)$), we can then use the eigenvector $S$ as an initial guess 
to stabilize the superconducting solution using the algorithm from section \ref{sec:forward_recursion}.
We have also examined other irreducible representations of the symmetry group
and found that this $d$-wave representation is 
the one with highest $T_{c}$, in agreement with Refs.~\onlinecite{OtsukiHafermannPRB2014,MaierPRB2015}.

%%%%%%%%%%%%%%%%%%%%%%%%%%%%%%%%%%%%%%%%%%%%%%%%%%%%%%%%%%%%%%%%%%%%%%%%%%%%%%%%%%%%%%%%
%%%%%%%%%%%%%%%%%%%%%%%%%%%%%%%%%%%%%%%%%%%%%%%%%%%%%%%%%%%%%%%%%%%%%%%%%%%%%%%%%%%%%%%%
%%%%%%%%%%%%%%%%%%%%%%%%%%%%%%%%%%%%%%%%%%%%%%%%%%%%%%%%%%%%%%%%%%%%%%%%%%%%%%%%%%%%%%%%
%%%%%%%%%%%%%%%%%%%%%%%%%%%%%%%%%%%%%%%%%%%%%%%%%%%%%%%%%%%%%%%%%%%%%%%%%%%%%%%%%%%%%%%%
%%%%%%%%%%%%%%%%%%%%%%%%%%%%%%%%%%%%%%%%%%%%%%%%%%%%%%%%%%%%%%%%%%%%%%%%%%%%%%%%%%%%%%%%
%%%%%%%%%%%%%%%%%%%%%%%%%%%%%%%%%%%%%%%%%%%%%%%%%%%%%%%%%%%%%%%%%%%%%%%%%%%%%%%%%%%%%%%%
%%%%%%%%%%%%%%%%%%%%%%%%%%%%%%%%%%%%%%%%%%%%%%%%%%%%%%%%%%%%%%%%%%%%%%%%%%%%%%%%%%%%%%%%
%%%%%%%%%%%%%%%%%%%%%%%%%%%%%%%%%%%%%%%%%%%%%%%%%%%%%%%%%%%%%%%%%%%%%%%%%%%%%%%%%%%%%%%%

\subsection{The AF instability} \label{sec:AF_instability}

As documented in Refs. \onlinecite{Ayral2015,Ayral2015c}, the TRILEX equations
present an instability towards antiferromagnetism below some temperature $T_\mathrm{AF}$ (see also Refs \onlinecite{AokiPRB2015,OtsukiHafermannPRB2014}).
The antiferromagnetic susceptibility $\chi^\mathrm{sp}$ is related 
to the propagator of the boson in the spin channel via
$$ W^\mathrm{sp}_\mathbf{q}(i\Omega) = U^\mathrm{sp} -U^\mathrm{sp}\chi^\mathrm{sp}_\mathbf{q}(i\Omega)U^\mathrm{sp}.$$
They both diverge at $T=T_\mathrm{AF}$ because the polarization becomes too large (the denominator in \eqref{eq:W_dyson} vanishes).
This instability, which is an artifact of the approximation for the two-dimensional
Hubbard model, violates the Mermin-Wagner theorem.
For many values of $t', t''$, this AF instability prevents us from 
reaching the superconducting temperature $T_c$.

This AF instability also exists in conventional
cluster DMFT methods (cellular DMFT, DCA) \cite{MaierPRB2014,MaierPRL2005,KotliarCaponePRB2006}.
Yet, in most works, it is simply ignored by enforcing 
a paramagnetic solution (by symmetrizing up and down spin components).
In TRILEX, however, we do not have this possibility.
Indeed, the antiferromagnetic susceptibility directly enters the
equations (via $W$), and its divergence makes it impossible to 
stabilize a paramagnetic solution of the TRILEX equations
at a temperature lower than $T_\mathrm{AF}$. For a precise definition of $T_\mathrm{AF}$ in the present context, see Appendix \ref{app:fit}.

In the following, we circumvent this issue in two ways: either by extrapolating the temperature dependence
of the eigenvalue of the linearized gap equation to low temperatures, despite the AF instability (section \ref{sec:phase_diagram}, with tight-binding values $t', t''$ relevant for cuprate physics), or, in section \ref{sec:weak_coupling}, by finding other values
of $t', t''$, where the Fermi surface shape is 
qualitatively similar to the cuprate case, but where the AF instability
occurs at a temperature lower than $T_c$. 

%%%%%%%%%%%%%%%%%%%%%%%%%%%%%%%%%%%%%%%%%%%%%%%%%%%%%%%%%%%%%%%%%%%%%%%%%%%%%%%%%%%%%%%%
%%%%%%%%%%%%%%%%%%%%%%%%%%%%%%%%%%%%%%%%%%%%%%%%%%%%%%%%%%%%%%%%%%%%%%%%%%%%%%%%%%%%%%%%
%%%%%%%%%%%%%%%%%%%%%%%%%%%%%%%%%%%%%%%%%%%%%%%%%%%%%%%%%%%%%%%%%%%%%%%%%%%%%%%%%%%%%%%%
%%%%%%%%%%%%%%%%%%%%%%%%%%%%%%%%%%%%%%%%%%%%%%%%%%%%%%%%%%%%%%%%%%%%%%%%%%%%%%%%%%%%%%%%
%%%%%%%%%%%%%%%%%%%%%%%%%%%%%%%%%%%%%%%%%%%%%%%%%%%%%%%%%%%%%%%%%%%%%%%%%%%%%%%%%%%%%%%%
%%%%%%%%%%%%%%%%%%%%%%%%%%%%%%%%%%%%%%%%%%%%%%%%%%%%%%%%%%%%%%%%%%%%%%%%%%%%%%%%%%%%%%%%
%%%%%%%%%%%%%%%%%%%%%%%%%%%%%%%%%%%%%%%%%%%%%%%%%%%%%%%%%%%%%%%%%%%%%%%%%%%%%%%%%%%%%%%%
%%%%%%%%%%%%%%%%%%%%%%%%%%%%%%%%%%%%%%%%%%%%%%%%%%%%%%%%%%%%%%%%%%%%%%%%%%%%%%%%%%%%%%%%

\section{Results and discussion}
\label{sec:results}
\subsection{Phase diagram} \label{sec:phase_diagram}

\begin{figure}[!ht] 
\centering
  \includegraphics[width=3.0in, trim=0.0cm 0.0cm 0.0cm 0.0cm, page=2]{trilex_vs_edmftgw_Tc}  
  \includegraphics[width=3.0in, trim=0.0cm 0.0cm 0.0cm 0.0cm, page=1]{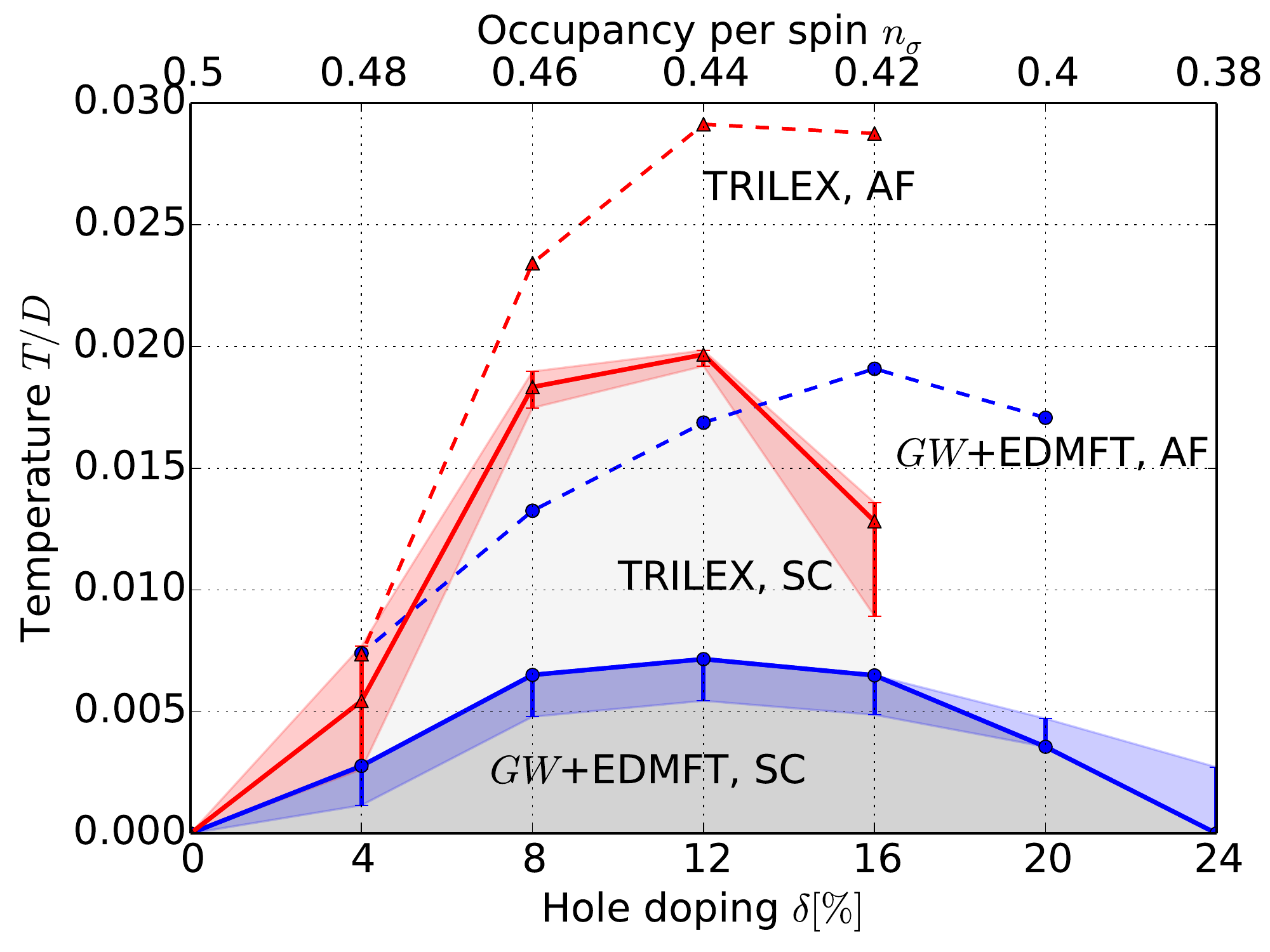}
\caption{
 Top panel: the leading eigenvalue of the linearized gap equation in TRILEX and $GW$+EDMFT.
 Bottom panel: SC critical temperature in both methods for $U/D=4$, $(t',t'')=(-0.2t,0)$. The dashed lines represents the AF instability, see text.  
}
\label{fig:SC_edmftgw_vs_trilex}
\end{figure}

First, using the linearized-gap equation (LGE) method described in Sect.~\ref{sec:lge}, we compute the SC phase boundary from high temperature, for $t' = -0.2t, t''=0$,
a physically relevant case for the physics of cuprates.
We set $U/D=4$ in order to be above the Mott transition threshold at half filling (we recall that for the square lattice, $U_c/D \approx 2.4$ within single-site DMFT\cite{Schafer2014}).
The results are presented on Fig. \ref{fig:SC_edmftgw_vs_trilex}.

The top panel presents the largest eigenvalue of the LGE as a function of temperature, for 
TRILEX and $GW$+EDMFT. The calculation becomes unstable due to AF instability before we can observe $\lambda_m>1$.
The extrapolation of $\lambda_m$ towards low temperature is not straightforward.
We use an empirical law 
\begin{equation} \label{eq:lev_law}
 \lambda_m(T)\approx a\exp(bT^{\gamma}+cT^{2\gamma})
\end{equation}
to fit the data and extrapolate to lower temperature.
This form can be shown (see Appendix \ref{app:fit})
to provide a very good fit to similar computations
in the DCA and DCA$^{+}$ methods, from the data of Refs.~\onlinecite{MaierPRL2006,MaierPRB2014}.
We perform the fit and extrapolation with $\gamma=0.3$ for $GW$+EDMFT and $\gamma=0.45$ for TRILEX,
and get the result for $T_c$ reported with solid lines on the bottom panel. The error bars shown are obtained by fitting and extrapolating with $\gamma$ varied in the window 0.3-0.6. The error bars coming from the uncertainty of the fit for a fixed $\gamma$ and a detailed discussion of the fitting procedure can be found
in Appendix \ref{app:fit}.
The dashed lines denote the temperature of the antiferromagnetic instability, below which no stable paramagnetic calculation can be made.

For all values of $\gamma$, the raw data at high temperature for both methods indicate 
a similar dome shape for $T_c$ vs $\delta$, 
where $\delta$ is the percentage of hole-doping, $\delta [\%]= (1-2n_\sigma)\times 100$ ($n_\sigma=\frac{1}{2}$ corresponds to half-filling).
The fact that $T_c$ vanishes at zero $\delta$ can be checked directly,
but we cannot exclude that it vanishes at a finite, small value of $\delta$.
The optimal doping in both methods is found to be around 12\%.
At half-filling, both methods recover a Mott insulating state, and $\lambda_m(T)$ is found to be very small.
We observe that TRILEX has a higher $T_c$ than $GW$+EDMFT, 
showing that the effects of the renormalization of the electron-boson vertex
are non-negligible in this regime.

These results for $T_c(\delta)$ are qualitatively comparable to the results of cluster DMFT methods, 
e.g. the 4-site CDMFT + ED computation of Refs~\onlinecite{KotliarCaponePRB2006, CivelliImadaPRB2016,TremblayPRB2008}, or the 8-site DCA results of Ref.~\onlinecite{GullMillisPRB2015}.
In particular, Ref.~\onlinecite{CivelliImadaPRB2016} reports a $T_c/D\approx0.0125$ at doping $\delta=13\%$ in a doped Mott insulator, which falls half-way between the TRILEX and $GW$+EDMFT results. 
Furthermore, the optimal doping in Ref.~\onlinecite{KotliarCaponePRB2006} seems to coincide with our result, while in Ref.~\onlinecite{CivelliImadaPRB2016} it is somewhat bigger (around $20\%$).
We emphasize however that here we solve only a {\it single-site} quantum impurity problem, 
and obtain the $d$-wave order, which is not possible in single-site DMFT due to symmetry reasons.

\begin{figure}[!ht] 
\centering
  \includegraphics[width=3.0in, trim=0.0cm 0.0cm 0.0cm 0.0cm, page=1]{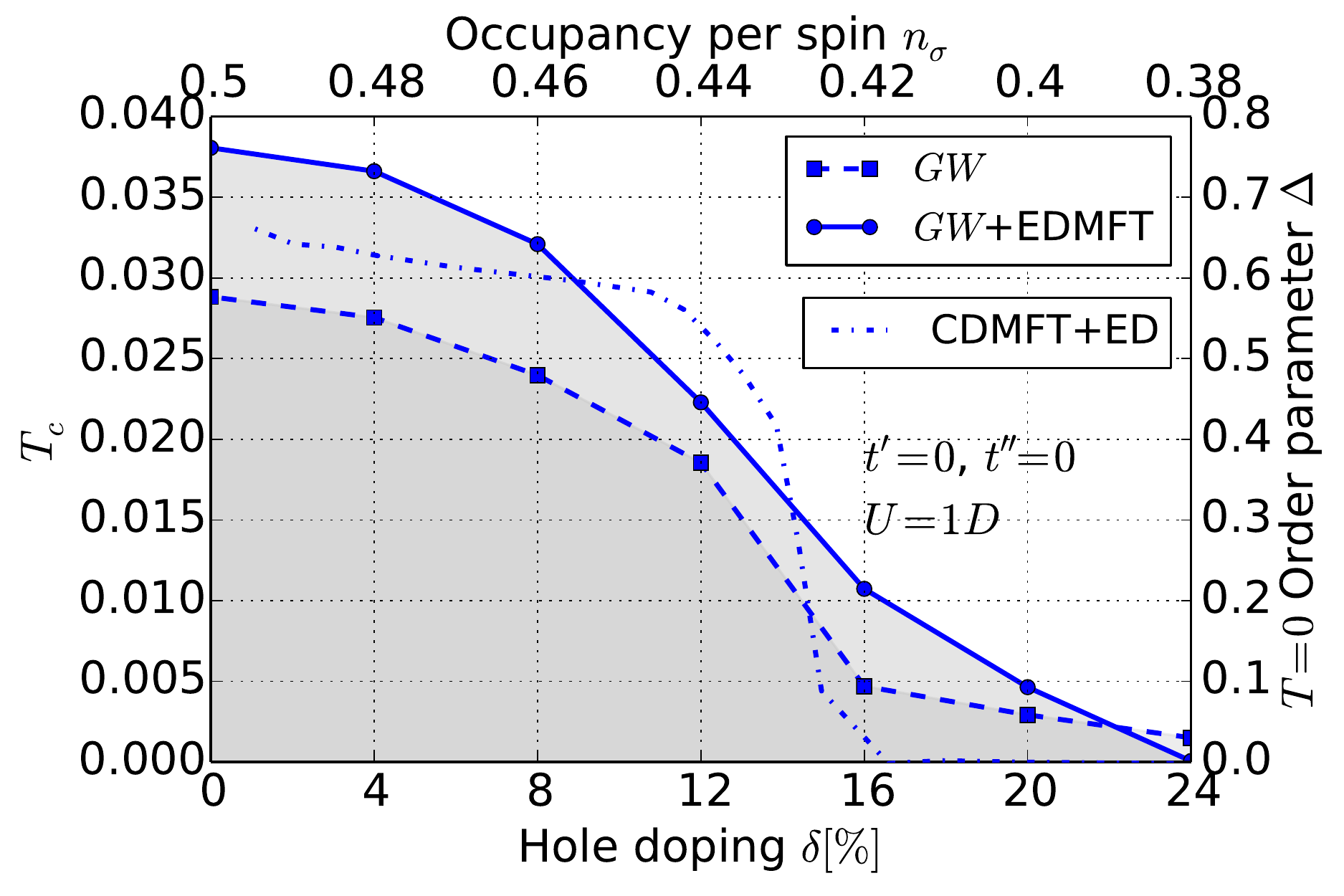}
\caption{ Comparison of $T_c(\delta)$ in $GW$ and $GW$+EDMFT methods at weak-coupling $U/D=1$, $t'=t''=0$. The dotted line is the order parameter $\Delta$ at $T=0$ from a $2\times2$ CDMFT+ED calculation, replotted from Ref.~\onlinecite{KotliarCaponePRB2006} (scale on the right).
}
\label{fig:gw_sc}
\end{figure}

Let us now turn to the weak-coupling regime ($U/D=1$). We present in Fig. \ref{fig:gw_sc} the SC temperature in
the $GW$ and $GW$+EDMFT approximation within the Ising decoupling (for the $\lambda(T)$ plot, see Appendix \ref{app:fit}).
Both methods give similar results, which justifies using the faster $GW$ at weak coupling.
In contrast to the larger-$U$ case, one does not obtain the
dome versus doping due to the absence of Mott insulator at $\delta=0$.

We compare our results with the order parameter at $T=0$ obtained from a $2\times2$ CDMFT+ED calculation\cite{KotliarCaponePRB2006}.
The general trend observed is similar: optimal doping is zero, and there is a quick reduction of $T_c$ between 12 and 16\% doping. 

As for the value of $T_c$, we compare to the
result presented in Ref.~\onlinecite{MaierPRB2014}. Here, a DCA$^{+}$ calculation
with a 52-site cluster impurity, at $U/D=1$,$t'=t''=0$, $\delta=10\%$, predicts
$T_c/D\approx0.06$. With the same parameters, $GW$ gives $T_c/D\approx0.21$, $GW$+EDMFT gives $T_c/D\approx0.27$,
hence overestimating $T_c$. 

%%%%%%%%%%%%%%%%%%%%%%%%%%%%%%%%%%%%%%%%%%%%%%%%%%%%%%%%%%%%%%%%%%%%%%%
%%%%%%%%%%%%%%%%%%%%%%%%%%%%%%%%%%%%%%%%%%%%%%%%%%%%%%%%%%%%%%%%%%%%%%%
%%%%%%%%%%%%%%%%%%%%%%%%%%%%%%%%%%%%%%%%%%%%%%%%%%%%%%%%%%%%%%%%%%%%%%%
%%%%%%%%%%%%%%%%%%%%%%%%%%%%%%%%%%%%%%%%%%%%%%%%%%%%%%%%%%%%%%%%%%%%%%%

\subsection{Weak coupling} \label{sec:weak_coupling}

As explained in Sec. \ref{sec:AF_instability}, in order to study the SC phase itself, 
we need to identify a dispersion for which $T_c$ is above $T_\mathrm{AF}$.
To achieve this, we first scan a large set of parameters $t',t''$ with the $GW$ approximation 
at weak coupling.
Indeed, at weak coupling, we can approximate TRILEX by $GW$,
which is faster to compute (there is no quantum impurity model to solve).
We look for a $(t',t'')$ point for which not only $T_\mathrm{AF}>T_c$, 
but also the shape of the Fermi surface is qualitatively compatible with cuprates.
We find a whole region of parameters where this is satisfied, and
then use these parameters in a strong-coupling computation with $GW$+EDMFT and TRILEX.
Whether a weak coupling computation is a reliable guide in the search for 
$t',t''$ with maximal $T_c$ at strong coupling 
remains open and would require a systematic exploration with cluster methods.
However, at least in one example (shown below), this assumption will provide us with an appropriate choice of hopping amplitudes
that allows us to stabilize a superconducting solution in the doped Mott insulator regime.

\begin{figure}[!ht]
\centering
  \includegraphics[width=2.55in, trim=0.0cm 4.0cm 0.0cm 0.0cm]{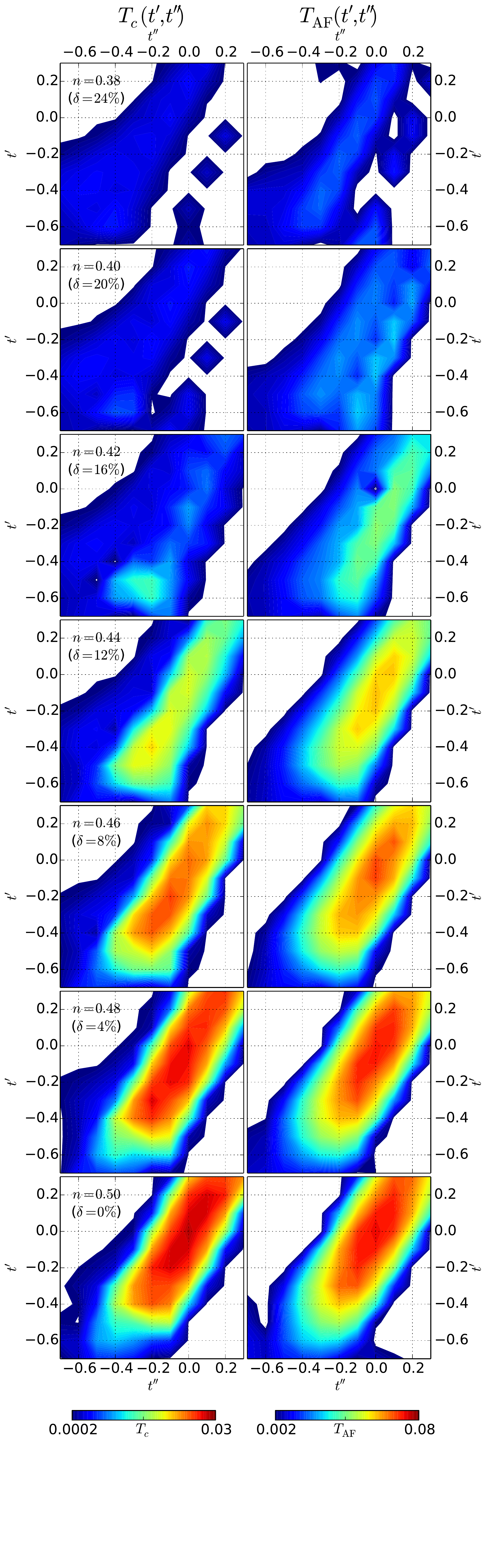}
\caption{$GW$ calculation of  $d$-wave $T_c$ (left panels) and $T_\mathrm{AF}$ (right panels) at $U/D=1$, $t=-1.0$,
for different values of $n$, as functions of $(t',t'')$.
$t'$ and $t''$ are sampled between (and including) $-0.7$ and $0.3$ with the
step $0.1$. 
$n$ is taken between (and including) $0.38$ and $0.5$
(i.e. the half-filling) with the step $0.02$. 
} 
 \label{fig:Tc_vs_Tneel}
\end{figure}

Fig. \ref{fig:Tc_vs_Tneel} presents the computation of the AF instability ($T_\mathrm{AF}$) and
the SC instability ($T_c$) in $GW$, for $U/D=1$ and various $t',t''$ ($t=-1.0$ is held fixed) and various dopings.
The temperature is taken from $0.2$ down to the lowest accessible temperature, but not below $0.01$ in cases where
the extrapolation of $\lambda(T)$ yielded no finite $T_c$. The temperature step
depends on $T$ (smaller step at lower $T$; see Appendix \ref{app:fit} for an example of raw data).

The first observation is that the region of high $T_c$ broadly coincides with the
region of high $T_\mathrm{AF}$. This is expected as
in $GW$ the attractive interaction comes from the spin-boson, and a high-valued
and sharply-peaked $W^\mathrm{sp}$ is clearly necessary for satisfying the gap
equation Eq.~\eqref{eq:trilex_lge} with $\lambda=1$. However, the maximum of
$T_c$ with respect to $(t',t'')$ at a fixed $n$ does not
coincide with the maximum of $T_\mathrm{AF}$, thus indicating that there
are factors other than sharpness (criticality) of the spin-boson which
contribute to the height of $T_c$. 
While the maximum of $T_\mathrm{AF}$ is
found rather close to $t'=t''=0$ at all dopings, the maximum in $T_c$ starts
from $(t',t'')=(-0.6,-0.4)$ at $n=0.38$ and gradually moves as $n$ is
increased. It is only at half-filling that the two maxima are found to
coincide. Furthermore, while at around $t'=t''=0$ and $t'\approx t''$ one sees
$T_\mathrm{AF}>T_c$, this trend is gradually reversed as $t''$ is made more
and more negative, such that around  $t'\approx t''+0.4$ one usually sees a
finite $T_c$ in the absence of a finite $T_\mathrm{AF}$.

\begin{figure*}[!ht]
\centering
  \includegraphics[width=6.4in, trim=2.0cm 1.0cm 2.0cm 2.0cm]{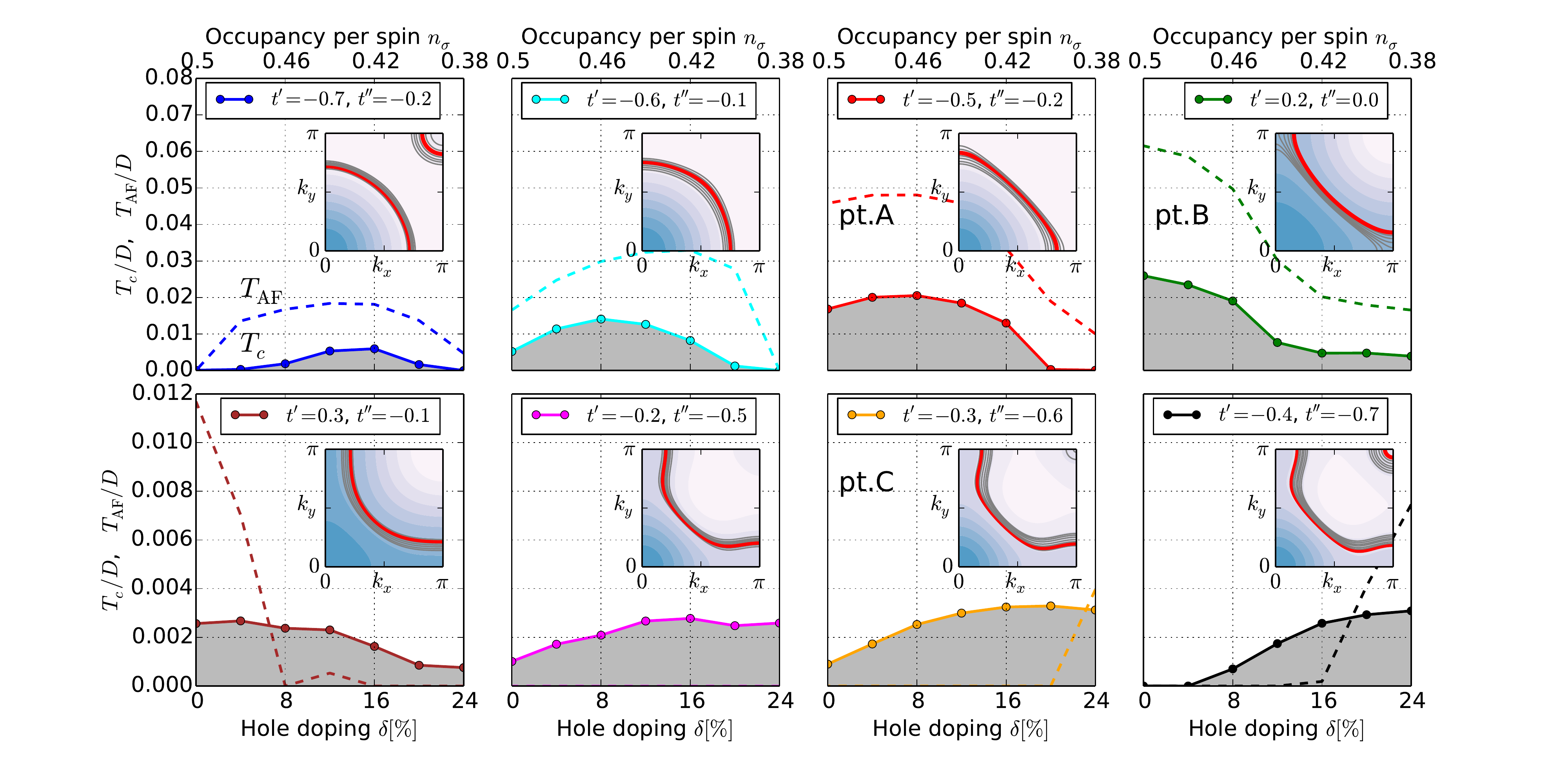}
  \caption{$GW$ calculations at $U/D=1$, $t=-1$. Dashed lines denote $T_\mathrm{AF}$, full lines $T_c$. 
Inset: color map for $\varepsilon_\mathbf{k}$. Gray contours denote bare Fermi
 surfaces at examined values of doping. The red line corresponds to the Fermi
 surface with maximum $T_c$.}
 \label{fig:Tc_vs_n_examples}
\end{figure*}

In Fig. \ref{fig:Tc_vs_n_examples}, we plot $T_\mathrm{AF}$ and $T_c$ vs doping 
for different values of $t',t''$. The corresponding dispersion (color map) and Fermi surfaces (gray contours; red for the maximal $T_c$)
are presented in the insets. 

\begin{figure}[!ht] 
\centering
  \includegraphics[width=3.2in, trim=1.0cm 1.0cm 1.0cm 0.0cm, page=1]{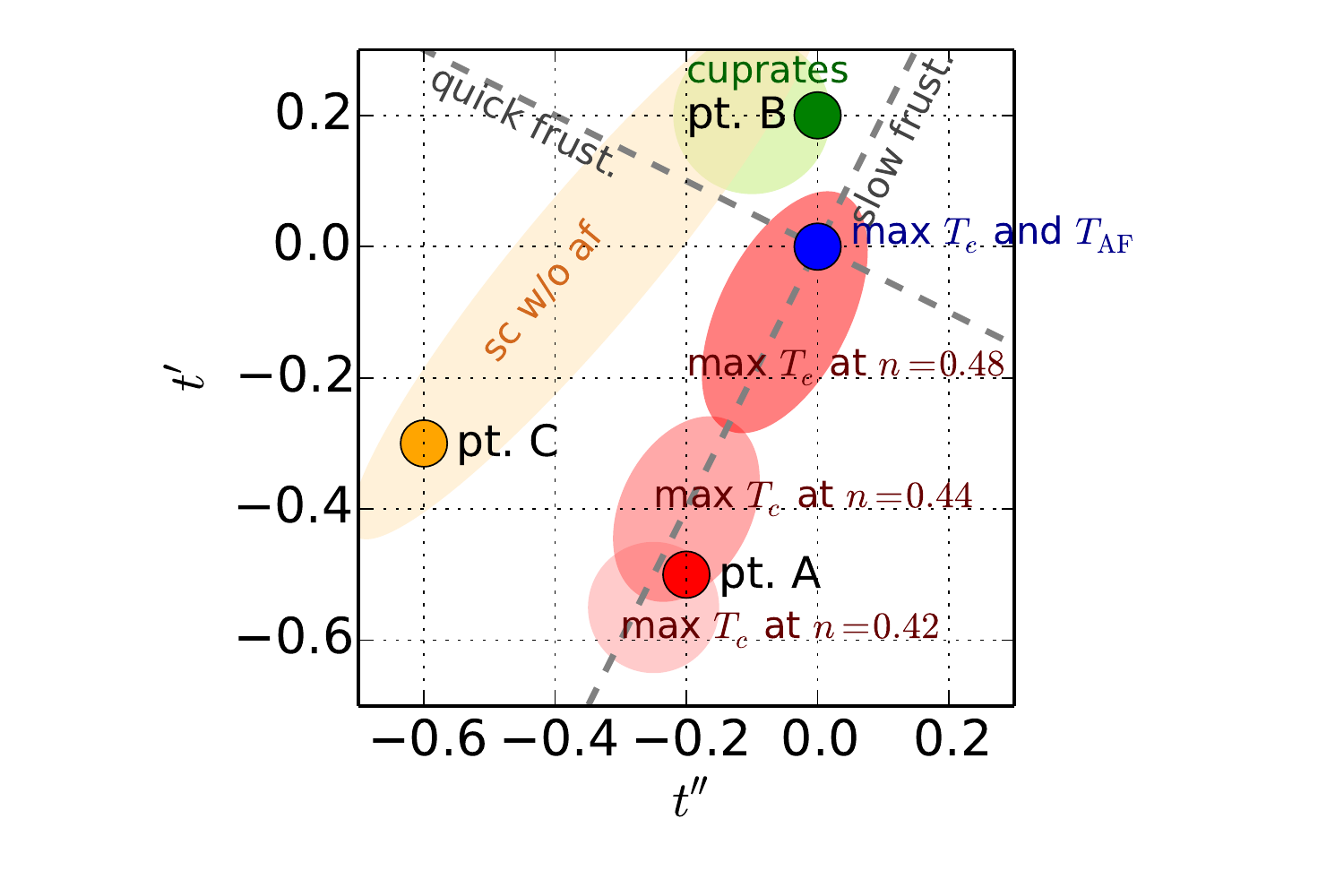}  
\caption{ Sketch of the $GW$ phase diagram at $U/D=1$, $t=-1.0$ based on Fig.\ref{fig:Tc_vs_Tneel}. Points A,B and C are of special interest, and are further studied at strong coupling. 
} 
\label{fig:tpts_phase_diagram_sketch}
\end{figure}

Finally, in Fig.\ref{fig:tpts_phase_diagram_sketch}, we summarize the observations from Fig.~\ref{fig:Tc_vs_Tneel}. The blue dot denotes the global
maximum of $T_c$ and $T_\mathrm{AF}$. The dashed gray lines denote the
directions of the slowest and quickest decay of antiferromagnetism. 
The red ellipses denote the
regions of maximal $T_c$, at various dopings. The yellow region is where one finds
little antiferromagnetism, but still a sizable $T_c$. The green region
corresponds to dispersions relevant for cuprates\cite{PavariniPRL2001}. 
The points A,B, and C are the dispersions that we focus on and for which we perform TRILEX and $GW$+EDMFT computations. 
Pt.~B is most relevant for the cuprates, and was analyzed in Fig.\ref{fig:SC_edmftgw_vs_trilex}. Pt.~C has $T_\mathrm{AF}<T_c$ which allows us to converge a superconducting solution at both weak and strong coupling. We analyze it in the next subsection. Pt.~A is where we observe an maximal $T_c$ at $16\%$ doping, and we focus on it in Section \ref{sec:ptA}.

%%%%%%%%%%%%%%%%%%%%%%%%%%%%%%%%%%%%%%%%%%%%%%%%%%%%%%%%%%%%%%%%%%%%%%%
%%%%%%%%%%%%%%%%%%%%%%%%%%%%%%%%%%%%%%%%%%%%%%%%%%%%%%%%%%%%%%%%%%%%%%%

\subsection{The nature of the superconducting phase at strong coupling} \label{sec:sc_phase}

In this section, we study the dispersion C $(t, t', t'')=  (-1, -0.3, -0.6)$.
In Figure \ref{fig:Tc_vs_n_examples}, we have determined that {\sl at weak coupling} ($U/D=1$), the superconducting temperature $T_c$ is larger than the 
AF temperature: we can therefore reach the superconducting phase numerically (see Appendix \ref{app:sc_weak_coupling}).
It turns out that at strong coupling, the AF instability is also absent. This allows us to stabilize superconducting solutions 
in the doped Mott insulator regime. 
We also perform a calculation restricted to the normal phase for all parameters in order to compare results to the ones in the SC phase.
For simplicity, in this section we will 
present only $GW$+EDMFT results for $U/D=4$. 

\begin{figure}[!ht] 
\centering
  \includegraphics[width=3.0in, trim=0.0cm 0.0cm 0.0cm 0.0cm, page=1]{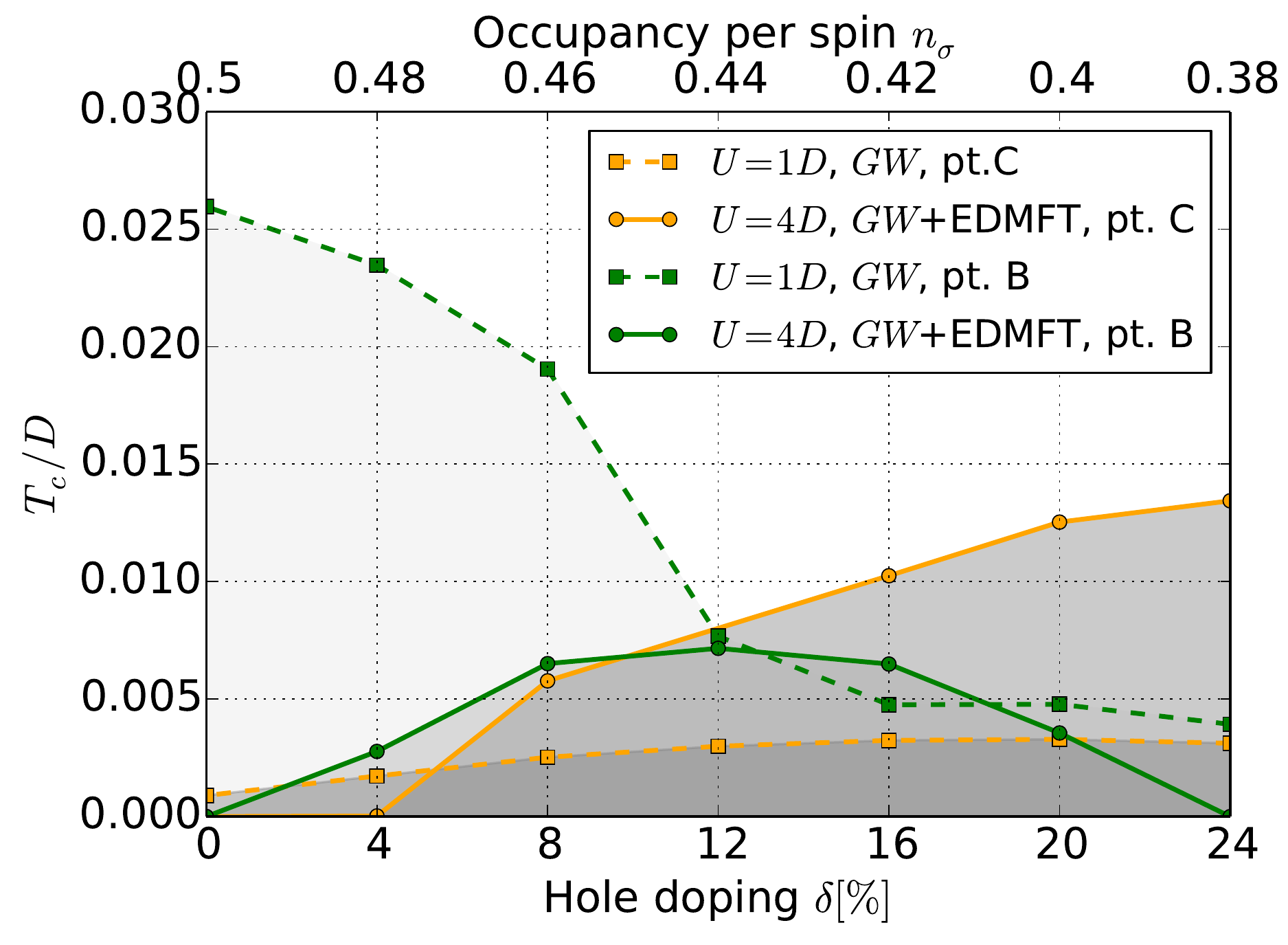}
\caption{ $T_c$ for dispersions B and C at weak and strong couplings.
}
\label{fig:ptC_Tcs}
\end{figure}

In Fig.~\ref{fig:ptC_Tcs}, we show the superconducting temperatures
at $U/D=1$ and $U/D=4$. Contrary to pt.B, in pt.C strong coupling seems to strongly enhance superconductivity. 
Also, the SC dome extends to higher dopings.

\begin{figure}[!ht]
%\centering
%  \includegraphics[width=3.0in, trim=0.0cm 0.0cm 0.0cm 0.0cm, page=2]{Figures/sc_phase_line_plots}
  \includegraphics[width=3.0in, trim=0.0cm 0.0cm 0.0cm 0.0cm, page=1]{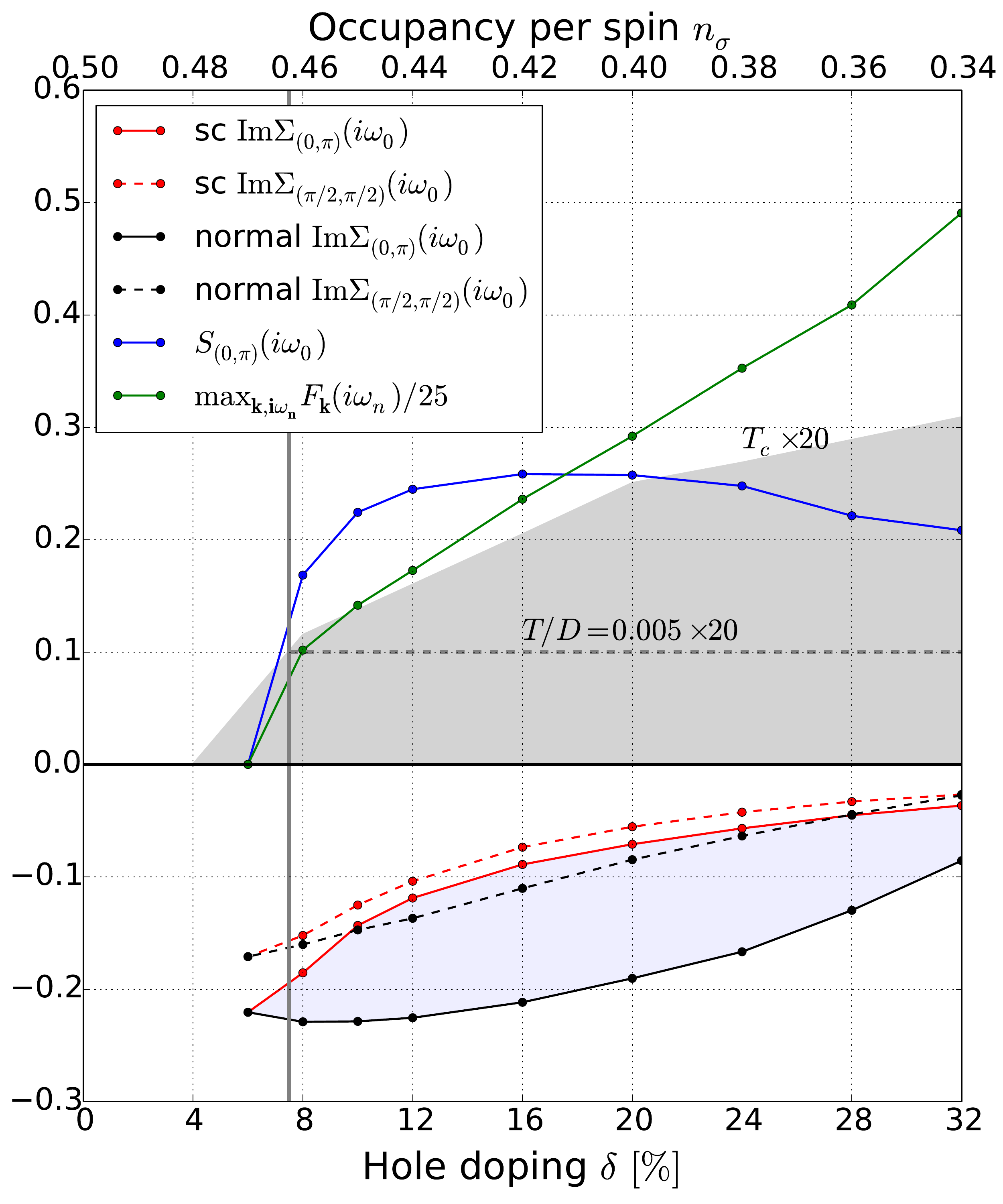}    
\caption{Evolution of various
quantities within the superconducting dome at dispersion pt.C, using 
$GW$+EDMFT, $U/D=4$, $T=0.005D$. 
The $T_c$, as obtained from $\lambda_m(T)$, is denoted by the gray area. 
Quantities are scaled to fit the same plot.
The gray dashed horizontal line denotes the temperature at
which the data is taken, relative to the (scaled) $T_c$. The vertical full line
denotes the end of the superconducting dome at the temperature denoted by the
dashed horizontal line, i.e. denotes the doping where all the anomalous
quantities are expected to go to zero.
 } 
 \label{fig:sc_phase_anomalous}
\end{figure}

In Fig.~\ref{fig:sc_phase_anomalous} we
show the results for the both the anomalous self-energy and Green's function, as well as the
imaginary part of the normal self-energy, in both the normal phase and
superconducting solution, anti-nodal and nodal regions. 

The imaginary part of the normal self-energy is larger at antinodes than at nodes
and is growing when approaching the Mott insulator.
When going from the normal phase to the SC phase, the imaginary part of the self-energy
is strongly reduced at the antinode and weakly reduced at the node. The difference between the normal and SC solution (light blue area) is roughly proportional to the anomalous self-energy in the SC phase (blue line).
Note that we observe a similar phenomenon even at weak coupling (see Appendix \ref{app:sc_weak_coupling}).

\begin{figure}[!ht]  
  \includegraphics[width=3.5in, trim=0.0cm 0.0cm 0.0cm 0.0cm, page=2]{spectral_plots_at_AN.pdf}    
  \includegraphics[width=3.5in, trim=0.0cm 0.0cm 0.0cm 0.0cm, page=1]{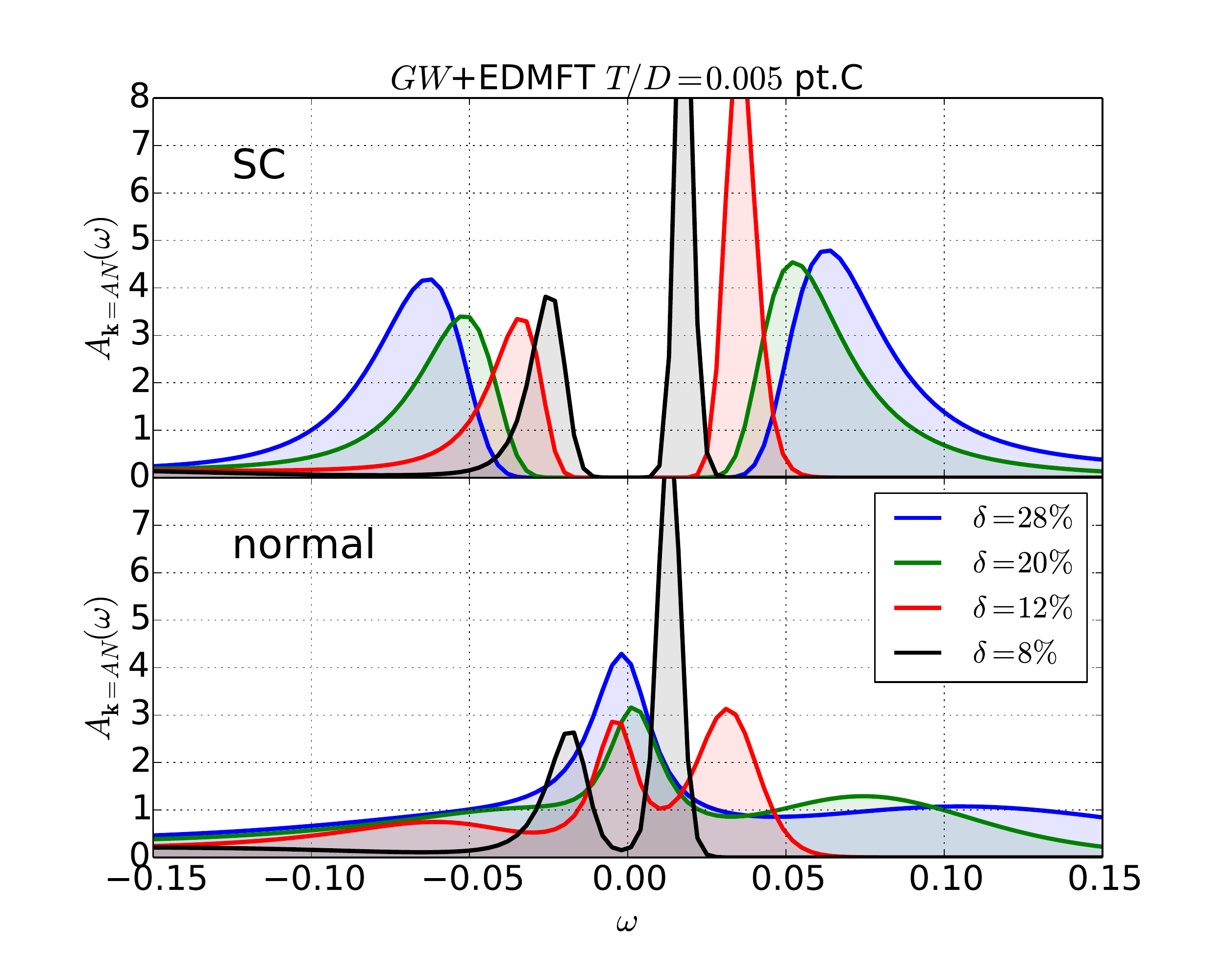}    
\caption{
Top panel: Spectral function versus frequency, at the anti-nodal wave vector, defined by $n_{\mathbf{k}_\mathrm{AN}=(\pi,k_x(\mathrm{AN}))} = 0.5$, 
obtained by maximum entropy method\cite{Bryan1990} from $G_\mathbf{k}(i\omega_n)$. 
$U/D=4$, $T/D=0.005$ for doping $\delta = 8,12,20,28 \%$.
Bottom panel: zoom in at low frequencies.
}
 \label{fig:mem}
\end{figure}

In Fig. \ref{fig:mem}, we plot the spectral function at the antinodes at low temperature, 
in the normal and in the superconducting phase.
At low doping,  we observe at low energy 
a pseudo-gap  in the normal phase and the superconducting gap in the SC phase.
The result obtained here is qualitatively different 
to the one obtained using 8 sites DCA cluster by Gull et al. \cite{GullParcolletMillisPRL2013, GullMillisPRB2015}.
In the cluster computations, the superconducting gap is smaller than the pseudogap, i.e. the 
quasi-particle peak at the edge of the SC gap appears within the pseudogap.
It is not the case here. %, where the pseudo gap {\it is} comparable to the SC gap.
Also, we do not see any ``peak-dip-hump'' structure. 
Note that we are however using different parameters (for the hoppings $t',t''$, the interaction $U$ and the doping $\delta$).
It is not clear at this stage whether these qualitative differences
are due to this different parameter regime 
or to an artifact of the single-site TRILEX method, e.g. the lack of local singlet physics in a single-site
impurity model. Further investigations with cluster-TRILEX methods are necessary in the SC phase.

\begin{figure}[!ht]
  \includegraphics[width=3.5in, trim=1.0cm 0.0cm 0.0cm 0.0cm]{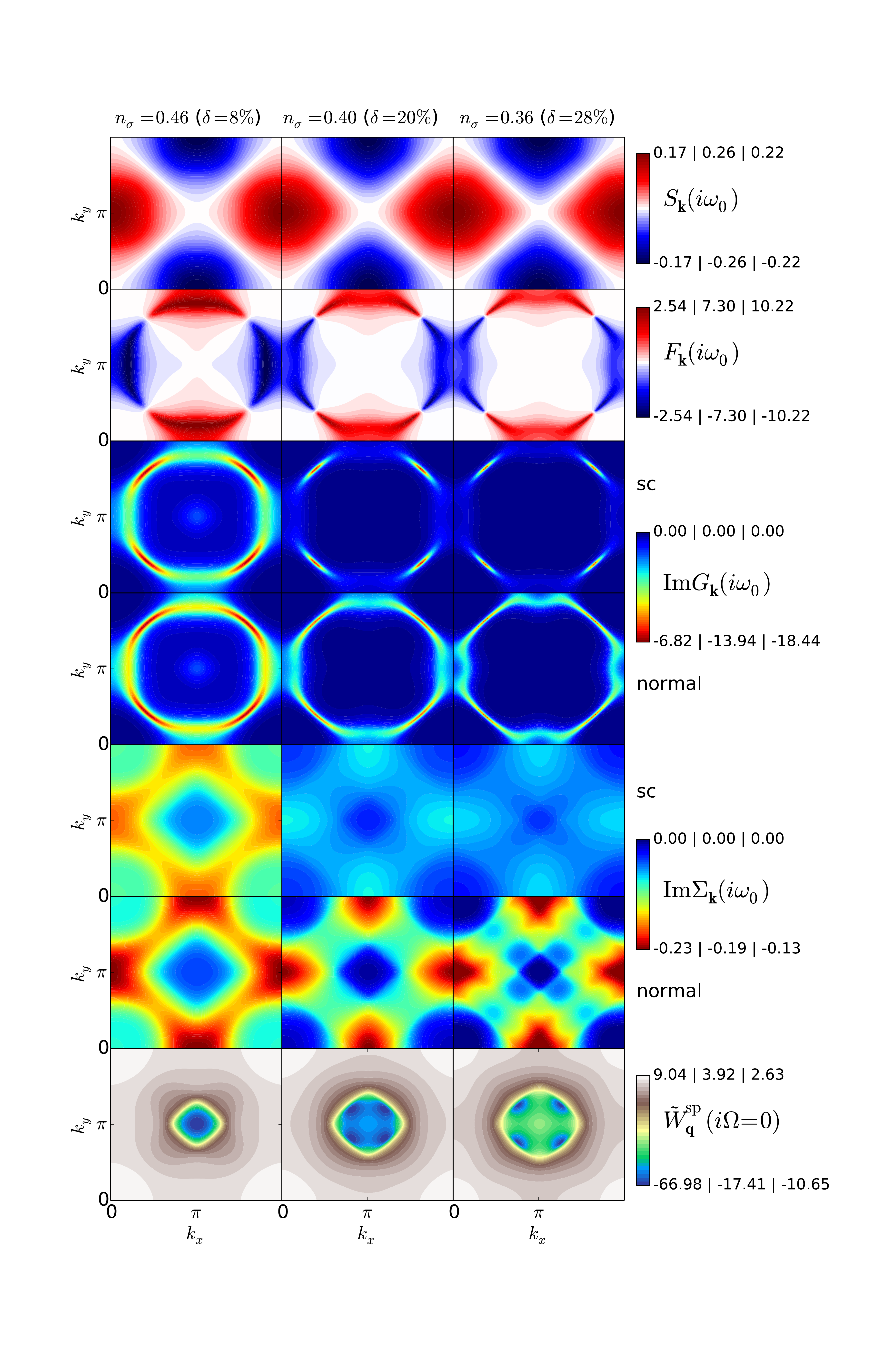}    
\caption{ Color plots of various quantities in the first Brillouin zone, at lowest Matsubara frequency. $GW$+EDMFT calculation at pt.~C dispersion, $U/D=4$. Temperature is below $T_c$, $T/D=0.005$. All plots correspond to the superconducting phase unless stated differently. The three numbers defining the colorbar range, correspond to 3 columns (different dopings) in the figure.
}
 \label{fig:SFG}
\end{figure}

In Fig. \ref{fig:SFG}, we plot various quantities at the lowest Matsubara frequency, as a function of $\mathbf{k}$.
In the first two rows we compare the anomalous self-energy and the pairing amplitude.
Both are clearly of $d$-wave symmetry. The pairing amplitude has a different
order of magnitude (see Appendix \ref{app:formalism} for an illustration of the
dependence between $F$,$G$,$\Sigma$ and $S$). In the third and fourth row we
show the imaginary part of the Green's function in the SC and normal phase. 
%In the superconducting phase, the spectral weight in the antinodal region is much
%more strongly suppressed. 
Due to the absence of long-lived quasiparticles in
this sector, the maximum of $F_\mathbf{k}$ is moved towards the nodes, and does
not coincide with the maximum of $S_\mathbf{k}$. At small doping, the Fermi
surface in both cases becomes less sharp and more featureless, due to proximity
to the Mott insulator. 
%In this regime, the details of the dispersion are the
%least expected to play a significant role, and a comparison to the cuprates is
%more justified. 
In the next two rows we show the imaginary part of the normal
self-energy. In the superconducting phase, $\mathrm{Im}\Sigma_\mathbf{k}$ is
strongly reduced in only anti-nodal regions, and thus flattened (made more
local). In the last row, we show the non-local part of the propagator for the
spin boson. At large doping we observe a splitting of resonance at $(\pi,\pi)$
which corresponds to incommensurate AF correlations (see
e.g. Ref.~\onlinecite{PfeutyOnufrievaPRB2002} for a similar phenomenon). 
Having that the Green's function at around $\mathbf{k}=(0,0)$ is quite featureless, and
that the boson is sharply peaked at zero frequency, the shape of the spin-boson
around $\mathbf{q}=(\pi,\pi)$ is similar to the self-energy at around
$\mathbf{k}=(\pi,\pi)$. This pattern is observed 
%in both quantities
at all three dopings.

%%%%%%%%%%%%%%%%%%%%%%%%%%%%%%%%%%%%%%%%%%%%%%%%%%%%%%%%%%%%%%%%%%%%%%%
%%%%%%%%%%%%%%%%%%%%%%%%%%%%%%%%%%%%%%%%%%%%%%%%%%%%%%%%%%%%%%%%%%%%%%%
%%%%%%%%%%%%%%%%%%%%%%%%%%%%%%%%%%%%%%%%%%%%%%%%%%%%%%%%%%%%%%%%%%%%%%%

\subsection{Strong-coupling $T_c$ at pt.A} \label{sec:ptA}

\begin{figure}[!ht] 
\centering  
  \includegraphics[width=3.2in, trim=0.0cm 0.0cm 0.0cm 0.0cm, page=1]{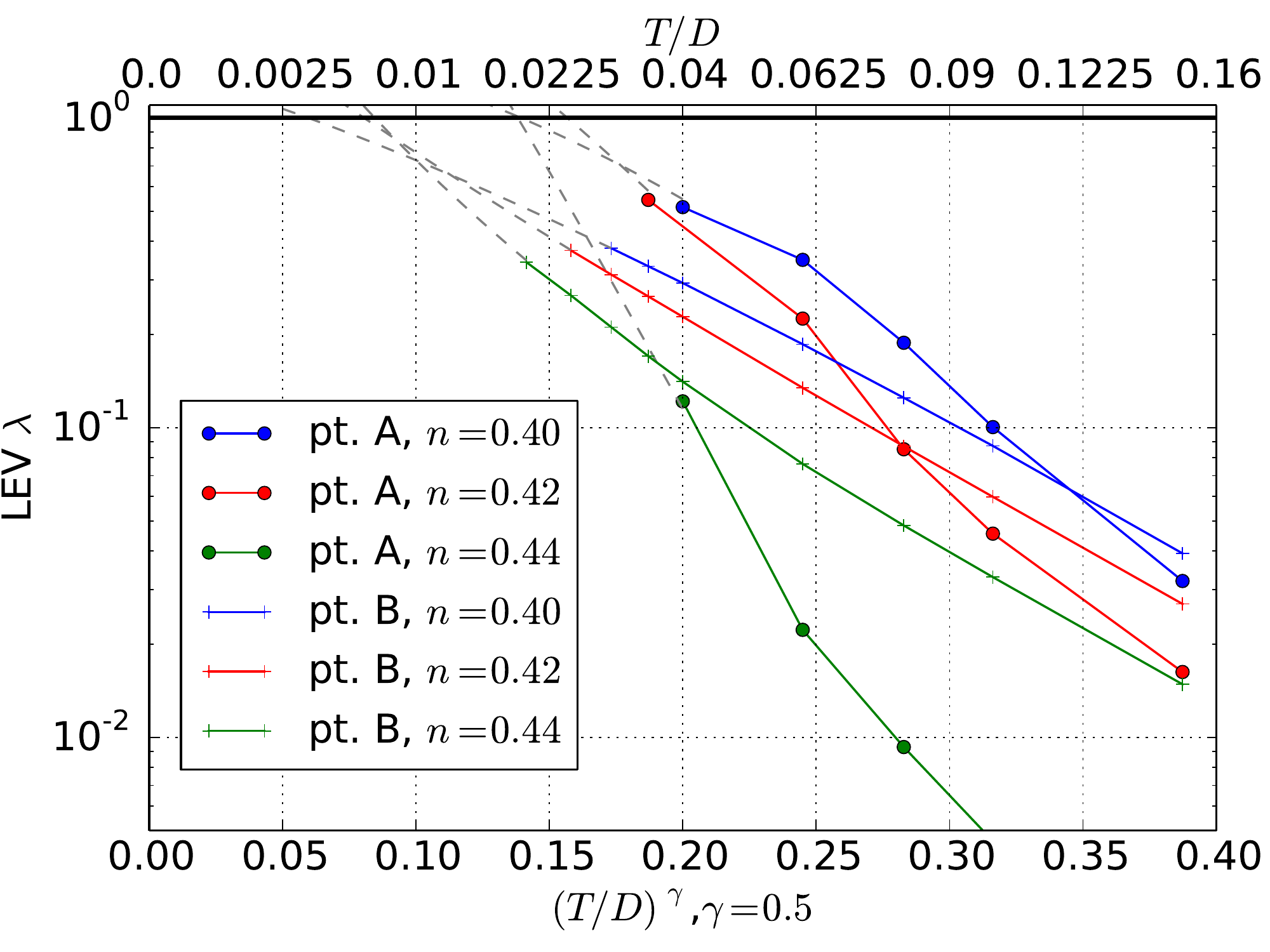}
  \includegraphics[width=3.2in, trim=0.0cm 0.0cm 0.0cm 0.0cm, page=1]{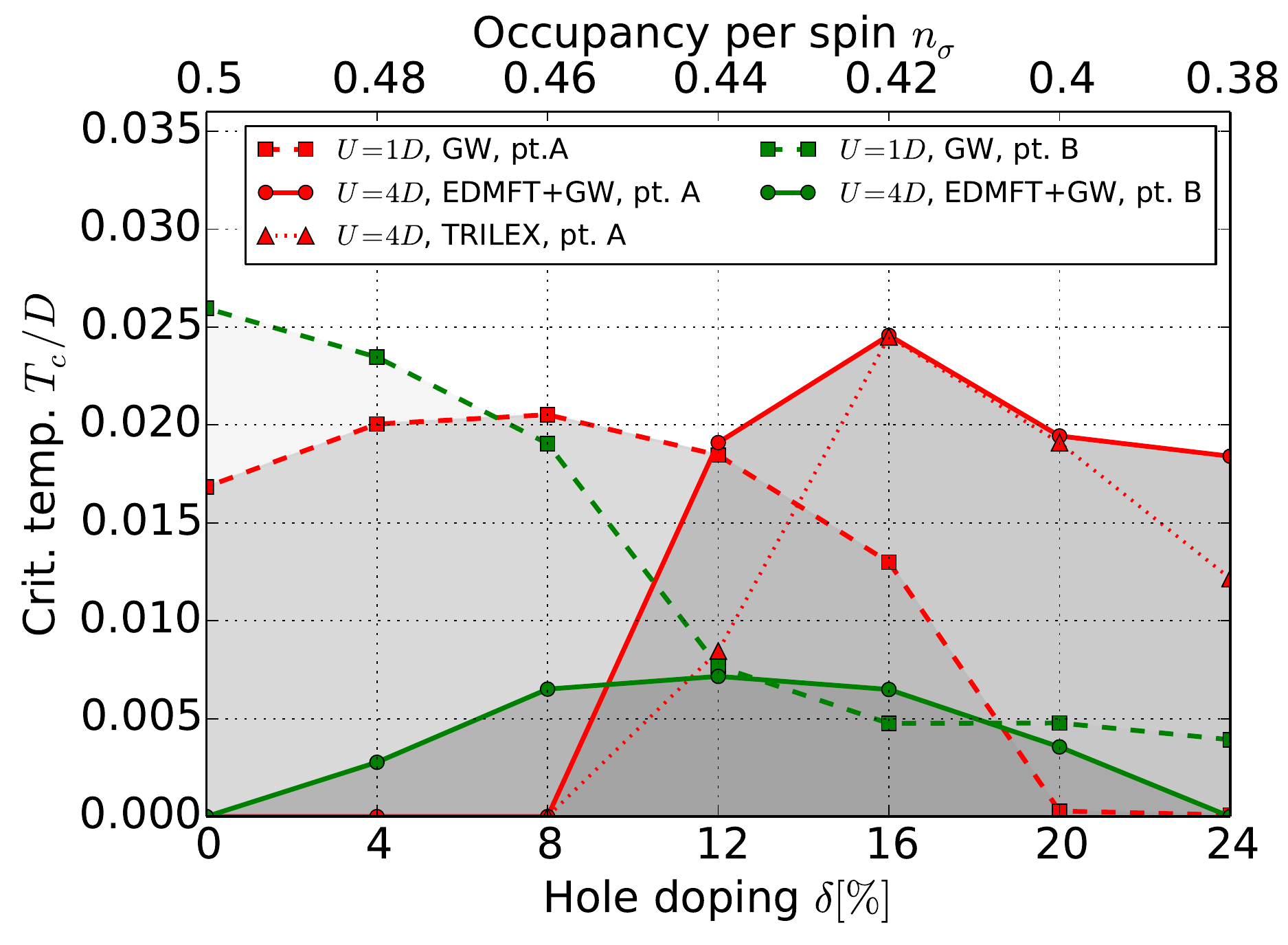}
\caption{ Top panel: evolution of the LGE leading eigenvalue $\lambda_m$ with temperature at pt.A and pt.B, in a $GW$+EDMFT calculation. Bottom panel: the extrapolated $T_c$ in both cases, including a TRILEX calculation at pt.A. } 
\label{fig:ptA}
\end{figure}

At weak coupling, we have observed in section \ref{sec:weak_coupling} that the
dispersion pt.A ($(t,t',t'')=(-1,-0.5,-0.2)$) presents a pronounced maximum in
$T_c(t',t'')$ at $16\%$ doping. Here, we investigate that point at strong
coupling using $GW$+EDMFT and TRILEX and find that also at $U/D=4$, the $T_c$
is substantially higher than in pt.B and pt.C. Here $T_c$ is below
$T_\mathrm{AF}$ and the result is again based on extrapolation of $\lambda$.
The proposed fitting function in this case does not perform as well and the
extrapolation is less reliable, but $GW$+EDMFT and TRILEX are in better
agreement than in the case of pt.B. A further investigation using cluster
methods is necessary since, apart from
Refs.~\onlinecite{TremblayPRB2008,JarrellPRB2013,MaierPRB2015}, little
systematic exploration of $T_c(t',t'')$ has been performed. 

%%%%%%%%%%%%%%%%%%%%%%%%%%%%%%%%%%%%%%%%%%%%%%%%%%%%%%%%%%%%%%%%%%%%%%%
%%%%%%%%%%%%%%%%%%%%%%%%%%%%%%%%%%%%%%%%%%%%%%%%%%%%%%%%%%%%%%%%%%%%%%%
%%%%%%%%%%%%%%%%%%%%%%%%%%%%%%%%%%%%%%%%%%%%%%%%%%%%%%%%%%%%%%%%%%%%%%%

\section{Conclusion}

In this work, we have generalized the TRILEX equations and their simplifications $GW$+EDMFT and $GW$ to the case of paramagnetic superconducting phases, using the Nambu formalism. We also generalized the corresponding Hedin equations.
We have then investigated within TRILEX, $GW$+EDMFT and $GW$ 
the doping-temperature phase diagram of the two-dimensional
single-band Hubbard model with various choices of hopping parameters.
In the case of a bare dispersion relevant for cuprates, in the doped Mott insulator regime, both TRILEX and $GW$+EDMFT yield a superconducting dome of $d_{x^2-y^2}$-wave symmetry, in qualitative agreement with earlier cluster DMFT calculations.
Let us emphasize that this was obtained at the  low cost of solving a \emph{single-site} impurity model.
At weak coupling, we have performed a systematic scan of
tight-binding parameter space within the $GW$ approximation. 
We have identified the region of parameter space
where superconductivity emerges at temperatures higher than antiferromagnetism. 
With one of those dispersions, we studied the properties of the
superconducting phase at strong coupling with $GW$+EDMFT. 
We also addressed the question of the optimal dispersion for superconductivity in
the Hubbard model at weak coupling. 
At $16\%$ doping we identify a candidate dispersion for the highest $d$-wave $T_c$, 
which remains to be investigated in detail at strong coupling (e.g. with cluster DMFT methods).

The next step will be to solve in the SC phase the recently developed cluster TRILEX methods \cite{Ayral2017c}.
Indeed, the single-site TRILEX method contains essentially 
an Eliashberg-like equation with a decoupling boson,
and a local vertex (computed from the self-consistent impurity model)
which has no anomalous components.
The importance of anomalous vertex components and 
the effect of local singlet physics (present in cluster methods)
is an important open question.
Note that the framework developed in this paper can also be used
to study more general pairings and decoupling schemes in TRILEX, e.g.
the effect of bosonic fluctuations in the particle-particle (i.e. superconducting) channel.

Finally, let us emphasize that the question of superconductivity in multi-orbital systems like iron-based superconductors
is another natural application of the TRILEX method, in particular in view of the strong AF fluctuations in these compounds.
In this multi-orbital case, being able to describe the SC phase without having to solve clusters 
(which are numerically very expensive within multi-orbital cluster DMFT\cite{Nomura2015b,Semon2016})
could prove to be very valuable.

\begin{acknowledgments}
We thank M. Kitatani for useful insights and discussion.
This work is supported by the FP7/ERC, under Grant Agreement No. 278472-MottMetals. Part
of this work was performed using HPC resources from GENCI-TGCC (Grant
No. 2016-t2016056112). The CT-HYB algorithm has been implemented using the TRIQS toolbox\cite{Parcollet2014}.

\end{acknowledgments}

%%%%%%%%%%%%%%%%%%%%%%%%%%%%%%%%%%%%%%%%%%%%%%%%%%%%%%%%%%%%%%%%%%%%%%%
%%%%%%%%%%%%%%%%%%%%%%%%%%%%%%%%%%%%%%%%%%%%%%%%%%%%%%%%%%%%%%%%%%%%%%%
%%%%%%%%%%%%%%%%%%%%%%%%%%%%%%%%%%%%%%%%%%%%%%%%%%%%%%%%%%%%%%%%%%%%%%%
%%%%%%%%%%%%%%%%%%%%%%%%%%%%%%%%%%%%%%%%%%%%%%%%%%%%%%%%%%%%%%%%%%%%%%%
%%%%%%%%%%%%%%%%%%%%%%%%%%%%%%%%%%%%%%%%%%%%%%%%%%%%%%%%%%%%%%%%%%%%%%%
%%%%%%%%%%%%%%%%%%%%%%%%%%%%%%%%%%%%%%%%%%%%%%%%%%%%%%%%%%%%%%%%%%%%%%%
%%%%%%%%%%%%%%%%%%%%%%%%%%%%%%%%%%%%%%%%%%%%%%%%%%%%%%%%%%%%%%%%%%%%%%%
%%%%%%%%%%%%%%%%%%%%%%%%%%%%%%%%%%%%%%%%%%%%%%%%%%%%%%%%%%%%%%%%%%%%%%%
%%%%%%%%%%%%%%%%%%%%%%%%%%%%%%%%%%%%%%%%%%%%%%%%%%%%%%%%%%%%%%%%%%%%%%%
%%%%%%%%%%%%%%%%%%%%%%%%%%%%%%%%%%%%%%%%%%%%%%%%%%%%%%%%%%%%%%%%%%%%%%%
%%%%%%%%%%%%%%%%%%%%%%%%%%%%%%%%%%%%%%%%%%%%%%%%%%%%%%%%%%%%%%%%%%%%%%%
%%%%%%%%%%%%%%%%%%%%%%%%%%%%%%%%%%%%%%%%%%%%%%%%%%%%%%%%%%%%%%%%%%%%%%%

\appendix

%%%%%%%%%%%%%%%%%%%%%%%%%%%%%%%%%%%%%%%%%%%%%%%%%%%%%%%%%%%%%%%%%%%%%%%
%%%%%%%%%%%%%%%%%%%%%%%%%%%%%%%%%%%%%%%%%%%%%%%%%%%%%%%%%%%%%%%%%%%%%%%
%%%%%%%%%%%%%%%%%%%%%%%%%%%%%%%%%%%%%%%%%%%%%%%%%%%%%%%%%%%%%%%%%%%%%%%
%%%%%%%%%%%%%%%%%%%%%%%%%%%%%%%%%%%%%%%%%%%%%%%%%%%%%%%%%%%%%%%%%%%%%%%
%%%%%%%%%%%%%%%%%%%%%%%%%%%%%%%%%%%%%%%%%%%%%%%%%%%%%%%%%%%%%%%%%%%%%%%
%%%%%%%%%%%%%%%%%%%%%%%%%%%%%%%%%%%%%%%%%%%%%%%%%%%%%%%%%%%%%%%%%%%%%%%
%%%%%%%%%%%%%%%%%%%%%%%%%%%%%%%%%%%%%%%%%%%%%%%%%%%%%%%%%%%%%%%%%%%%%%%
%%%%%%%%%%%%%%%%%%%%%%%%%%%%%%%%%%%%%%%%%%%%%%%%%%%%%%%%%%%%%%%%%%%%%%%
%%%%%%%%%%%%%%%%%%%%%%%%%%%%%%%%%%%%%%%%%%%%%%%%%%%%%%%%%%%%%%%%%%%%%%%
%%%%%%%%%%%%%%%%%%%%%%%%%%%%%%%%%%%%%%%%%%%%%%%%%%%%%%%%%%%%%%%%%%%%%%%
%%%%%%%%%%%%%%%%%%%%%%%%%%%%%%%%%%%%%%%%%%%%%%%%%%%%%%%%%%%%%%%%%%%%%%%
%%%%%%%%%%%%%%%%%%%%%%%%%%%%%%%%%%%%%%%%%%%%%%%%%%%%%%%%%%%%%%%%%%%%%%%

\section{Details of derivations}

\subsection{Relation between $\boldsymbol{\chi}^3$ and $\tilde{\boldsymbol{\chi}}^3$}
\label{sec:chi3_relations}
Let us define the following correlation functions:
\begin{subequations}
\begin{align}
  \boldsymbol{\chi}^{3}_{uv\alpha} & \equiv \Big\langle\bPsi_u\bPsi_v\phi_\alpha\Big\rangle\label{eq:chi3_def} \\
\boldsymbol{\chi}^{3,\mathrm{disc}}_{uv\alpha}(\tau) & \equiv \Big\langle\bPsi_u\bPsi_v\Big\rangle\Big\langle\phi_\alpha\Big\rangle \\
\tilde{\boldsymbol{\chi}}^{3}_{uv\alpha} & \equiv \Big\langle\bPsi_u\bPsi_v
\Big(\bPsi_x\boldsymbol{\lambda}_{xw\alpha}\bPsi_w\Big)\Big\rangle \\
\tilde{\boldsymbol{\chi}}^{3,\mathrm{disc}}_{uv\alpha}(\tau) & \equiv  \Big\langle\bPsi_u\bPsi_v\Big\rangle\Big\langle\bPsi_x\boldsymbol{\lambda}_{xw\alpha}\bPsi_w\Big\rangle \\
%\boldsymbol{\chi}^{3,\mathrm{conn}}_{uv\alpha} & \equiv \boldsymbol{\chi}^{3}_{uv\alpha} - \boldsymbol{\chi}^{3,\mathrm{disc}}_{uv\alpha}\\
\tilde{\boldsymbol{\chi}}^{3,\mathrm{conn}}_{uv\alpha} & \equiv \tilde{\boldsymbol{\chi}}^{3}_{uv\alpha} - \tilde{\boldsymbol{\chi}}^{3,\mathrm{disc}}_{uv\alpha}
\end{align}
\end{subequations}
In this section, we derive useful relations between these quantities.

Let us introduce source fields in the electron-boson action, Eq.~\ref{eq:S_eb}:
\begin{eqnarray} \nonumber
 S^\mathrm{Nambu}_{\mathrm{eb}}[\bPsi,\phi] &=&-\frac{1}{2}\bPsi_{u}\left[\boldsymbol{G}_{0}^{-1}-\boldsymbol{F}\right]_{uv}\bPsi_{v} \\ \nonumber
 &&-\frac{1}{2}\phi_{\alpha}\left[W_{0}^{-1}\right]_{\alpha\beta}\phi_{\beta}\nonumber \\
  &&+\frac{1}{2}\phi_{\alpha}\bPsi_{u}\boldsymbol{\lambda}_{uv\alpha}\bPsi_{v}-H_{\alpha}\phi_{\alpha}\label{eq:S_eb_with_sources}
\end{eqnarray}
We may now write
\begin{eqnarray}
  \boldsymbol{\chi}^{3}_{uv\alpha} &=& - \frac{2}{Z}\left.\frac{\partial^2 Z}{\partial \boldsymbol{F}_{uv} \partial H_\alpha}\right|_{\boldsymbol{F},H=0}\label{eq:chi3_derivative} \\
  \boldsymbol{\chi}^{3,\mathrm{disc}}_{uv\alpha}&=&- \frac{2}{Z^2}\left.\frac{\partial Z}{\partial \boldsymbol{F}_{uv}}\right|_{\boldsymbol{F},H=0}\left.\frac{\partial Z}{\partial H_\alpha}\right|_{\boldsymbol{F},H=0}\label{eq:chi3tilde_derivative}
\end{eqnarray}
Let us now integrate out the bosonic degrees of freedom in Eq.~\ref{eq:S_eb_with_sources}. We obtain:
\begin{equation}
 Z = \int {\cal D}[\boldsymbol{\Psi}] e^{-S_{\mathrm{ee}}^{\mathrm{Nambu}}[\boldsymbol{\Psi}]}
 \label{eq:Z_with_ee_Nambu}
\end{equation}
with 
\begin{align} \label{eq:latt_ee_action_Psi}
 & S_{\mathrm{ee}}^{\mathrm{Nambu}}[\boldsymbol{\Psi}]= \frac{1}{2}\boldsymbol{\Psi}_u[-\boldsymbol{G}_{0}^{-1}+\boldsymbol{F}]_{uv}\boldsymbol{\Psi}_v\nonumber \\ 
 &+\frac{1}{2} W_{0,\alpha\beta} \Bigg( H_\alpha - \frac{\boldsymbol{\Psi}_u \boldsymbol{\lambda}_{uv\alpha} \boldsymbol{\Psi}_v}{2} \Bigg)
 \Bigg( H_\beta - \frac{\boldsymbol{\Psi}_x \boldsymbol{\lambda}_{xw\beta} \boldsymbol{\Psi}_w}{2} \Bigg)
\end{align}

We now perform the derivatives of Eqs.~\ref{eq:chi3_derivative} and \ref{eq:chi3tilde_derivative} using the new expression Eq.~\ref{eq:Z_with_ee_Nambu}, yielding:
\begin{eqnarray}
 \boldsymbol{\chi}^{3}_{uv\alpha} &=&
 -2\Big\langle\frac{1}{2}\bPsi_u\bPsi_v\frac{1}{2}W_{0\alpha\beta}(-2)\frac{\bPsi_x \boldsymbol{\lambda}_{xw\beta} \bPsi_{w}}{2} \Big\rangle \\
 \boldsymbol{\chi}^{3,\mathrm{disc}}_{uv\alpha} &=&
 -2\Big\langle\frac{1}{2}\bPsi_u\bPsi_v\Big\rangle\Big\langle\frac{1}{2}W_{0,\alpha\beta}(-2)\frac{\bPsi_x \boldsymbol{\lambda}_{xw\beta} \bPsi_w }{2} \Big\rangle
\end{eqnarray}
Thus, we have, for the full correlator, as well as for the connected and disconnected parts:
\begin{equation}
  \boldsymbol{\chi}^3_{uv\alpha} = \frac{1}{2}W_{0,\alpha\beta}\tilde{\boldsymbol{\chi}}^3_{uv\beta} \label{eq:chi3_chi3tilde_rel}
\end{equation}

\subsection{Derivation of Hedin equations from Equations of motion}
\label{sec:eom}
In this section, we derive the Hedin equations of the main text using the Dyson-Schwinger equation of motion technique\cite{Zinn-Justin2002} already used in Ref.~\onlinecite{Ayral2015c}.
\subsubsection{E.O.M. for the self-energy}

Since the functional integral of a total derivative
vanishes:
\begin{equation}
\int{\cal D}[\boldsymbol{\Psi}]\frac{\partial(f[\boldsymbol{\Psi}]g[\boldsymbol{\Psi}])}{\partial\boldsymbol{\Psi}_{x}}=0
\end{equation}
for any $f$ and $g$, we have 
\begin{equation} \label{eq:Leibniz}
-(-)^{\deg f}\int{\cal D}[\boldsymbol{\Psi}]f[\boldsymbol{\Psi}]\frac{\partial g[\boldsymbol{\Psi}]}{\partial\boldsymbol{\Psi}_{x}}=\int{\cal D}[\boldsymbol{\Psi}]\left(\frac{\partial f[\boldsymbol{\Psi}]}{\partial\boldsymbol{\Psi}_{x}}\right)g[\boldsymbol{\Psi}]
\end{equation}
which comes directly from the Leibniz derivation rule for Grassmann variables. $\deg f$ denotes the degree of the polynomial $f$ in the variable $\bPsi$.
Let us now assume $f[\Psi]=e^{-S_{0}[\boldsymbol{\Psi}]}=e^{\frac{1}{2}\bPsi_u\boldsymbol{G}_{0,uv}^{-1}\bPsi_v}$
and $g[\boldsymbol{\Psi}]=h[\Psi]e^{-V[\boldsymbol{\Psi}]}$, with $h$ containing an odd number of Grassmann fields. $f$ has an infinite number of terms but all are products of an even number of $\bPsi$ fields. We obtain
\begin{eqnarray*}
-\int{\cal D}\left\{ \frac{\partial h}{\partial\boldsymbol{\Psi}_{x}}-h\left(-\frac{\partial V}{\partial\boldsymbol{\Psi}_{x}}\right)\right\} e^{-(S_{0}+V)}=\\
\left[\boldsymbol{G}_{0}^{-1}\right]_{xw}\int{\cal D}[\boldsymbol{\Psi}]\boldsymbol{\Psi}_{w}he^{-(S_{0}+V)}
\end{eqnarray*}
On the l.h.s. we have again used the Leibniz rule with $\deg h$ assumed to be odd, hence the extra minus sign. 
On the r.h.s similarly, $\deg \bPsi = 1$, and $\boldsymbol{G}_{0,uv}^{-1} = -\boldsymbol{G}_{0,vu}^{-1}$, so the $\frac{1}{2}$ prefactor is canceled.
Both integrals are now averages with respect to the action $S=S_{0}+V$,
namely 
\begin{equation}
\Big\langle\frac{\partial h}{\partial\boldsymbol{\Psi}_{x}}+h[\boldsymbol{\Psi}]\frac{\partial V}{\partial\boldsymbol{\Psi}_{x}}\Big\rangle=-\left[\boldsymbol{G}_{0}^{-1}\right]_{xw}\Big\langle\boldsymbol{\Psi}_{w}h[\boldsymbol{\Psi}]\Big\rangle\label{eq:general_eom}
\end{equation}

Let us now consider the case when $h\equiv\boldsymbol{\Psi}_{v}$,
and $V$ is the interacting part of the electron-electron action \eqref{eq:latt_ee_action_Psi},
with the source field $H$ set to zero, i.e. $V\equiv\frac{1}{8}\left[W_{0}\right]_{\alpha\beta}\left(\boldsymbol{\Psi}_{u}\boldsymbol{\lambda}_{uw\alpha}\boldsymbol{\Psi}_{w}\right)\left(\boldsymbol{\Psi}_{y}\boldsymbol{\lambda}_{yz\beta}\boldsymbol{\Psi}_{z}\right)$.
We get 
\begin{align}
 & \delta_{xv}+\frac{1}{8}\left[W_{0}\right]_{\alpha\beta}\boldsymbol{\lambda}_{xw\alpha}\cdot4\langle\boldsymbol{\Psi}_{v}\boldsymbol{\Psi}_{w}\left(\bPsi_{y}\boldsymbol{\lambda}_{yz\beta}\bPsi_{z}\right)\rangle\\
 & =-\left[\boldsymbol{G}_{0}^{-1}\right]_{xw}\langle\boldsymbol{\Psi}_{w}\boldsymbol{\Psi}_{v}\rangle
\end{align}
Multiplying both sides by $\boldsymbol{G}_{0}$ and using Eqs.~\eqref{eq:chi3_def} and \eqref{eq:chi3_chi3tilde_rel}: 
\begin{eqnarray}
\boldsymbol{G}_{uv} & = & \boldsymbol{G}_{0,uv}-\frac{1}{2}\boldsymbol{G}_{0,ux}W_{0,\alpha\beta}\boldsymbol{\lambda}_{xw\alpha}\tilde{\boldsymbol{\chi}}_{wv\beta}^{3}\\
 & = & \boldsymbol{G}_{0,uv}-\boldsymbol{G}_{0,ux}\boldsymbol{\lambda}_{xw\alpha}\boldsymbol{\chi}_{wv\alpha}^{3}\nonumber \\
 & = & \boldsymbol{G}_{0,uv}-\boldsymbol{G}_{0,ux}\boldsymbol{\lambda}_{xw\alpha}\left(\boldsymbol{\chi}_{wv\alpha}^{3,\mathrm{conn}}+\frac{1}{2}W_{0,\alpha\beta}\tilde{\boldsymbol{\chi}}_{wv\alpha}^{3,\mathrm{disc}}\right)\nonumber \\
 & = & \boldsymbol{G}_{0,uv}-\boldsymbol{G}_{0,ux}\boldsymbol{\lambda}_{xw\alpha}\boldsymbol{G}_{wy}W_{\alpha\beta}\boldsymbol{\Lambda}_{yz\beta}\boldsymbol{G}_{zv}\nonumber \\
 &  & -\boldsymbol{G}_{0,ux}\boldsymbol{\lambda}_{xw\alpha}\frac{1}{2}W_{0,\alpha\beta}\langle\bPsi_{y}\boldsymbol{\lambda}_{yz\beta}\bPsi_{z}\rangle(-\boldsymbol{G}_{wv})\nonumber 
\end{eqnarray}
Since the self-energy is defined as 
\begin{equation}
\boldsymbol{G}_{uv}=\boldsymbol{G}_{0,uv}+\boldsymbol{G}_{0,ux}\boldsymbol{\Sigma}_{xw}\boldsymbol{G}_{wv}\label{eq:Dyson_G}
\end{equation}
we obtain 
\begin{equation}\label{eq:Sigma_final_with_indices}  
\boldsymbol{\Sigma}_{uv}=-\boldsymbol{\lambda}_{uw\alpha}\boldsymbol{G}_{wx}W_{\alpha\beta}\boldsymbol{\Lambda}_{xv\beta}+\boldsymbol{\lambda}_{uv\alpha}\frac{1}{2}W_{0,\alpha\beta}\langle\bPsi_{y}\boldsymbol{\lambda}_{yz\beta}\bPsi_{z}\rangle 
\end{equation}
The second term is the Hartree term (note the 1/2 factor). The Fock term is included in the first term.

\subsubsection{E.O.M. for the polarization}

Real fields $\phi$ commute with the derivative, so the Leibniz rule is simpler. Analogously to Eq.~\eqref{eq:Leibniz}
\begin{equation} \label{eq:bosonic_Leibniz}
-\int{\cal D}[\phi,\bPsi]f[\phi,\boldsymbol{\Psi}]\frac{\partial g[\phi,\boldsymbol{\Psi}]}{\partial\phi_{\gamma}}=\int{\cal D}[\phi,\boldsymbol{\Psi}]\left(\frac{\partial f[\phi,\boldsymbol{\Psi}]}{\partial\phi_{\gamma}}\right)g[\phi,\boldsymbol{\Psi}]
\end{equation}
Similarly to Eq. \eqref{eq:general_eom}, by taking $f[\phi,\bPsi]=e^{-S_{0}[\boldsymbol{\Psi},\phi]}$,
where $S_{0}$ is the non-interacting part of the electron-boson action
\eqref{eq:eb_action}, and $V[\boldsymbol{\Psi},\phi]=\frac{1}{2}\boldsymbol{\Psi}_{u}\boldsymbol{\lambda}_{uv\delta}\boldsymbol{\Psi}_{v}\phi_{\delta}$,
one has 
\begin{equation}
\Bigg\langle\frac{\partial h}{\partial\phi_{\gamma}}-\frac{1}{2}\boldsymbol{\Psi}_{u}\boldsymbol{\lambda}_{uv\gamma}\boldsymbol{\Psi}_{v}h[\phi]\Bigg\rangle=-\left[W_{0}^{-1}\right]_{\gamma\beta}\big\langle\phi_{\beta}h[\phi]\big\rangle\label{eq:Dyson_Schwinger_bosonic}
\end{equation}
Again, note the minus sign in the left-hand side (to be compared with Eq.~\eqref{eq:general_eom}) coming from the bosonic nature of the field $\phi$.
For $h\equiv\phi_{\alpha}-\langle\phi_{\alpha}\rangle$,
\begin{align*}
 & \delta_{\gamma\alpha}-\frac{1}{2}\boldsymbol{\lambda}_{uv\gamma}\langle\boldsymbol{\Psi}_{u}\boldsymbol{\Psi}_{v}(\phi_{\alpha}-\langle\phi_{\alpha}\rangle)\rangle=\\
 & -\left[W_{0}^{-1}\right]_{\gamma\beta}\langle(\phi_{\beta}-\langle\phi_{\beta}\rangle)(\phi_{\alpha}-\langle\phi_{\alpha}\rangle)\rangle
\end{align*}
Multiplying by $W_{0}$ and using Eqs.~\ref{eq:chi3conn_def} and \ref{eq:Lambda_def}, we obtain
\begin{eqnarray}
W_{\delta\alpha}&=&W_{0,\delta\alpha}+W_{0,\delta\gamma}\frac{1}{2}\boldsymbol{\lambda}_{uv\gamma}\boldsymbol{\chi}^{3,\mathrm{conn}}_{vu\alpha}\nonumber\\
&=&W_{0,\delta\alpha}+W_{0,\delta\gamma}\frac{1}{2}\boldsymbol{\lambda}_{uv\gamma}\boldsymbol{G}_{vx}\boldsymbol{G}_{wu} \boldsymbol{\Lambda}_{xw,\beta} W_{\beta\alpha}\nonumber
\end{eqnarray}
With the definition of $P$ as 
\begin{equation}
W_{\delta\alpha}=W_{0,\delta\alpha}+W_{0,\delta\gamma}P_{\gamma\beta}W_{\beta\alpha}\label{eq:Dyson_W}
\end{equation}
we identify
\begin{equation}
  P_{\gamma\beta} = \frac{1}{2}\boldsymbol{\lambda}_{uv\gamma}\boldsymbol{G}_{vx}\boldsymbol{G}_{wu} \boldsymbol{\Lambda}_{xw\beta}
\end{equation}
Note the extra prefactor $\frac{1}{2}$ compared to the normal-case expression.

\subsection{Proof that $P$ is real} \label{app:realness_of_P}

In the derivation of Eq.~\eqref{eq:P_Nambu} we have used the symmetries of $G$,$F$ and $\Lambda$. It turns out that the imaginary part of $\Lambda$ does not play a role in the summation and that the polarization is strictly real.

The renormalized vertex has the following symmetries\cite{Ayral2015c}
\begin{subequations}
\begin{eqnarray}
 \Lambda(i\omega,-i\Omega) &=& \Lambda(i\omega-i\Omega,i\Omega) \\
 \Lambda^*(i\omega,-i\Omega) &=& \Lambda(-i\omega,i\Omega)
\end{eqnarray}
\end{subequations}
Under the present assumptions, all components of the Green's function ($G$ and $F$) have the property
$$X_\mathbf{k}(-i\omega) = X^*_\mathbf{k}(i\omega)$$
$$X_\mathbf{k}(i\omega) = X_{-\mathbf{k}}(i\omega)$$
Therefore
\begin{align} \label{eq:no_star}
 & \sum_{\mathbf{k},i\omega} X_{\mathbf{k}}(i\omega)X_{\mathbf{k}+\mathbf{q}}(i\omega+i\Omega)\Lambda(i\omega,i\Omega)  \\ \nonumber
 = & \sum_{\mathbf{k},i\omega} X_{\mathbf{k}}(-i\omega)X_{\mathbf{k}+\mathbf{q}}(-i\omega+i\Omega)\Lambda(-i\omega,i\Omega)\\ \nonumber
 = & \sum_{\mathbf{k},i\omega} X_{\mathbf{k}}(-i\omega)X_{\mathbf{k}+\mathbf{q}}(-i\omega+i\Omega)\Lambda^*(i\omega,-i\Omega)\\ \nonumber
 = & \sum_{\mathbf{k},i\omega} X_{\mathbf{k}}(-i\omega)X_{\mathbf{k}+\mathbf{q}}(-i\omega+i\Omega)\Lambda^*(i\omega-i\Omega,i\Omega) \\ \nonumber
 = & \sum_{\mathbf{k},i\omega'}X_{\mathbf{k}}(-i\omega'-i\Omega)X_{\mathbf{k}+\mathbf{q}}(-i\omega')\Lambda^*(i\omega',i\Omega)\\  \nonumber
 = & \sum_{\mathbf{k},i\omega'}X^*_{\mathbf{k}}(i\omega'+i\Omega)X^*_{\mathbf{k}+\mathbf{q}}(i\omega')\Lambda^*(i\omega',i\Omega)\\ \nonumber
 = & \Bigg[\sum_{\mathbf{k}',i\omega'}X_{-\mathbf{k}'-\mathbf{q}}(i\omega'+i\Omega)X_{-\mathbf{k}'}(i\omega')\Lambda(i\omega',i\Omega)\Bigg]^* \\
 = & \Bigg[\sum_{\mathbf{k}',i\omega'}X_{\mathbf{k}'+\mathbf{q}}(i\omega'+i\Omega)X_{\mathbf{k}'}(i\omega')\Lambda(i\omega',i\Omega)\Bigg]^* \label{eq:star_makes_no_diff}
\end{align}
which proves that the polarization is real. In the derivation of the first term in Eq.~\eqref{eq:P_Nambu}, we have used the equality between Eq.~\eqref{eq:no_star} and Eq.~\eqref{eq:star_makes_no_diff}. Then, for any real-valued $F$, we furthermore have $F_\mathbf{k}(i\omega)=F_\mathbf{k}(-i\omega)$, which gives us
\begin{align}
 & \sum_{\mathbf{k},i\omega} F_{\mathbf{k}}(i\omega)F_{\mathbf{k}+\mathbf{q}}(i\omega+i\Omega)\Lambda(i\omega,i\Omega)  \\ \nonumber
 = & \sum_{\mathbf{k},i\omega}F_{\mathbf{k}+\mathbf{q}}(i\omega+i\Omega)F_{\mathbf{k}}(i\omega)\Lambda^*(i\omega,i\Omega) \\ \nonumber
 = & \sum_{\mathbf{k},i\omega}F_{\mathbf{k}+\mathbf{q}}(i\omega+i\Omega)F_{\mathbf{k}}(i\omega)\mathrm{Re}\Lambda(i\omega,i\Omega) \\ \nonumber
\end{align}
which is what we use in the derivation of the second term in Eq.~\eqref{eq:P_Nambu}.

\subsection{Fourier transforms: Hedin equations with translational symmetry} \label{app:translational}
Here, we derive Eq.~\eqref{eq:Lambda_translational}. A completely analogous derivation can be used for Eqs.~\eqref{eq:Nambu_Hedin}.

For the sake of clarity, we omit the spatial indices, as the spatial Fourier transform (FT) is completely analogous to the temporal FT.
\begin{eqnarray}
 \boldsymbol{\Lambda}_{uv\alpha} 
 &=& [\boldsymbol{G}^{-1}]_{uw}[\boldsymbol{G}^{-1}]_{xv} [W^{-1}]_{\alpha\beta} \boldsymbol{\chi}^{3,\mathrm{conn}}_{wx\beta} \\ \nonumber
 &=& \sum_{\omega,\omega',\omega'',\Omega,\Omega'}
 e^{i\omega(\tau_u-\tau_w)}[\boldsymbol{G}^{-1}(i\omega)]_{a_ua_w} \\ \nonumber
 && \times e^{i\omega'(\tau_x-\tau_v)}[\boldsymbol{G}^{-1}(i\omega')]_{a_xa_v}
 e^{i\Omega(\tau_\alpha-\tau_\beta)}\big(W^{I_\alpha}(i\Omega)\big)^{-1}  \\ \nonumber
 && \times e^{i\omega''(\tau_w-\tau_x)+i\Omega'(\tau_\beta-\tau_x)}\boldsymbol{\chi}^{3,\mathrm{conn},I_\alpha}_{a_wa_x}(i\omega'',i\Omega') \\ \nonumber
 &=& \sum_{\omega,\omega',\omega'',\Omega,\Omega'} e^{i\omega\tau_u-i\omega'\tau_v+i\Omega\tau_\alpha} \\ \nonumber
 && \times e^{i\tau_x(\omega'-\omega''-\Omega')} e^{i\tau_w(\omega''-\omega)} e^{i\tau_\beta(\Omega'-\Omega)} \\ \nonumber
 && \times [\boldsymbol{G}^{-1}(i\omega)]_{a_ua_w}[\boldsymbol{G}^{-1}(i\omega')]_{a_xa_v}
 \big(W^{I_\alpha}(i\Omega)\big)^{-1} \\ \nonumber
 && \times  \boldsymbol{\chi}^{3,\mathrm{conn},I_\alpha}_{a_wa_x}(i\omega'',i\Omega')
\end{eqnarray}
Applying the (implicit) integration over times produces Kronecker delta functions at $\omega''=\omega$, $\omega'=\omega+\Omega$ and $\Omega=\Omega'$. Therefore
\begin{eqnarray}
&&\sum_{\omega\Omega} e^{i\omega(\tau_u-\tau_v)+i\Omega(\tau_\alpha-\tau_v)} \boldsymbol{\Lambda}^{I_\alpha}_{a_ua_v}(\omega,\Omega) \\ \nonumber 
 &=& \sum_{\omega\Omega} e^{i\omega(\tau_u-\tau_v)+i\Omega(\tau_\alpha-\tau_v)} \\ \nonumber 
 && \times [\boldsymbol{G}^{-1}(i\omega'+\Omega)]_{a_ua_w}[\boldsymbol{G}^{-1}(i\omega')]_{a_xa_v}
 \big(W^{I_\alpha}(i\Omega)\big)^{-1} \\ \nonumber
 && \times  \boldsymbol{\chi}^{3,\mathrm{conn},I_\alpha}_{a_wa_x}(i\omega,i\Omega)
\end{eqnarray}
We now reinstate the momentum indices, and obtain Eq.~\eqref{eq:Lambda_translational} by identifying the summands on both sides of the equation
\begin{eqnarray}
\boldsymbol{\Lambda}^{I}_{\mathbf{kq},ab}(i\omega,i\Omega) &=&
  [\boldsymbol{G}_{\mathbf{k}+\mathbf{q}}^{-1}(i\omega+\Omega)]_{ac}[\boldsymbol{G}_\mathbf{k}^{-1}(i\omega)]_{db} \\ \nonumber 
 && \times \big(W^{I}_{\mathbf{q}}(i\Omega)\big)^{-1} \boldsymbol{\chi}^{3,\mathrm{conn},I}_{\mathbf{kq},cd}(i\omega,i\Omega)
\end{eqnarray}
Here, summation over $c,d$ is implicit.

\subsection{$\boldsymbol{\Lambda}_\mathrm{imp}$ from $\Lambda_\mathrm{imp}$} \label{app:Lambda_imp}

Here we prove Eq.~\eqref{eq:boldLambda_imp}. In the Hubbard model we'll have
\begin{equation} \label{eq:avg_Psi_lambda_Psi}
 \sum_{yz} \bPsi_y \boldsymbol{\lambda}_{yz\beta} \bPsi_z = 2 n^{I_\beta}_{i_\beta}(\tau_\beta)  
\end{equation}
On the impurity \eqref{eq:imp_eb_action_Psi}, where we have no anomalous components
\begin{widetext}
\begin{equation}
 \boldsymbol{\chi}^{3,I}_\mathrm{imp}(\tau,\tau') = \int_{\tau''} {\cal U}^I(\tau'-\tau'') \left[
 \begin{array}{cccc}
   &\langle c^*_\up(\tau)  c_\dn(0) n^I(\tau'') \rangle&& \langle c^*_\up(\tau)  c_\up(0) n^I(\tau'') \rangle \\
   \langle c_\dn(\tau)  c^*_\up(0) n^I(\tau'') \rangle&&\langle c_\dn(\tau)  c^*_\dn(0) n^I(\tau'') \rangle& \\
   &\langle c^*_\dn(\tau)  c_\dn(0) n^I(\tau'') \rangle&& \langle c^*_\dn(\tau)  c_\up(0) n^I(\tau'') \rangle \\
   \langle c_\up(\tau)  c^*_\up(0) n^I(\tau'') \rangle&&\langle c_\up(\tau)  c^*_\dn(0) n^I(\tau'') \rangle&
 \end{array}
 \right]
\end{equation}
\end{widetext}
The $\frac{1}{2}$ prefactor in \eqref{eq:chi3_chi3tilde_rel} cancels the prefactor $2$ in \eqref{eq:avg_Psi_lambda_Psi}.
If we define 
\begin{eqnarray} \nonumber
\tilde{\chi}^{3,I=0,z}_\mathrm{imp}(\tau,\tau') \equiv \langle c_\up(\tau)  c^*_\up(0) n^I(\tau') \rangle
 = \frac{1}{2}
\tilde{\boldsymbol{\chi}}^{3,I}_{\mathrm{imp},30}(\tau,\tau')
\end{eqnarray}
\begin{eqnarray} \nonumber
\tilde{\chi}^{3,I=x,y}_\mathrm{imp}(\tau,\tau') \equiv \langle c_\up(\tau)  c^*_\dn(0) n^I(\tau') \rangle 
 = \frac{1}{2}
\tilde{\boldsymbol{\chi}}^{3,I}_{\mathrm{imp},32}(\tau,\tau')
\end{eqnarray}
we can rewrite

\begin{subequations}
\begin{align} \nonumber
  & \boldsymbol{\chi}^{3,I=0,z}_\mathrm{imp}(i\omega,i\Omega) = {\cal U}^I(i\Omega) \times \\
& \;\;\; \times \left[
\begin{array}{cccc}
  &&& \big(-\tilde{\chi}^{3,I}_\mathrm{imp}\big)^* \\
  &&\pm\tilde{\chi}^{3,I}_\mathrm{imp}& \\ 
  & \pm\big(-\tilde{\chi}^{3,I}_\mathrm{imp}\big)^*&& \\
   \tilde{\chi}^{3,I}_\mathrm{imp}&&&   
\end{array}
\right](i\omega,i\Omega)
\label{eq:boldchi3imp_0z}   
\end{align}
\begin{align} \nonumber
  & \boldsymbol{\chi}^{3,I=x,y}_\mathrm{imp}(i\omega,i\Omega) = (-i)^{\delta_{I,y}}{\cal U}^I(i\Omega) \times\\ 
  & \;\;\; \times \left[
\begin{array}{cccc}
  &\pm\big(-\tilde{\chi}^{3,I}_\mathrm{imp}\big)^*&&  \\
  \pm\tilde{\chi}^{3,I}_\mathrm{imp}&&& \\ 
  & && \big(-\tilde{\chi}^{3,I}_\mathrm{imp}\big)^* \\
  &&\tilde{\chi}^{3,I}_\mathrm{imp}&
\end{array}
\right](i\omega,i\Omega)
 \label{eq:boldchi3imp_xy}
\end{align}
\end{subequations}

More compactly
\begin{align}
  & \boldsymbol{\chi}^{3,I}_\mathrm{imp}(i\omega,i\Omega)=\mathcal{U}^{I}(i\Omega)\times\nonumber\\
  & \;\;\; \times (\boldsymbol{\lambda}^I)^\mathsf{T} \circ \left[\begin{array}{cccc}
       & (\tilde{\chi}_{\mathrm{imp}}^{3,I})^{*} &  & (\tilde{\chi}_{\mathrm{imp}}^{3,I})^{*}\\
       \tilde{\chi}_{\mathrm{imp}}^{3,I} &  & \tilde{\chi}_{\mathrm{imp}}^{3,I}\\
        & (\tilde{\chi}_{\mathrm{imp}}^{3,I})^{*} &  & (\tilde{\chi}_{\mathrm{imp}}^{3,I})^{*}\\
        \tilde{\chi}_{\mathrm{imp}}^{3,I} &  & \tilde{\chi}_{\mathrm{imp}}^{3,I}
    \end{array}\right](i\omega,i\Omega) \label{chi3_bold_vs_chitilde3}
\end{align}
% \begin{align}
%  & \boldsymbol{\chi}^{3,I}(i\omega,i\Omega)=\frac{1}{2}\mathcal{U}^{I}(i\Omega)\times\nonumber\\
%   & \;\;\left[\begin{array}{cccc}
%        & \sigma_{\uparrow\downarrow}^{I}(-\tilde{\chi}_{\mathrm{imp}}^{3,I})^{*} &  & \sigma_{\uparrow\uparrow}^{I}(-\tilde{\chi}_{\mathrm{imp}}^{3,I})^{*}\\
%        \sigma_{\uparrow\downarrow}^{I}\tilde{\chi}_{\mathrm{imp}}^{3,I} &  & \sigma_{\downarrow\downarrow}^{I}\tilde{\chi}_{\mathrm{imp}}^{3,I}\\
%         & \sigma_{\downarrow\downarrow}^{I}(-\tilde{\chi}_{\mathrm{imp}}^{3,I})^{*} &  & \sigma_{\downarrow\uparrow}^{I}(-\tilde{\chi}_{\mathrm{imp}}^{3,I})^{*}\\
%         \sigma_{\uparrow\uparrow}^{I}\tilde{\chi}_{\mathrm{imp}}^{3,I} &  & \sigma_{\downarrow\uparrow}^{I}\tilde{\chi}_{\mathrm{imp}}^{3,I}
%     \end{array}\right](i\omega,i\Omega) \label{chi3_bold_vs_chitilde3}
% \end{align}
where $\boldsymbol{\lambda}^I$ and $\circ$ have been defined in main text. For $I=0,z$, we have used
\begin{align}
 &\int_{\tau,\tau',\tau''} e^{i\omega(\tau-\tau')+i\Omega(\tau''-\tau')} \langle c^*_\up(\tau)  c_\up(\tau') n^I(\tau'') \rangle \\ \nonumber
 &=-\int_{\tau,\tau',\tau''} e^{i\omega(\tau-\tau')+i\Omega(\tau''-\tau')} \langle c_\up(\tau') c^*_\up(\tau) n^I(\tau'') \rangle \\ \nonumber
 &=-\int_{\tau,\tau',\tau''} e^{-i\omega(\tau'-\tau)+i\Omega(\tau''-\tau+\tau-\tau')} \langle c_\up(\tau') c^*_\up(\tau) n^I(\tau'') \rangle \\ \nonumber
 &=-\int_{\tau,\tau',\tau''} e^{-i(\omega+\Omega)(\tau'-\tau)+i\Omega(\tau''-\tau)} \langle c_\up(\tau') c^*_\up(\tau) n^I(\tau'') \rangle \\ \nonumber
 &=-\tilde{\chi}^{3,I}_\mathrm{imp}(-i\omega-i\Omega,i\Omega) \\ \nonumber
 &=-\tilde{\chi}^{3,I}_\mathrm{imp}(-i\omega,-i\Omega) \\ \nonumber
 &=-(\tilde{\chi}^{3,I}_\mathrm{imp}(i\omega,i\Omega))^*
\end{align}
and
\begin{align}
  &\Big\langle c^*_\up(\tau) c_\up(0) \big(n_\up(\tau') \pm n_\dn(\tau')\big) \Big\rangle \\ \nonumber
  &= \pm\Big\langle c^*_\dn(\tau) c_\dn(0) \big(n_\up(\tau') \pm n_\dn(\tau')\big) \Big\rangle
\end{align}
and similar considerations for $I=x,y$.
Expressions completely analogous to \eqref{eq:boldchi3imp_0z} and \eqref{eq:boldchi3imp_xy} hold for the connected part of $\boldsymbol{\chi}^3$.
Plugging these in Eq.~\eqref{eq:Lambda_translational} together with Eq.~\eqref{eq:chi3_chi3tilde_rel},

\begin{equation}
\boldsymbol{G}^{-1}_\mathrm{imp}(i\omega) =
\left[ \begin{array}{cccc}
  & & & G_\mathrm{imp}^{-1} \\
 &&-\big(G^{-1}_\mathrm{imp}\big)^{*}&  \\
& G^{-1}_\mathrm{imp}& &\\
-\big(G^{-1}_\mathrm{imp}\big)^{*}&& & \\
\end{array}\right](i\omega) \\ \nonumber
\end{equation}
and 
\begin{align} \label{eq:su2_symm_for_chi3}
  &\Big\langle c_\up(\tau) c^*_\up(0) (n_\up(\tau') - n_\dn(\tau')) \Big\rangle \\ \nonumber
  &= \Big\langle c_\up(\tau) c^*_\dn(0) c^*_\up(\tau') c_\dn(\tau') \Big\rangle
\end{align}
immediately yields Eq.~\eqref{eq:boldLambda_imp}. Eq.~\eqref{eq:su2_symm_for_chi3} holds in presence of SU(2) symmetry. It can be proven by applying a $\pi/2$ rotation around the $y$ axis ($n^z \rightarrow -n^x$,$n^x \rightarrow n^z$, $n^y \rightarrow n^y$), i.e $c_\sigma \rightarrow [\exp(-\frac{i}{2}\frac{\pi}{2}\sigma^y)]_{\sigma,\sigma'} c_{\sigma'} = \frac{1}{\sqrt{2}}(c_{\sigma} +(-)^{\delta_{\sigma,\up}} c_{\bar{\sigma}})$:
\begin{align}
  &\Big\langle c_\up(\tau) c^*_\up(0) (n_\up(\tau') - n_\dn(\tau')) \Big\rangle \\ \nonumber
  &= \frac{1}{2}\Big\langle (c_\up(\tau)-c_\dn(\tau)) (c^*_\up(0)-c^*_\dn(0)) \\ \nonumber
  & \;\;\;\;\;\;   \times  (-c^*_\up(\tau') c_\dn(\tau')-c^*_\dn(\tau') c_\up(\tau')) \Big\rangle \\ \nonumber
  &= \frac{1}{2}\Bigg[\Big\langle (-c_\dn(\tau)c^*_\up(0))(-c^*_\dn(\tau') c_\up(\tau'))\Big\rangle \\ \nonumber
  &\;\;\;\;\;\;+ \Big\langle(-c_\up(\tau)c^*_\dn(0)) (-c^*_\up(\tau') c_\dn(\tau')) \Big\rangle \Bigg]   
\end{align}
and then rotating the operators of the first term on the r.h.s. by $\pi$ around the $y$ axis ($c_\sigma \rightarrow [\exp(-\frac{i}{2}\pi\sigma^y)]_{\sigma,\sigma'} c_{\sigma'} = (-)^{\delta_{\up,\sigma}}c_{\bar{\sigma}}$).

\begin{figure}[!ht]
\centering
  \includegraphics[width=3.2in, trim=1.0cm 0.0cm 0.0cm 0.0cm, page=1]{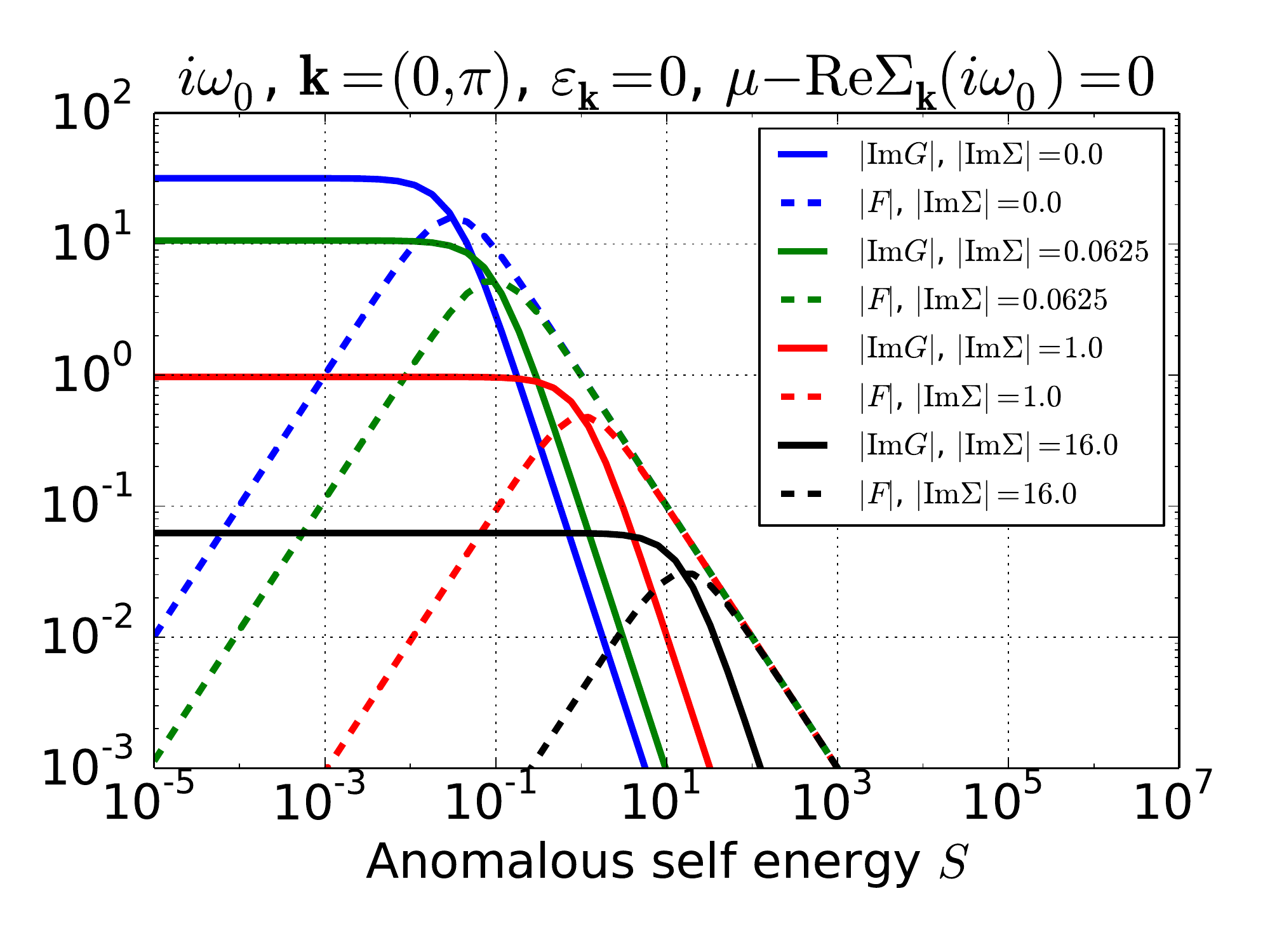}
\caption{ The anomalous Green's function (or pairing amplitude $F$) and the normal Green's function $G$ as functions of the anomalous self-energy $S$ at various values of fixed normal self-energy $\Sigma$. All quantities are taken at the lowest Matsubara frequency $i\omega_0$, at the anti-nodal wave-vector $\mathbf{k}=(0,\pi)$, assuming particle-hole symmetry ( $\epsilon_{\mathbf{k}=(0,\pi)}=0$ and $\mu - \mathrm{Re}\Sigma_{\mathbf{k}=(0,\pi)}(i\omega_n) = 0$). The anti-node in this case is precisely at the Fermi-surface.} 
 \label{fig:F_vs_Xi}
\end{figure}

\subsection{Relation between $S$, $F$, $\Sigma$ and $G$}
\label{app:formalism}
Here we emphasize that the order of magnitude of the anomalous self-energy $S$ and that of the pairing amplitude $F$ are not the same, as illustrated on Fig.~\ref{fig:F_vs_Xi}. The pairing amplitude has a strongly non-monotonous dependence on the anomalous self-energy. At a given normal self-energy, there is a ``sweet spot'' where a small anomalous self-energy produces a very strong superconducting pairing. As soon as the anomalous self-energy starts gapping out the Green's function, this affects also the pairing amplitude as no pairing is possible in the absence of long-lived quasi-particles. In general, strong superconducting gap and normal self-energy diminish both the Green's function and the pairing amplitude.

%%%%%%%%%%%%%%%%%%%%%%%%%%%%%%%%%%%%%%%%%%%%%%%%%%%%%%%%%%%%%%%%%%%%5
%%%%%%%%%%%%%%%%%%%%%%%%%%%%%%%%%%%%%%%%%%%%%%%%%%%%%%%%%%%%%%%%%%%%5
%%%%%%%%%%%%%%%%%%%%%%%%%%%%%%%%%%%%%%%%%%%%%%%%%%%%%%%%%%%%%%%%%%%%5
%%%%%%%%%%%%%%%%%%%%%%%%%%%%%%%%%%%%%%%%%%%%%%%%%%%%%%%%%%%%%%%%%%%%5
%%%%%%%%%%%%%%%%%%%%%%%%%%%%%%%%%%%%%%%%%%%%%%%%%%%%%%%%%%%%%%%%%%%%5
%%%%%%%%%%%%%%%%%%%%%%%%%%%%%%%%%%%%%%%%%%%%%%%%%%%%%%%%%%%%%%%%%%%%5
%%%%%%%%%%%%%%%%%%%%%%%%%%%%%%%%%%%%%%%%%%%%%%%%%%%%%%%%%%%%%%%%%%%%5

\section{Numerical details}

The numerical parameters in our calculations include
\begin{itemize}
\item the number of $\mathbf{k}$-points in the irreducible Brillouin zone, $N_k$; we take it to be temperature dependent, growing as temperature is lowered, to be able to capture increasingly sharp Fermi surface, and gain extra precision when the spin boson is nearly critical.

\begin{tabular}{l*{6}{c}r}
$T$              & $N_k$ \\
\hline
0.06+  & 32  \\
0.03-0.06  & 48  \\
0.005-0.03 & 64  \\
0-0.005     & 96 \\
\end{tabular}

 \item the cutoff frequency $i\omega_{\mathrm{max}}$ for the Green's functions, and the frequency above which the data is replaced by the high-frequency tail fit $i\omega_{\mathrm{fit}}$. Throughout the paper we use $i\omega_{\mathrm{fit}}=14.0$ and $i\omega_{\mathrm{max}}=30$. The actual number of Matsubara frequencies taken is therefore temperature dependent.
 \item the number of $\tau$-points is taken simply as the number of frequencies times 3.
 \item the mixing ratio for the polarization between iterations; in $GW$ we take $P^\mathrm{old}:P^\mathrm{new}=0.95:0.05$. In $GW$+EDMFT and TRILEX, we use $P^\mathrm{old}:P^\mathrm{new}=0.7:0.3$.
 \item number of iterations performed and the level of convergence reached; in $GW$ we start from the non-interacting solution, and perform up to 70 iterations. In the superconducting phase, we perform 150 iterations. In $GW$+EDMFT and TRILEX, we start from DMFT solution at the highest temperature, and then use the $GW$+EDMFT solution as the initial guess at lower temperature, and perform up to 30 iterations. In all cases, we reach convergence level $\mathrm{max}_{i\omega_n} |G^\mathrm{loc,new}(i\omega_n)-G^\mathrm{loc,old}(i\omega_n)|\lesssim10^{-3}$.
 \item the parameter $\gamma$ used in the LEV extrapolation; in $GW$ for Fig.\ref{fig:Tc_vs_Tneel} we use $\gamma=0.5$.
 
\end{itemize}

\section{Extrapolation of the lowest eigenvalue} \label{app:fit}
\begin{figure}[!ht] 
\centering
  \includegraphics[width=3.2in, trim=1.0cm 0.0cm 0.0cm 0.0cm, page=1]{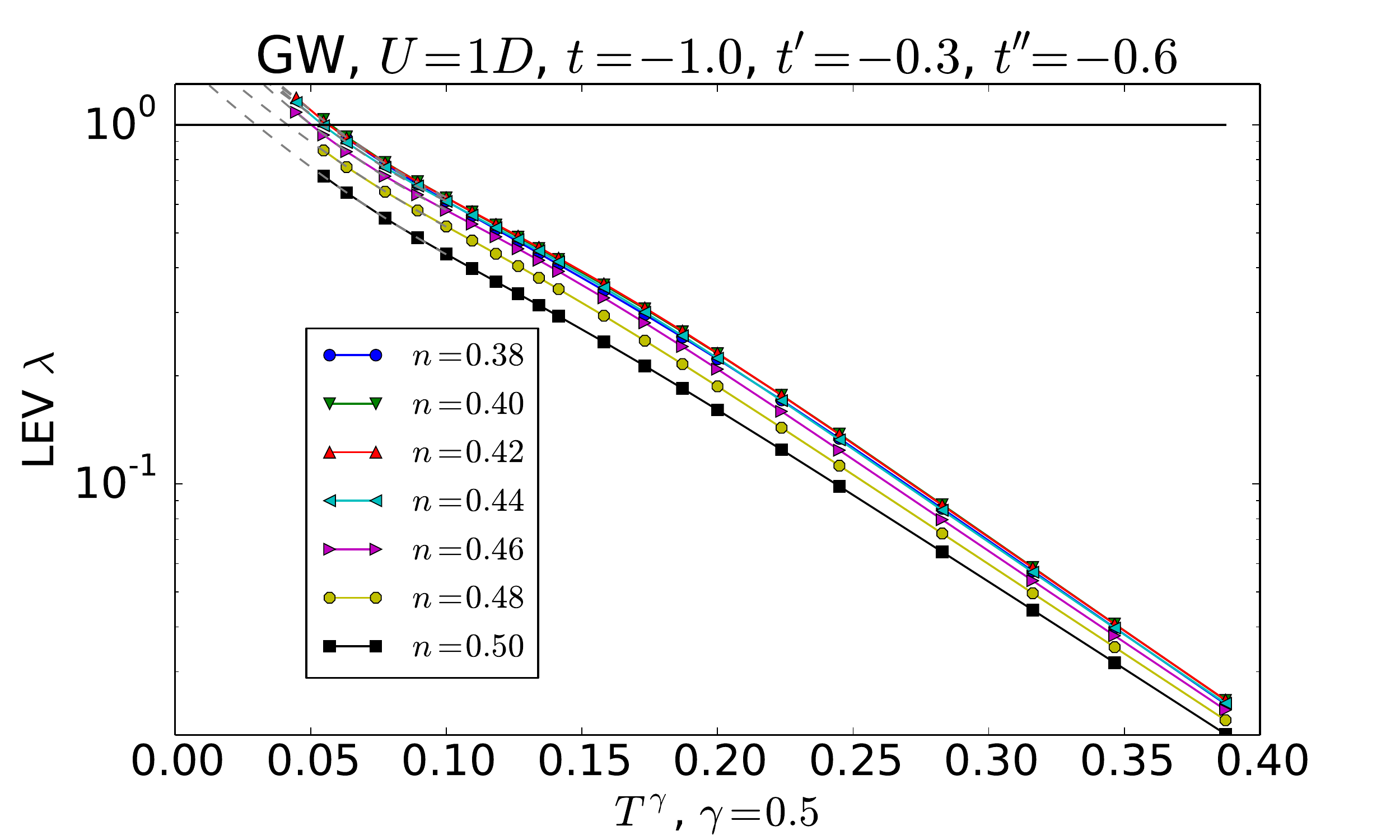}
  \includegraphics[width=3.2in, trim=1.0cm 0.0cm 0.0cm 0.0cm, page=2]{lev_and_maxpu}
  \includegraphics[width=3.2in, trim=1.0cm 0.0cm 0.0cm 0.0cm, page=3]{lev_and_maxpu}
  \includegraphics[width=3.2in, trim=1.0cm 0.0cm 0.0cm 0.0cm, page=4]{lev_and_maxpu}
\caption{ Extrapolation of $\lambda(T)$ (see text). } 
\label{fig:lev_and_maxpu}
\end{figure}

\begin{figure}[!ht]
\centering
  \includegraphics[width=3.2in, trim=1.0cm 0.0cm 0.0cm 0.0cm, page=1]{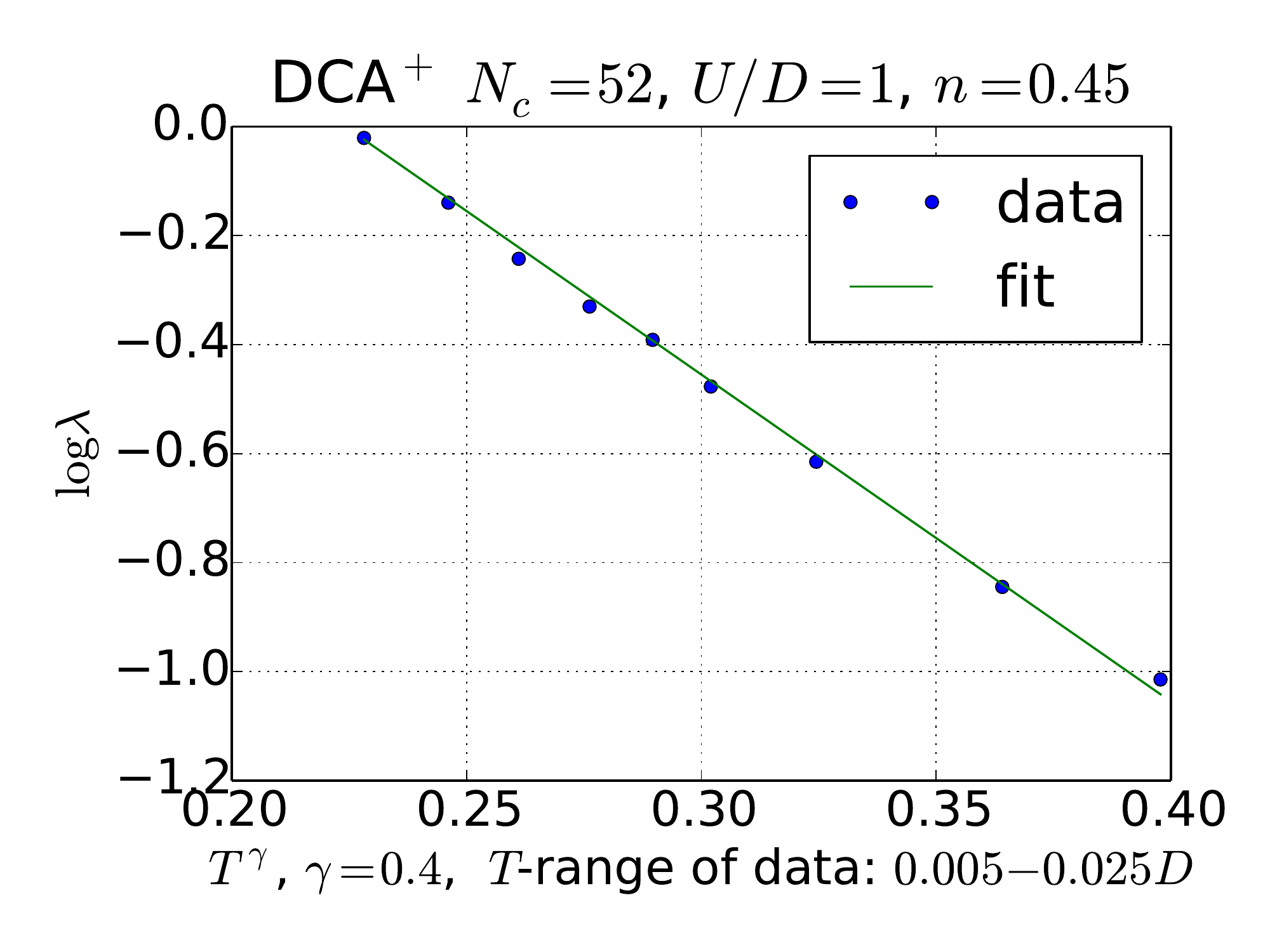}
  \includegraphics[width=3.2in, trim=1.0cm 0.0cm 0.0cm 0.0cm, page=1]{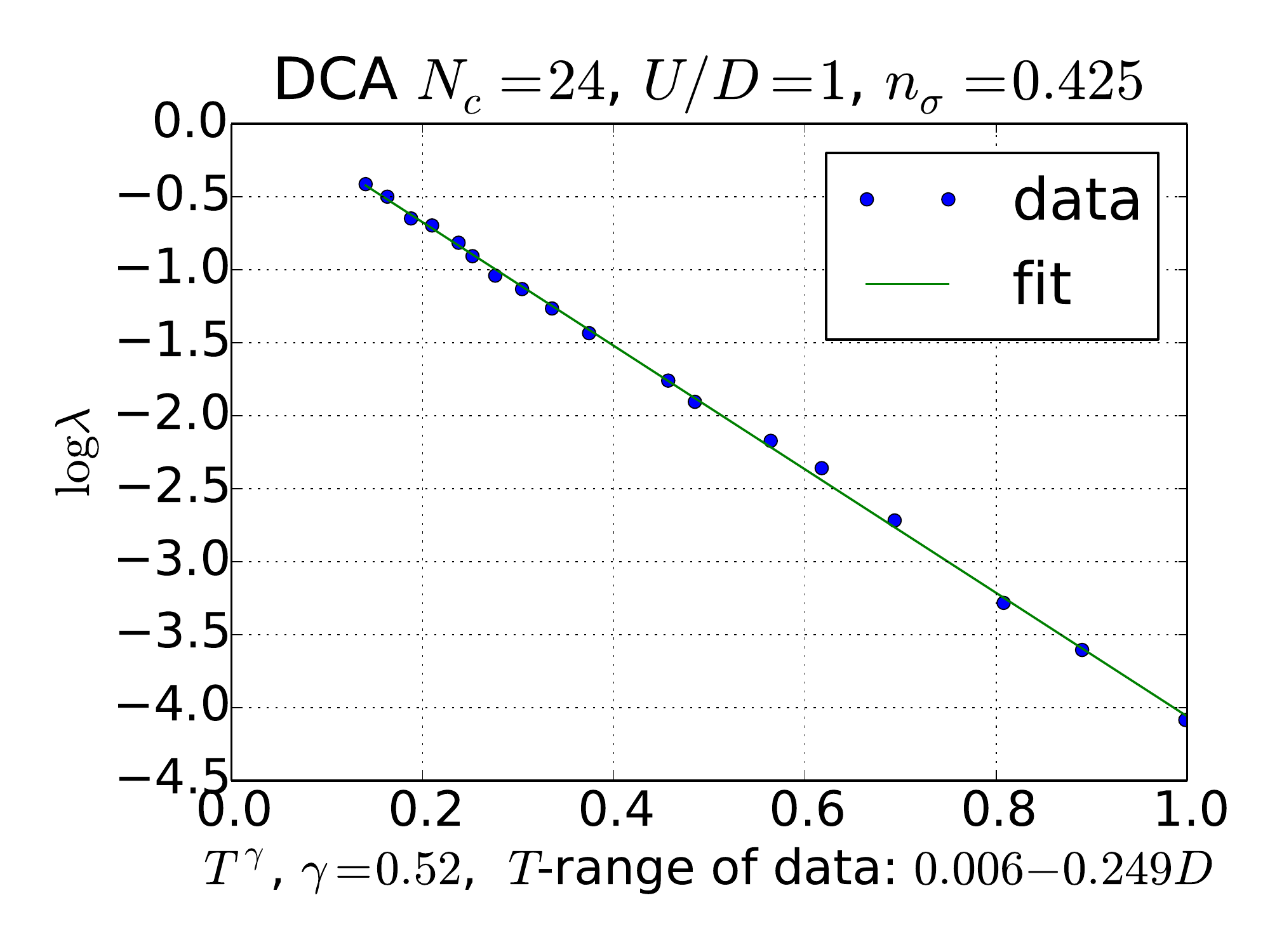}
\caption{ In DCA and DCA$^+$, one observes a behavior very similar to what is seen in $GW$. Data are replotted from Refs.~\onlinecite{MaierPRL2006,MaierPRB2014} and fitted to the phenomenological form Eq.~\eqref{eq:fit_form} with $c=0$. See text for a more detailed discussion. } 
 \label{fig:maier_lev_fits}
\end{figure}

\begin{figure}[!ht] 
\centering
  \includegraphics[width=3.2in, trim=0.0cm 0.0cm 0.0cm 0.0cm, page=2]{trilex_vs_edmftgw_Tc_bayesian.pdf}
  \includegraphics[width=3.2in, trim=0.0cm 0.0cm 0.0cm 0.0cm, page=1]{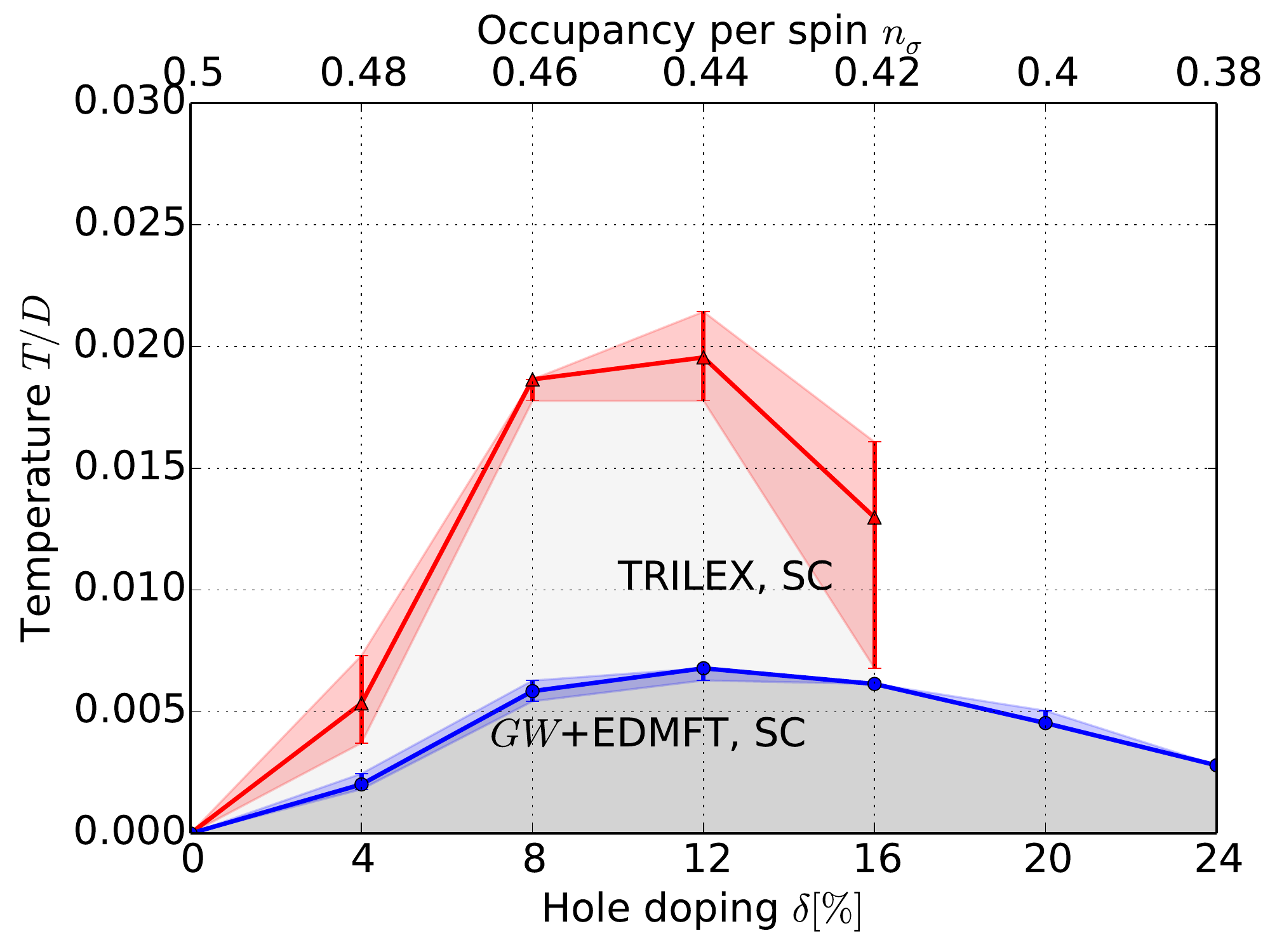}
  \caption{ Error bars determined by standard Bayesian statistics method at a fixed $\gamma=0.45$. } 
\label{fig:bayesian}
\end{figure}

Because of the AF instability in the methods used in the present paper, there is a need for extrapolating the results for the leading eigenvalue (LEV, $\lambda(T)$) in the linearized gap equation (LGE) to lower temperatures. In Fig.~\ref{fig:lev_and_maxpu} we show some examples of this procedure. The $\lambda(T)$ results are contrasted with $\mathrm{max}_{\mathbf{q},i\nu_m} U^\mathrm{sp} P^\mathrm{sp}_\mathbf{q}(i\nu_m)$ which is shown to approach $1$ at finite temperature. Below this temperature, a stable calculation is not possible. For the precise definition of $T_\mathrm{AF}$ shown in figures in \ref{sec:weak_coupling} and \ref{sec:phase_diagram}, we follow Ref.\onlinecite{AokiPRB2015}, and identify it with the condition $\mathrm{max}_{\mathbf{q},i\nu_m} U^\mathrm{sp} P^\mathrm{sp}_\mathbf{q}(i\nu_m) = 0.99$ (this value is denoted with a horizontal black line in the bottom two panels of Fig.~\ref{fig:lev_and_maxpu}).

The LEV $\lambda(T)$ is found to follow a simple law and we perform a parabola fit
\begin{equation}\label{eq:fit_form}
  \log \lambda(T) \approx a + bT^\gamma + cT^{2\gamma} \equiv f(T, \hat{\theta}),
\end{equation}
with $\hat{\theta}=a,b,c$, to extrapolate it to lower temperatures.

Interestingly, a similar $\lambda(T)$ behavior is observed in DCA and DCA$^+$ calculations (see Fig.~\ref{fig:maier_lev_fits}).
The fact that the general temperature-dependent behavior of the LEV (as found in the LGE) is captured correctly with respect to DCA, indicates that the leading contribution to $\Gamma^{pp}_{\sigma\bar\sigma}$, and therefore the superconducting glue, is indeed bosonic-like, dominated by the RPA-like processes. Otherwise, one would expect a slower decay of $\lambda(T)$ with temperature in DCA than observed in $GW$, as here the decay is determined primarily by the gradual decondensation of the spin boson. This notion has been investigated thoroughly in Ref.~\onlinecite{MaierArXiv2015} where the authors have found both the spin-spin correlation and the $pp$-irreducible vertex from a full DCA calculation to be in excellent agreement with simple random-phase approximation estimates. 

In the main text (section~\ref{sec:lge}), we have estimated the error bar on the extrapolation of the lowest eigenvalue by varying the parameter $\gamma$ (see Fig.~\ref{fig:SC_edmftgw_vs_trilex}). Here, we give a method to determine the prediction interval for the extrapolation at fixed $\gamma$. We choose the parameters corresponding to pt.B (Fig.\ref{fig:tpts_phase_diagram_sketch}) to illustrate this method. 

Following standard statistics (see e.g. Ref.~\onlinecite{Ryan1997}, sec. 13.8.1), we proceed as follows: 

(i) for a given doping $n$, we carry out a least-squares fit of the $N$ data points ($T_i$,$\lambda_i$) to Eq.~\eqref{eq:fit_form}: this yields optimal least-square parameters $\hat{\theta}=a^*,b^*,c^*$.

(ii) for a given temperature $T_0$ (not necessarily in the same range as the data points), the prediction interval at $100\cdot(1-\alpha)$\% is given by the two extremal values

$$ f_{\alpha,\pm}(T_0)=f(T_0,\hat{\theta})\pm \overline{\sigma} t_{\alpha/2,N-3}\sqrt{1+v_{0}^{t}\left[V^{t}V\right]^{-1}v_{0}}  $$
where $\overline{\sigma}$ is the empirical variance
$$ \overline{\sigma}=\frac{1}{N-3}\sum_{i=1}^{N}\left(\log \lambda_{i}-f(T_{i},\hat{\theta})\right)^{2},$$
$t_{\alpha,k}$ is defined as 
$$ \int_{t_{\alpha,N}}^{\infty}P_{N}(t)dt =\alpha,  $$
where $P_{N}(t)$ is the probability density function of the Student distribution function.
$V$ is the $N\times 3$ matrix 
$$V_{ij}=\frac{\partial f}{\partial\theta_{j}}\Big|_{T=T_{i}} $$
and $v_{0}$ the column vector:
\begin{figure}[!ht]
  \includegraphics[width=3.0in, trim=0.0cm 0.0cm 0.0cm 0.0cm, page=1]{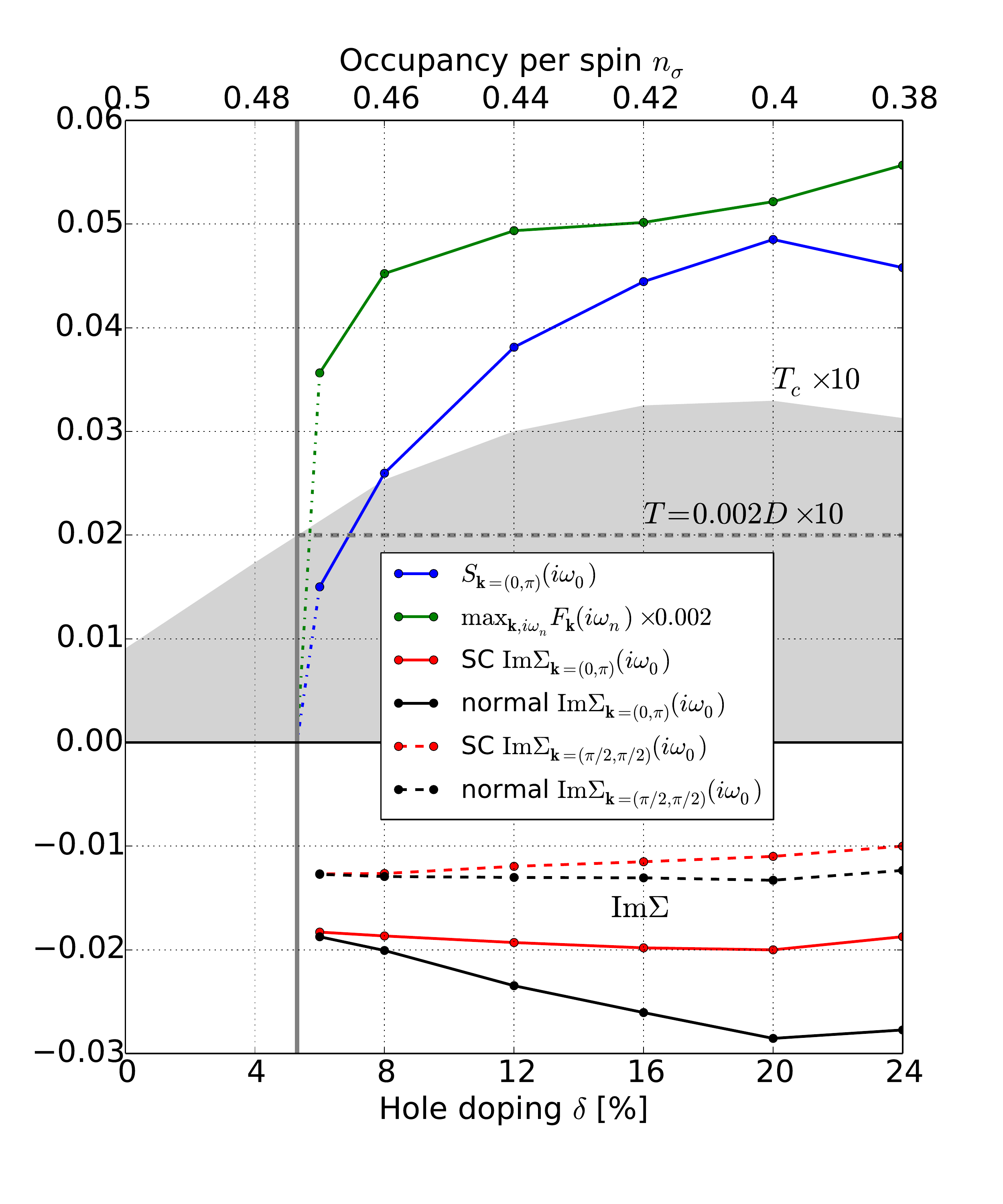}
\caption{Evolution of various
quantities within the superconducting dome at dispersion pt.C.
, $GW$ calculation , $U/D=1$, $T/D=0.002$.
The $T_c$, as obtained from $\lambda_m(T)$, is denoted by the gray area. 
Quantities are scaled to fit the same plot.
The gray dashed horizontal line denotes the temperature at
which the data is taken, relative to the (scaled) $T_c$. The vertical full line
denotes the end of the superconducting dome at the temperature denoted by the
dashed horizontal line, i.e. denotes the doping where all the anomalous
quantities are expected to go to zero.
 } 
\label{fig:weak_coupling_sc_phase_anomalous} 
\end{figure}

$$ v_{0j}=\frac{\partial f}{\partial\theta_{j}}\Big|_{T=T_{0}} $$
The corresponding prediction intervals (at 68\%) are shown in the upper panel of Fig.~\ref{fig:bayesian}. They are used to compute the error bars shown in the lower panel of the same figure.

Especially in $GW$+EDMFT, the fit is found to be of high quality and as the extrapolation is not carried far away from the range of data points, the prediction intervals are found to be small. In TRILEX, the fit is of poorer quality and the prediction intervals are comparable to the uncertainty due to free parameter $\gamma$.

\section{Superconducting phase at weak coupling} \label{app:sc_weak_coupling}

Here, we compare the results of the below-$T_c$ calculation: $GW$ at weak coupling (Fig.\ref{fig:weak_coupling_sc_phase_anomalous}) vs. $GW$+EDMFT at strong coupling (Fig.\ref{fig:sc_phase_anomalous}), at the same dispersion, pt.C. 
We observe that in the weak coupling case, the normal self-energy remains constant with doping, while at strong coupling it grows by a factor of about 5 in a similar range of doping, as Mott insulating phase at half-filling is approached. In the normal phase and at weak-coupling, the self-energy becomes smaller as half-filling is approached, while the trend is the opposite at strong coupling. On the other hand, the onset of the anomalous self-energy in the anti-nodal regions also seems to reduce the normal self-energy in these regions, therefore making the normal-self energy more local. This seems to be a generic feature, not only associated with the doped-Mott insulator regime. It is particularly interesting that the reduction in $\mathrm{Im}\Sigma$ seems proportional to $S$ in both cases.

\bibliography{extracted}
\bibliographystyle{apsrev4-1}

\end{document}